\documentclass[aps,prb,amsmath,amssymb,twocolumn, superscriptaddress, floatfix,longbibliography]{revtex4-2}

\usepackage{graphicx}
\usepackage{overpic}
\usepackage[subrefformat = parens, caption = false, labelformat=parens]{subfig}
\usepackage{adjustbox}
\usepackage{bm}
\usepackage{color}
\usepackage{braket}
\usepackage{standalone}
\usepackage{multirow}
\usepackage{tikz}
\usepackage{mathrsfs}
\usepackage[utf8]{inputenc}
\usepackage{dsfont}
\usepackage[colorlinks,bookmarks=true,citecolor=blue,linkcolor=red,urlcolor=blue]{hyperref}

\newcommand{\eq}[1]{Eq.~(\ref{#1})}

\usepackage{bbm}

\usepackage{floatrow}
\usepackage{verbatim}

\makeatletter
\renewcommand\p@subfigure{\thefigure}
\makeatother

\usepackage[acronym]{glossaries}

\makenoidxglossaries

\newacronym{qhe}{QHE}{quantum Hall effect}
\newacronym{qah}{QAHE}{quantum anomalous Hall effect}
\newacronym{qsh}{QSHE}{quantum spin Hall effect}
\newacronym{tbg}{TBG}{Twisted Bilayer Graphene}   
\newacronym{ab}{AB}{Aharonov-Bohm}
\newacronym{jde}{JDE}{Josephson diode effect}
\newacronym{cpr}{CPR}{Current-Phase Relation}
\newacronym{negf}{NEGF}{Non-Equilibrium Green's Function}
\newacronym{mgf}{MGF}{Matsubara Green's Function}
\newacronym{mtbg}{MTBG}{minimally twisted bilayer graphene}
\newacronym{hwhm}{$\Delta B$}{half-width at half-maximum}
\newacronym[
  longplural={Josephson junctions},
  shortplural={JJs},
  first={Josephson junction (JJ)},
  firstplural={Josephson junctions (JJs)}
]{jj}{JJ}{Josephson junction}
\newacronym[
  longplural={domain walls},
  shortplural={DWs},
  first={domain wall (DW)},
  firstplural={domain walls (DWs)}
]{dw}{DW}{domain wall}
\newacronym[
  longplural={superconducting quantum interference devices},
  shortplural={SQUIDs},
  first={superconducting quantum interference device (SQUID)},
  firstplural={superconducting quantum interference devices (SQUIDs)}
]{squid}{SQUID}{superconducting quantum interference device}

\glsdisablehyper

\begin{document}

\title{Domain Wall Engineering in Graphene-Based Josephson Junctions}
\author{Xia'an Du}
\affiliation{International Center for Quantum Materials, School of Physics, Peking University, Beijing 100871, China}
\author{Junjie Qi}
\affiliation{Beijing Academy of Quantum Information Sciences, Beijing 100193, China}
\author{Hua Jiang}
\email{jianghuaphy@fudan.edu.cn}
\affiliation{Interdisciplinary Center for Theoretical Physics and Information Sciences, Fudan University, Shanghai 200433, China}
\affiliation{State Key Laboratory of Surface Physics, Fudan University, Shanghai 200433, China}
\author{X. C. Xie}
\affiliation{International Center for Quantum Materials, School of Physics, Peking University, Beijing 100871, China}
\affiliation{Interdisciplinary Center for Theoretical Physics and Information Sciences, Fudan University, Shanghai 200433, China}
\affiliation{Hefei National Laboratory, Hefei 230088, China}

\date{\today}

\begin{abstract}
Recent progress has enabled the controlled fabrication of \glspl{dw} in graphene, which host topological kink states. Meanwhile, reliable techniques for constructing graphene-based Josephson junctions have been established. While the experimental prerequisites for combining DWs with Josephson junctions are now available, this direction remains largely unexplored. In this work, we theoretically investigate transport properties in graphene-based Josephson junctions mediated by topological kink states and propose three DW engineering strategies. (i) DW number engineering uncovers a continuous evolution of critical current interference pattern from Aharonov-Bohm oscillation to Fraunhofer diffraction with increasing DW number, reproducing experimental observations [Barrier et al., Nature 628, 741 (2024)] and suggesting enhanced sensitivity for magnetometry applications. (ii) DW symmetry engineering demonstrates that an asymmetric configuration of DWs under magnetic field yields an ideal Josephson diode characterized by pronounced nonreciprocal transport. (iii) DW geometry engineering reveals that intersecting DWs enable controllable supercurrent splitting with ratios among leads tunable through the intersection angle, magnetic field, and superconducting phase difference. Our findings elucidate the rich physics of DW-based Josephson junctions and establish a versatile platform for next-generation quantum devices.
\end{abstract}

\maketitle

\section{Introduction}

\glspl{jj}, whose supercurrent is governed by the superconducting phase, manifest macroscopic quantum coherence and exhibit high sensitivity to external fields and temperature \cite{JJ_B_dependence_1963,michael1996introduction,barone1982magnitude}, rendering them ideal candidates for precision sensing and low-dissipation applications \cite{Pankratov2025Detection}. Two-dimensional (2D) material-based \glspl{jj}, featuring atomic-scale thickness and superior interface quality \cite{calado2015ballistic_graphene_SC_coupling}, offer tunable electronic structures via electric fields, magnetic fields \cite{heersche2007bipolar2Dgatecontrol}, or strain, thereby establishing an ideal platform for quantum interference and novel superconducting research \cite{Ke2019Ballistic,Yuan2021Epitaxial2D,Wu2021Programmable}. Particularly, 2D topological \glspl{jj} have attracted considerable attention \cite{Fatemi2018Electrically}, as dissipationless edge states enhance supercurrent stability and quantum coherence, while their interplay with superconducting pairing facilitates the emergence of Majorana fermions relevant for topological quantum computation \cite{Pribiag2015Edge,Cheng2020Majorana}. Experimental realizations of \glspl{jj} based on \gls{qhe} \cite{Liu2017Superconductor, vignaud2023evidenceQH, Amet2016Supercurrent}, \gls{qah} \cite{uday2024inducedQAH, Anomalous2024Qi, Yan2020Anomalous, Zhao2023Chiral, Chen2018Emergent}, and \gls{qsh} \cite{bocquillon2017gaplessQSH,Pribiag2015} have demonstrated controllable quantum states and transport properties under external field modulation.

Nevertheless, in 2D topological \glspl{jj}, the supercurrent is confined to the edges \cite{hart2014induced}, limiting both the critical current and the tunability of its spatial distribution. More recently, researchers have sought to overcome these limitations by engineering \glspl{dw} to construct topological channels within the bulk material \cite{PhysRevLett.100.036804,Ju2015Topological,Wang2021TopologicalKink,Huang2024QVH}. Minimally twisted bilayer graphene (MTBG) offers a promising platform \cite{Rickhaus2018Transport, Yin2016Direct}. In 2024, Barrier et al. \cite{barrier2024One} constructed MTBG-based 2D topological \glspl{jj} with \gls{dw}s in the central region. These bulk \gls{dw}s, ranging from zero to more than ten per sample, form topological kink states that serve as supercurrent channels inside the bulk. By measuring the critical current response to magnetic fields ($I_{c}$-$B$), they observed exotic current patterns that vary with \gls{dw} number. Nevertheless, a comprehensive theoretical framework for how interference patterns evolve with \gls{dw} number is still lacking, raising the first question addressed in our work.

One major challenge is that MTBG-based topological \glspl{jj} demand high-precision twist-angle control, leading to complex, poorly reproducible fabrication \cite{yoo2019TBGmanufacture}. Moreover, the number and positions of \gls{dw}s are hard to predetermine \cite{Lebedeva2020Energetics, deVries2021GateDefined}. Therefore, alternative \gls{dw}-based approaches for topological \glspl{jj} must be explored. Two strategies have emerged for bilayer and monolayer graphene, respectively. Dual-gate voltages in bilayer graphene allow precise writing and erasing of \gls{dw}s \cite{bilayerchannel2016gate, Lee2017Realisation, Chen2020Gate, Kim2016ValleySymmetry, Hou2020Metallic, Pan2025Topological,Huang2024QVH,Li2018Valley}, yielding micrometer-scale topological kink states that retain ballistic transport \cite{Qiao2011Electronic}. In monolayer graphene, \gls{dw}s arise under a staggered onsite potential on both A and B sublattices \cite{Yao2009Edge,Hunt2013Massive,Woods2014Commensurate, Xiao2007Valley, Zhou2007Substrate, Wang2016Gaps, Hunt2013Massive}, breaking sublattice symmetry \cite{stagsublattice2015role, Gorbachev2014Detecting, He2022Graphene, Song2015Topological, Han2021Accurate}. Specialized engineering further enables \gls{dw} splitting and the formation of interference loops \cite{jiang2018DWloopexperimental, Mania2019Topological, Zhang2022Domino}. The mature \gls{dw} engineering methods and proven graphene-superconductor coupling \cite{vignaud2023evidenceQH,Amet2016Supercurrent,calado2015ballistic_graphene_SC_coupling,barrier2024One,BenShalom2016Quantum,Turini2022Josephson} motivate a second question addressed in this work: although the fabrication of graphene-based \glspl{jj} with DW-induced topological kink states in superconducting devices is now feasible, their transport mechanisms and device applications remain insufficiently understood. 

In this work, we investigate transport properties in graphene-based \glspl{jj} mediated by topological kink states and present three distinct \gls{dw} engineering strategies: (i) 
\gls{dw}-number engineering involves numerical simulations of the critical current response to magnetic field ($I_{c}$--$B$) which reveal a continuous evolution from \gls{ab} pattern to Fraunhofer diffraction with increasing \gls{dw} number, reproducing the observations reported by Barrier et al. \cite{barrier2024One} and suggesting that parallel \gls{dw} arrays enable high-field-resolution magnetometers. (ii) \gls{dw} symmetry engineering breaks inversion symmetry and induces unidirectional critical current under magnetic field \cite{Yuan2022Supercurrent,Davydova2022Universal}, showing potential applications as superconducting diodes for dissipationless rectification in quantum circuits. (iii) \gls{dw} geometry engineering enables controllable supercurrent splitting, with splitting ratios effectively tunable through the intersection angle, external magnetic field, and superconducting phase difference. Collectively, these \gls{dw} engineering strategies provide a versatile platform and new design principles for high-sensitivity magnetometry, programmable superconducting logic devices, and low-dissipation information processing.

The rest of this paper is organized as follows: In Sec.~\ref{sec:model}, we introduce the Hamiltonian of monolayer and bilayer graphene-based \glspl{jj}, and present the methods for calculating the supercurrent. The numerical results for three kinds of \gls{dw} engineering are presented in Sec.~\ref{sec:results}. Finally, a brief summary is given in Sec.~\ref{sec:conclusions}.

\section{Model and Numerical Methods}\label{sec:model}

\begin{figure}[!htb]
\centering
\includegraphics[width=1\columnwidth, trim=150 490 260 120, clip]{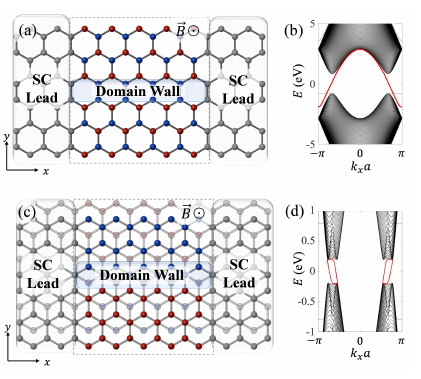}
\vbox to 0pt {
    \raggedright
    \textcolor{white}{
        \subfloatlabel[1][fig:setup_spectrum_a]
        \subfloatlabel[2][fig:setup_spectrum_b]
        \subfloatlabel[3][fig:setup_spectrum_c]
        \subfloatlabel[4][fig:setup_spectrum_d]
    }
}
\caption{Schematic plots of the graphene-based \gls{jj} devices and the corresponding electronic band structure of the central region with a single \gls{dw}. (a) Schematic of monolayer graphene-based \gls{jj} with a single \gls{dw}. $N_x$ and $N_y$ represent the number of sites in the $x$ and $y$ directions. Here, $N_x = 9$ and $N_y = 6$. (b) Energy band structure $E$ for this monolayer graphene. Here, $a$ refers to the lattice constant. (c) Schematic of the bilayer graphene-based \gls{jj} with a single \gls{dw}. Layer 1 (top) and layer 2 (bottom) are distinguished by opacity. Here, $N_x = 9$ and $N_y = 6$. (d) Energy band structure $E$ for this bilayer graphene. The parameters in (b) and (d) are $N_y = 80$ and $V_{\text{stag}} = 0.8$ eV.}
\label{fig:setup_spectrum}
\end{figure}

The \gls{jj} in our study consists of a graphene central region containing \gls{dw}s that host topological kink states, with superconducting leads attached at both ends. The presence of superconductivity necessitates the Nambu representation to account for both electron and hole excitations. The total Hamiltonian is given by
\begin{equation}
H = H_C + \sum_{l=L,R} H_l + H_T,
\end{equation}
\noindent where $H_C$ describes the central region, $H_l$ with $l \in {L, R}$ describes the superconducting lead $l$, and $H_T$ describes the tunneling between the central region and the lead.

We study the transport properties of \glspl{jj} in both monolayer and bilayer graphene systems. The Hamiltonian construction differs in these two cases, and we first introduce the monolayer Hamiltonian:

\begin{equation}
\begin{split}
H_{C}^{\text{mono}} &= \sum_{\langle i,j \rangle} \Psi_i^\dagger \begin{pmatrix} t_{ij} & 0 \\ 0 & -t_{ij}^* \end{pmatrix} \Psi_j + \text{H.c.} \\
&\quad + \sum_i \Psi_i^\dagger \begin{pmatrix} U_i & 0 \\ 0 & -U_i \end{pmatrix} \Psi_i,
\end{split}
\end{equation}

\noindent where $\Psi_i = [c_i, c_i^\dagger]^T$, with $c_i$ ($c_i^\dagger$) denoting the electron annihilation (creation) operator at site $i$ in the central region. The term $t_{ij}$ represents the hopping matrix element between nearest-neighbor sites in graphene. To investigate the $I_{c}$--$B$ response, we apply a perpendicular magnetic field to the device by using the gauge $\mathbf{A} = (-yB, 0, 0)$. This vector potential $\mathbf{A}$ is encoded into the hopping term in the central region through the Peierls substitution $t_{ij} = t e^{i\frac{e}{\hbar} \int \mathbf{A} \cdot d\mathbf{l}}$, where the hopping parameter $t = -2.75$ eV.

$U_i$ denotes the onsite potential containing \gls{dw} structural information. We consider a \gls{jj} device with one \gls{dw}, as illustrated in Fig.~\ref{fig:setup_spectrum_a}. Here, all A (B) sublattices in the central region above the \gls{dw} satisfy $U_i = +(-)V_{\text{stag}}$, while those below the \gls{dw} satisfy $U_i = -(+)V_{\text{stag}}$. The sign change of $U_i$ induces topologically protected kink states along the \gls{dw}. The energy band structure in Fig.~\ref{fig:setup_spectrum_b} reveals that these kink states form conducting channels inside the bulk gap \cite{semenoff2008Domain}. For \glspl{jj} with multiple \gls{dw}s, the $U_i$ configuration alternates between neighboring domains, resulting in conducting channels along each \gls{dw}, as shown by the red curves in Fig.~\ref{fig:setup_spectrum_b}.

For monolayer graphene-based \glspl{jj}, the Hamiltonian of the $s$-wave superconducting lead is given by:

\begin{equation}
\begin{split}
H_{l}^{\text{mono}} &= \sum_{\langle i,j \rangle} \Psi_{il}^\dagger \begin{pmatrix} t & 0 \\ 0 & -t \end{pmatrix} \Psi_{jl} + \text{H.c.} \\
&\quad + \sum_i \Psi_{il}^\dagger \begin{pmatrix} U_S & \Delta e^{i\theta_l} \\ \Delta e^{-i\theta_l} & -U_S \end{pmatrix} \Psi_{il},
\end{split}
\end{equation}

\noindent where $\Psi_{il} = [c_{il}, c_{il}^\dagger]^T$, with $c_{il}$ ($c_{il}^\dagger$) denoting the electron annihilation (creation) operator at site $i$ in superconducting lead $l = L, R$. The parameter $\theta_l$ represents the superconducting phase of lead $l$. For both left and right leads, the superconducting pairing potential is set to $\Delta = 10$ meV, and the onsite potential $U_S$ is typically chosen around $-1.2$ eV, with the chemical potential set to $\mu = 0$. Under this choice of $U_S$ and $\mu$, the leads remain metallic for $\Delta = 0$ and become $s$-wave superconducting for $\Delta \neq 0$. 

The tunneling Hamiltonian $H_T$ describes nearest-neighbor coupling across the contact interfaces:

\begin{equation}
  H_T^{\text{mono}} = \sum_{\substack{\langle il,j \rangle,\, i \in l, \\ j \in \text{central}}} \Psi_{il}^\dagger \begin{pmatrix} t & 0 \\ 0 & -t \end{pmatrix} \Psi_j + \text{H.c.},
\end{equation}

\noindent where the sum runs over all nearest pairs between lead $l$ and the central region.

For Bernal-stacked bilayer graphene, the Hamiltonian of central region includes an interlayer coupling term in addition to $H_C^{\text{mono}}$:

\begin{equation}
\begin{split}
H_{C}^{\text{bi}} &= H_{C}^{\text{mono},1} + H_{C}^{\text{mono},2} \\
&\quad + \sum_{\langle i_1,i_2 \rangle} \Psi_{i_1}^\dagger \begin{pmatrix} t_\perp & 0 \\ 0 & -t_\perp \end{pmatrix} \Psi_{i_2} + \text{H.c.}.
\end{split}
\end{equation}

Here, subscripts 1 and 2 denote the upper and lower graphene monolayers, respectively. $\Psi_{i_{1}} = [c_{i_{1}}, c_{i_{1}}^\dagger]^T$ denotes the Nambu spinor for the central site $i$ in layer 1. The interlayer hopping involves only vertically aligned atoms between the two monolayers, with coupling strength $t_\perp = -0.4$ eV.

To introduce \gls{dw}s in bilayer graphene, the onsite potential $U_i$ is modulated differently from the monolayer case. For a \gls{jj} device with a single \gls{dw}, as shown in Fig.~\ref{fig:setup_spectrum_c}, all sites in layer 1 above (below) the \gls{dw} satisfy $U_i = +(-)V_{\text{stag}}$, while all sites in layer 2 above (below) the \gls{dw} satisfy $U_i = -(+)V_{\text{stag}}$. This configuration induces conducting channels within the bulk gap along this \gls{dw}, as highlighted by the red curves in the band structure shown in Fig.~\ref{fig:setup_spectrum_d}.

Turning to the bilayer case, the Hamiltonian of the superconducting lead $H_l$ ($l = L, R$) in Bernal-stacked bilayer graphene-based \glspl{jj} is constructed from $H_l$ ($l = L, R$) is constructed from $H_{l}^{\text{mono}}$ together with the interlayer coupling, and can be written as

\begin{equation}
\begin{split}
H_{l}^{\text{bi}} &= H_{l}^{\text{mono},1} + H_{l}^{\text{mono},2} \\
&\quad + \sum_{\langle i_1,i_2 \rangle} \Psi_{i_1}^\dagger \begin{pmatrix} t_\perp & 0 \\ 0 & -t_\perp \end{pmatrix} \Psi_{i_2} + \text{H.c.}.
\end{split}
\end{equation}

In the bilayer system, the tunneling Hamiltonian is generalized from the monolayer case to account for the coupling of each graphene layer between the central region and the superconducting lead, and can be written as $H_T^{\text{bi}} = H_T^{\text{mono,1}}+H_T^{\text{mono,2}}$.

We next introduce the numerical methods to calculate the supercurrent. In \glspl{jj}, the supercurrent remains time-independent when the voltage biases of the left and right leads are equal \cite{Perfetto2009Equilibrium}. To compute this supercurrent, we employ the Keldysh Green’s function formalism, in which the current expectation value is expressed as an energy integral \cite{Kamenev2011Field,Furusaki1991Current}. For computational efficiency, this energy integration is replaced by a Matsubara frequency summation \cite{Haug2008Quantum}.

Following the methods in Ref.~\cite{Ryndyk2017Theory,Asano2001Numerical}, the total rightward supercurrent through a transverse cut line perpendicular to the $x$-axis in the central region is given by
\begin{equation}
\label{eq:current_integral}
\begin{split}
I &= \frac{e}{h} \int dE \text{Tr} [\Gamma_z (G_{CC}(E)\Sigma(E))^< \\
&\quad - (\Sigma(E) G_{CC}(E))^< \Gamma_z]. 
\end{split}
\end{equation}
Here, $G_{CC}$ is the Green's function matrix of sites in this transverse cut line, with dimensions $2 N_y \times 2 N_y$, where the factor 2 accounts for both electrons and holes, and $N_y$ represents the number of sites in the $y$ direction. $\Gamma_z = \sigma_z \otimes I_{N_y}$ accounts for the electron-hole difference. $\Sigma$ denotes the effective self-energy to the right of the transverse cut line, obtained by iterating from the right electrode leftward to this line \cite{song2016Quantum}. The number of iteration steps is proportional to $N_x$, the number of sites in the $x$ direction \cite{Golubov2004Current}.

The direct evaluation of \eq{eq:current_integral} is computationally expensive. Fortunately, the expression can be reformulated as a Matsubara frequency summation \cite{Li2022Quantum,Beenakker2024Josephson}
\begin{equation}
  \begin{split}
  I &= \frac{ieT}{\hbar} \sum_n \text{Tr}[\Gamma_z(G_{CC}(i\omega_n)\Sigma(i\omega_n))^< \\
  &\quad - (\Sigma(i\omega_n)G_{CC}(i\omega_n))^< \Gamma_z],
  \end{split}
  \end{equation}

\noindent with Matsubara frequencies defined as $\omega_n = (2n+1)\pi T$, and the temperature fixed at $T = \Delta/10$ with Boltzmann constant $k_B = 1$. This reformulation avoids the costly energy integration and significantly improves the computational efficiency.

\begin{figure*}[!htb]
  \centering
  \subfloat[]{\label{fig:dw_number_bilayer0}
    \begin{overpic}[width=0.31\textwidth,trim=230 150 260 150,clip]{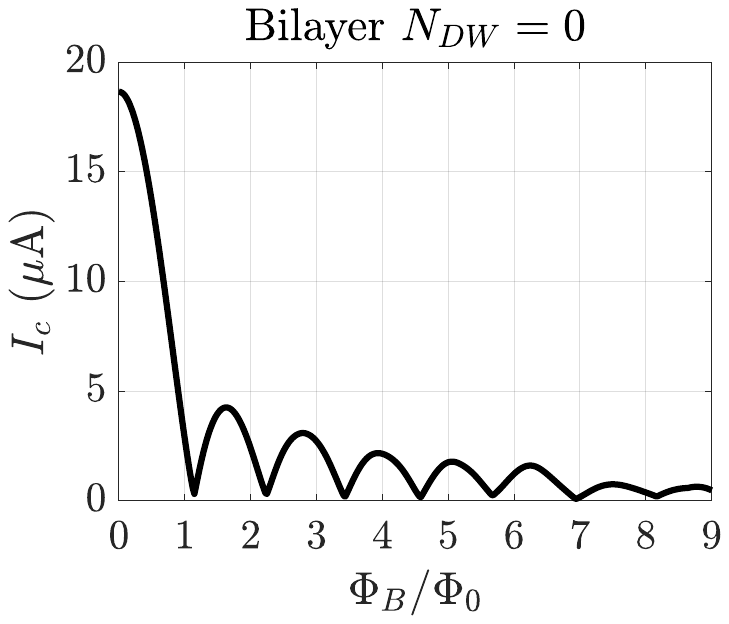}
      \put(89,68){(a)}
    \end{overpic}
  }\hfill
  \subfloat[]{\label{fig:dw_number_bilayer1}
    \begin{overpic}[width=0.31\textwidth,trim=230 150 260 150,clip]{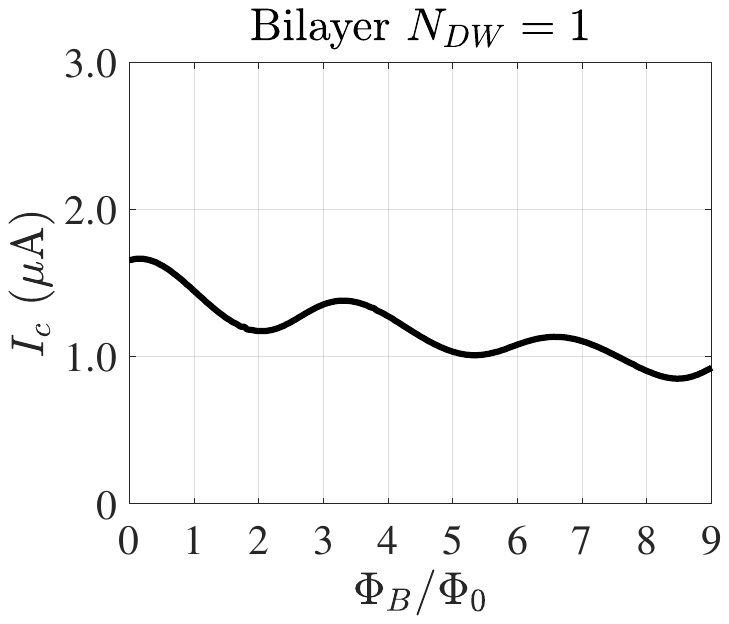}
      \put(89,68){(b)}
    \end{overpic}
  }\hfill
  \subfloat[]{\label{fig:dw_number_bilayer2}
    \begin{overpic}[width=0.31\textwidth,trim=230 150 260 150,clip]{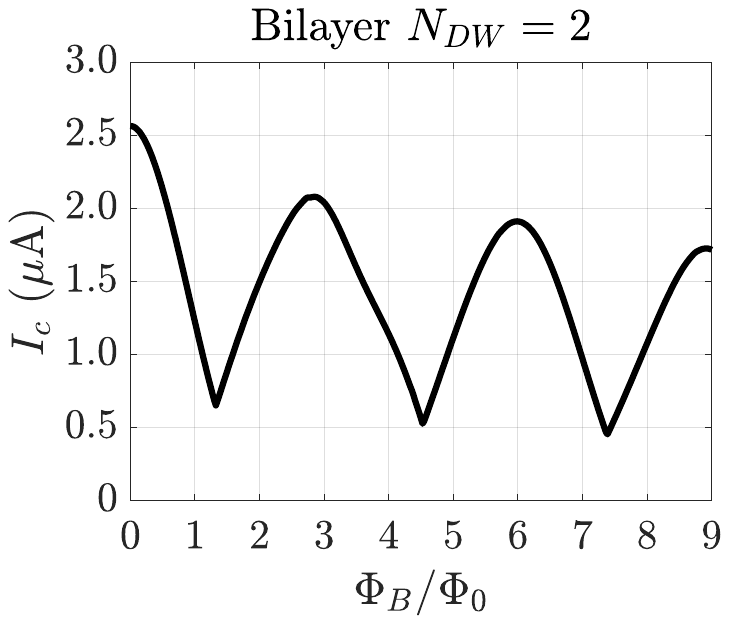}
      \put(89,68){(c)}
    \end{overpic}
  }\\[-1.4em]
  \subfloat[]{\label{fig:dw_number_mono0}
    \begin{overpic}[width=0.31\textwidth,trim=230 150 260 150,clip]{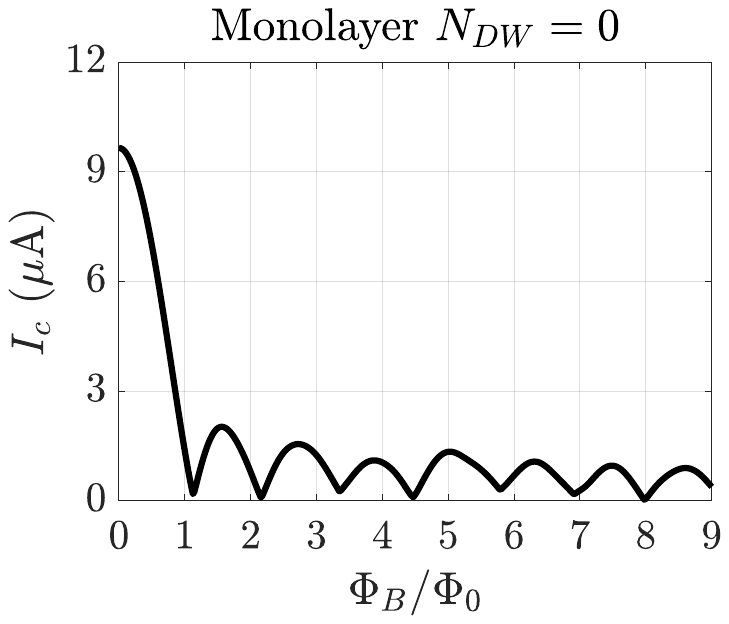}
      \put(89,68){(d)}
    \end{overpic}
  }\hfill
  \subfloat[]{\label{fig:dw_number_mono1}
    \begin{overpic}[width=0.31\textwidth,trim=230 150 260 150,clip]{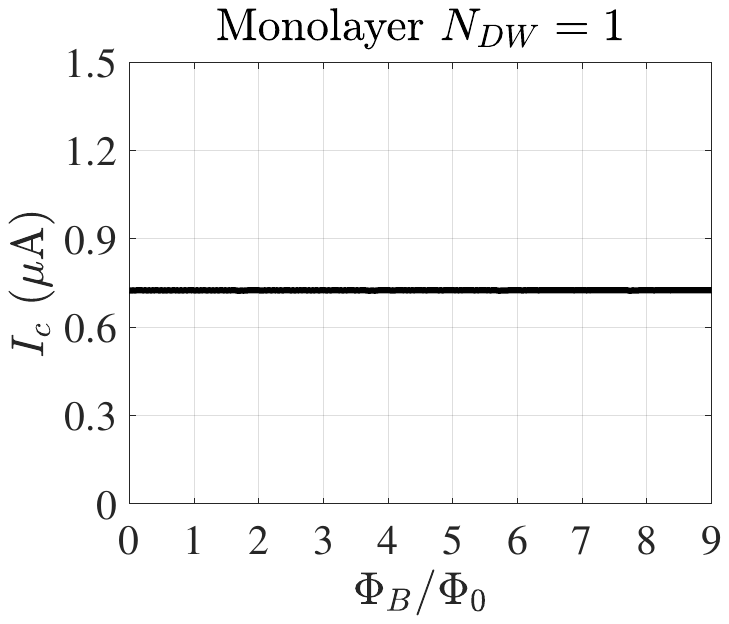}
      \put(89,68){(e)}
    \end{overpic}
  }\hfill
  \subfloat[]{\label{fig:dw_number_mono2}
    \begin{overpic}[width=0.31\textwidth,trim=230 150 260 150,clip]{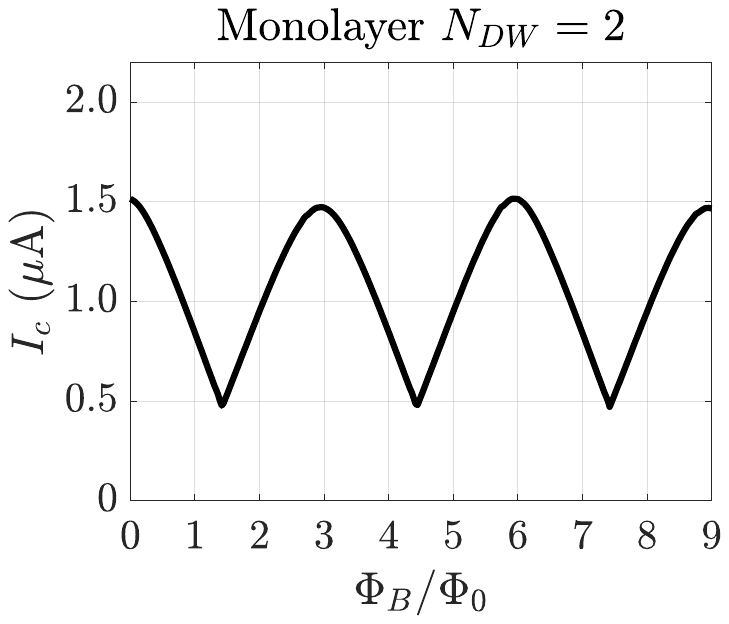}
      \put(89,68){(f)}
    \end{overpic}
  }\\[-1.4em]
  \subfloat[]{\label{fig:dw_number_mono3}
    \begin{overpic}[width=0.31\textwidth,trim=230 150 260 150,clip]{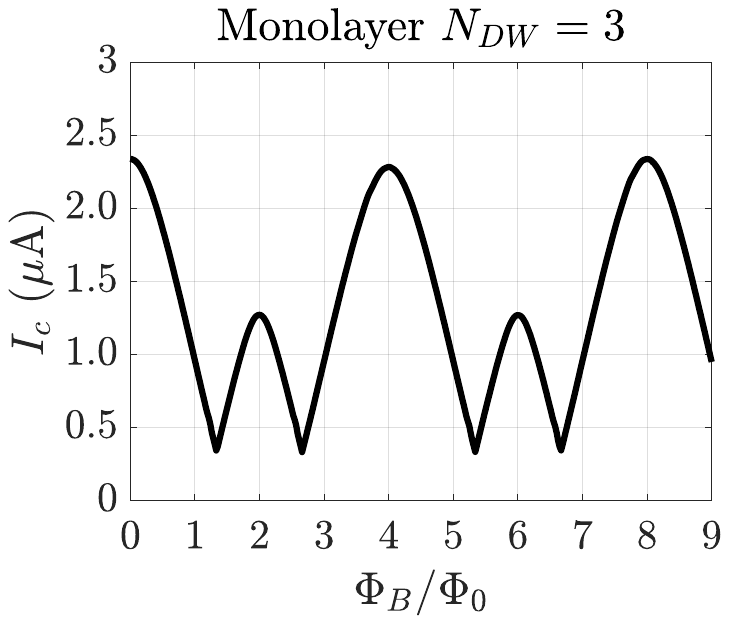}
      \put(89,68){(g)}
    \end{overpic}
  }\hfill
  \subfloat[]{\label{fig:dw_number_mono5}
    \begin{overpic}[width=0.31\textwidth,trim=230 150 260 150,clip]{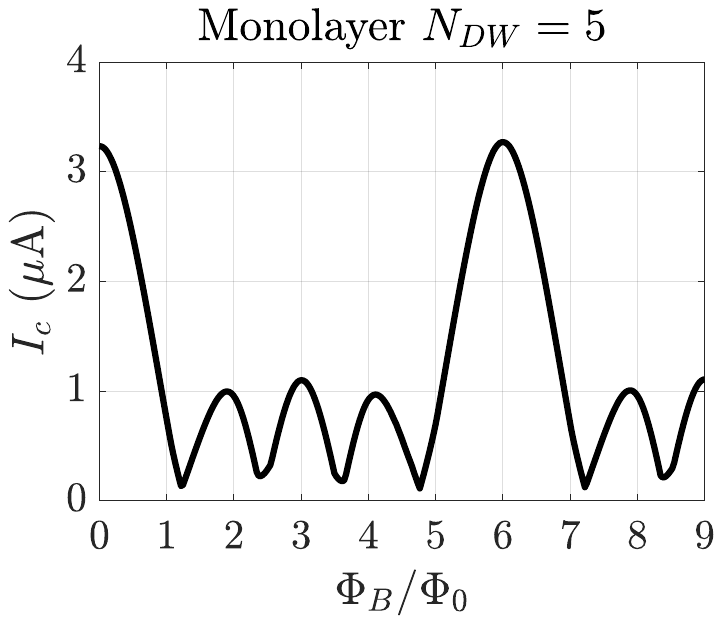}
      \put(89,68){(h)}
    \end{overpic}
  }\hfill
  \subfloat[]{\label{fig:dw_number_mono19}
    \begin{overpic}[width=0.31\textwidth,trim=230 150 260 150,clip]{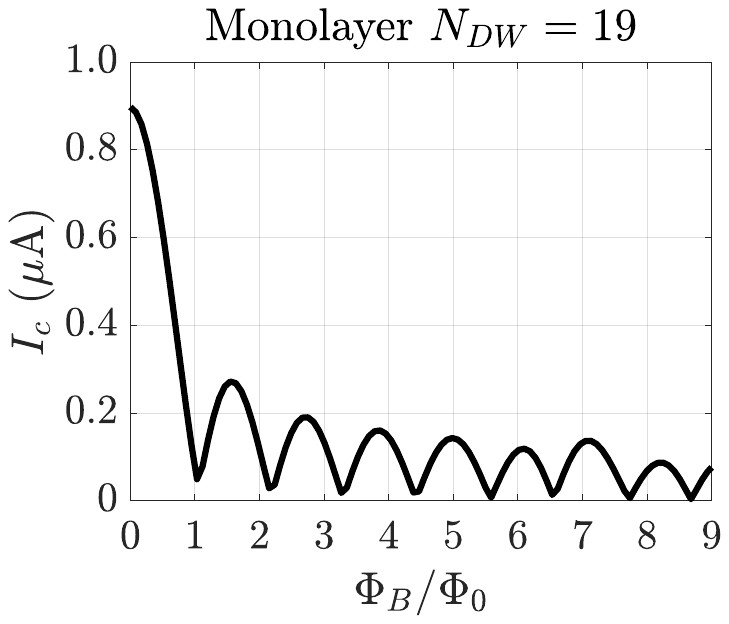}
      \put(89,68){(i)}
    \end{overpic}
  }
  \caption{$I_{c}$--$B$ responses in bilayer and monolayer graphene-based \glspl{jj} for different \gls{dw} numbers. (a)--(c) Bilayer graphene-based \glspl{jj} with 0, 1, and 2 \gls{dw}s constructed by engineering dual-gate voltage. The horizontal axis $\Phi_{B}/\Phi_{0}$ represents the total magnetic flux through the central region. The parameters are $N_{x} = 61$, $N_{y} = 80$ for $N_{\text{DW}} = 0, 1$, and $N_{y} = 84$ for $N_{\text{DW}} = 2$. (d)--(i) Monolayer graphene-based \glspl{jj} with 0, 1, 2, 3, 5, and 19 \gls{dw}s formed by staggered sublattice potentials. The parameters are $N_{x} = 61$, $N_{y} = 80$ for $N_{\text{DW}} = 0, 1, 3, 19$, and $N_{y} = 84$ for $N_{\text{DW}} = 2, 5$. For \glspl{jj} with \gls{dw}s, other parameters are $V_{\text{stag}} = \pm 1$ eV, $U_S = -1.5$ eV. For the \glspl{jj} without \gls{dw}s (Figs.~\ref{fig:dw_number_bilayer0} and \ref{fig:dw_number_mono0}), $V_{\text{stag}} = 0$ eV, $U_S = -1.5$ eV. }

  \label{fig:domain_wall_number_engineering}
  \end{figure*}

\section{Results and Discussion}\label{sec:results}
To explore the potential of \gls{dw}-induced topological kink states in superconducting devices, we investigate three \gls{dw} engineering strategies in graphene-based \glspl{jj}. First, \gls{dw} number engineering involves numerical simulations of $I_c$--$B$ curves, which reveal a continuous evolution from \gls{ab} patterns to Fraunhofer diffraction as DW number increases, consistent with the experimental observations of Barrier et al.~\cite{barrier2024One}. Fourier transform analysis of the simulation data provides additional insight by reconstructing the spatial distribution of supercurrent, confirming that the transport is dominated by topological kink states. A power-law dependence of magnetic field resolution on \gls{dw} number is also established, offering design guidelines for high-sensitivity magnetometers. Second, \gls{dw} symmetry engineering breaks inversion symmetry and generates asymmetric critical current behavior under magnetic field \cite{Yuan2022Supercurrent,Davydova2022Universal}. The resulting forward and backward critical current $I_c$--$B$ curves are periodically offset, yielding superconducting diode efficiency up to $24\%$ in our example, showing potential for dissipationless rectification in quantum circuits. Third, \gls{dw} geometry engineering investigates configurations of \glspl{jj} with single and double crossing points, revealing that intersection angles provide strong control of transmission and reflection supercurrents. For double intersections, magnetic field and superconducting phase differences offer additional tunability of supercurrent splitting among multiple leads.

\begin{figure}[!htb]
  \centering  
  \subfloat[]{\label{fig:fourier_bilayer}
    \begin{overpic}[width=0.95\textwidth,trim=200 150 200 150,clip]{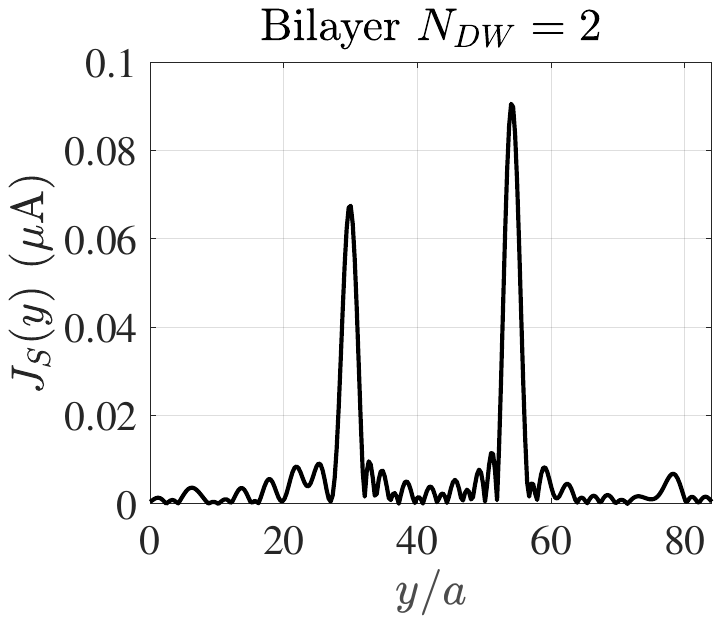}
      \put(76,55){(a)}
    \end{overpic}
  }\hfill
  \subfloat[]{\label{fig:fourier_monolayer}
    \begin{overpic}[width=0.95\textwidth,trim=200 150 200 150,clip]{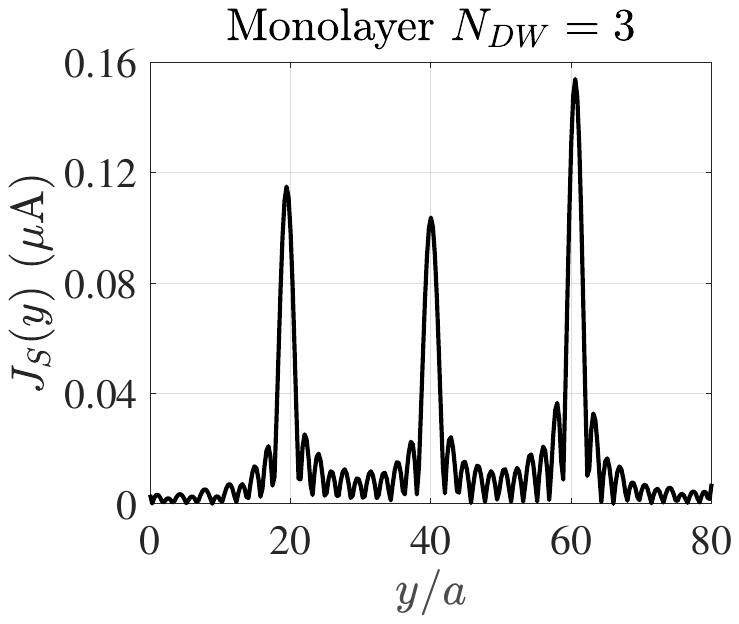}
      \put(76,55){(b)}
    \end{overpic}
  }
  \caption{Spatial distribution of the supercurrent $J_S(y)$ along the $y$-axis in graphene-based \glspl{jj} with \glspl{dw}, derived from Fourier analysis of the $I_c$--$B$ response. (a) $J_S(y)$ for a bilayer graphene-based \gls{jj} with two \gls{dw}s, extracted from Fig.~\ref{fig:dw_number_bilayer2}. (b) $J_S(y)$ for a monolayer graphene-based \gls{jj} with three engineered \gls{dw}s, extracted from Fig.~\ref{fig:dw_number_mono3}. The coordinate $y$ is the site index along the junction width, in units of the lattice constant $a$.}
  \label{fig:fourier_to_realspace}
\end{figure}

\begin{figure}[!htb]
  \centering
  \includegraphics[width=0.95\columnwidth, trim=220 150 250 150, clip]{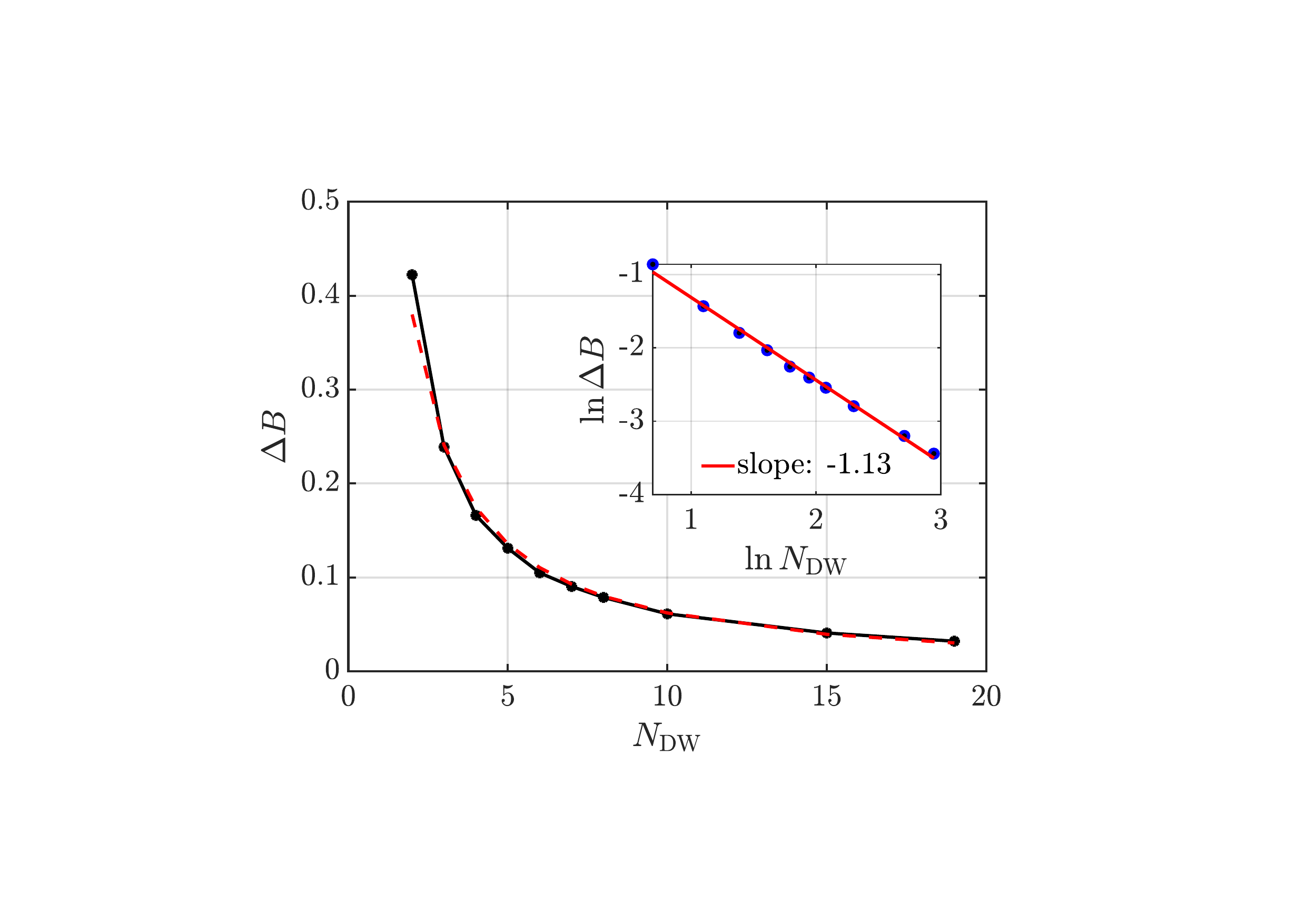}
  \caption{The half-width at half-maximum $\Delta B$ as a function of $N_{\text{DW}}$. To maintain a fixed spacing between adjacent \gls{dw}s, the system sizes are set to $N_x = 61$ and $N_y = 10 \times (N_{\text{DW}} + 1)$. Here, the quantity $\Delta B$ is calculated as the applied magnetic flux between adjacent DWs. The inset illustrates a log--log plot of the same data, where a linear fit yields a slope of $-1.13$. Other parameters are $U_S = -1$ eV and $V_{\text{stag}} = \pm 0.8$ eV. }
  \label{fig:squid_sensitivity}
  \end{figure} 
\subsection{\label{sec:dw_number}DW Number Engineering}

We first investigate the $I_{c}$--$B$ response through \gls{dw} number engineering. Equally spaced parallel \gls{dw}s are constructed in the central region of graphene-based \glspl{jj} through gate-voltage modulation in bilayer graphene \cite{Oostinga2008} or through staggered potential manipulation in monolayer graphene. The origin of \gls{dw}s presented here differs from that in the experiments of Barrier et al. \cite{barrier2024One}, where \gls{dw}s arise from AB/BA stacking domain boundaries induced by twist angles, with their number and locations requiring post-fabrication imaging for precise characterization \cite{deVries2021GateDefined, barrier2024One,Alden2013}. Nevertheless, the underlying physics governing the $I_{c}$--$B$ response remains identical regardless of the specific origin of the \gls{dw}s, as demonstrated below.

Numerical results demonstrate that the \gls{dw} number $N_{\text{DW}}$ significantly influences the $I_{c}$--$B$ response in both bilayer and monolayer graphene-based \glspl{jj}. In the absence of \gls{dw}s, the central region is tuned into a metallic state, rendering the junction bulk-conducting. The system then exhibits a Fraunhofer diffraction pattern in which the characteristic width of the side lobes is approximately one flux quantum $\Phi_{0} = h/2e$, consistent with standard Fraunhofer interference behavior, as shown in Figs.~\ref{fig:dw_number_bilayer0} and \ref{fig:dw_number_mono0}. For \glspl{jj} with two \gls{dw}s, interference between the conducting channels gives rise to sinusoidal oscillations. As shown in Figs.~\ref{fig:dw_number_bilayer2} and \ref{fig:dw_number_mono2}, the oscillation period is approximately $3\Phi_{0}$, as the $x$-axis represents the total magnetic flux through the junction area. Since the effective flux enclosed within the two DWs corresponds to one-third of the total flux, the interference period is then reduced to $\Phi_{0}$, in agreement with theoretical results \cite{Qi2025EdgeSupercurrent}. In contrast, for a single DW, the absence of additional conducting channels suppresses interference, rendering the critical current insensitive to magnetic field \cite{barrier2024One}, as shown in Figs.~\ref{fig:dw_number_bilayer1} and \ref{fig:dw_number_mono1}; this behavior is a distinctive feature absent in \gls{qhe} and \gls{qsh} systems \cite{Qi2025EdgeSupercurrent}. Our findings are consistent with the observations of Barrier et al. \cite{barrier2024One}, with $N_{\mathrm{DW}} = 1$ yielding a critical current insensitive to magnetic field and $N_{\mathrm{DW}}=2$ displaying distinct \gls{ab} oscillations. This confirms that the $I_{c}$--$B$ response exhibits identical interference characteristics regardless of the mechanism by which the \gls{dw}s are formed, since the topological kink states [see Figs.~\ref{fig:setup_spectrum_b} and \ref{fig:setup_spectrum_d}] provide conducting channels that are largely independent of the specific material details. Owing to the similarity between monolayer and bilayer results, our subsequent numerical calculations focus on monolayer graphene-based \glspl{jj}, which reduces computational cost without loss of generality.

\begin{figure*}[!htb]
  \centering
  \includegraphics[width=1.05\textwidth,trim=150 550 10 100,clip]{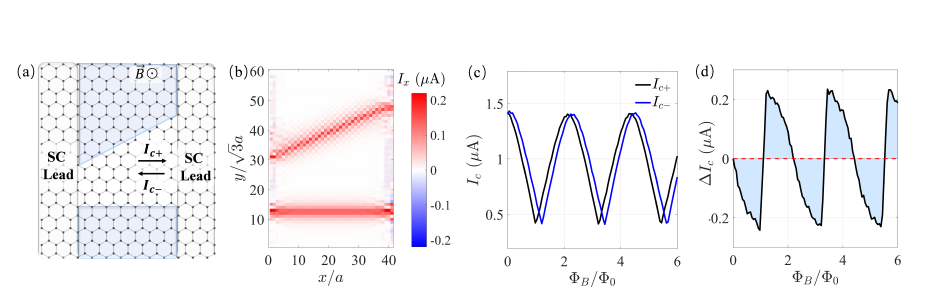}
  \vbox to 0pt {
      \raggedright
      \textcolor{white}{
          \subfloatlabel[1][fig:diode_setup_and_results_a]
          \subfloatlabel[2][fig:diode_setup_and_results_b]
          \subfloatlabel[3][fig:diode_setup_and_results_c]
          \subfloatlabel[4][fig:diode_setup_and_results_d]
      }
  }
  \caption{Monolayer graphene-based \gls{jj} with two asymmetric \glspl{dw}, showing \gls{jde}. (a) Schematic of device setup. There are two \gls{dw}s in the central region of \gls{jj}, with the upper \gls{dw} tilted. A perpendicular magnetic field is applied to the central region. (b) The $x$-component of the supercurrent density at $\Phi_B/\Phi_0 = 0.06$ with the superconducting phase difference set to $5\pi/12$. Coordinates $x \in N_x$ and $y \in N_y$ denote lattice indices in units of the lattice constant $a$ and $\sqrt{3}a$, respectively. (c) Forward and backward critical current magnitudes $I_{c+}$ and $I_{c-}$ versus $\Phi_B/\Phi_0$. (d) The current difference $\Delta I_c$ versus $\Phi_B/\Phi_0$, highlighting the diode response. Other parameters are $N_x = 41$, $N_y = 60$, $V_{\text{stag}} = \pm 0.8$ eV and $U_S = -1$ eV.}
  \label{fig:diode_setup_and_results}
  \end{figure*}

Different from conventional topological systems restricted to either zero or two edge modes, \gls{dw}-based systems permit the number of topological kink channels to be precisely tuned across a broad range by controlling the \gls{dw} numbers. Fig.~\ref{fig:dw_number_mono3} shows that introducing a third \gls{dw} generates one secondary peak within a full period of $4\Phi_{0}$. When the \gls{dw} number increases to five, three secondary peaks appear within a period of $6\Phi_{0}$, as depicted in Fig.~\ref{fig:dw_number_mono5}, demonstrating that each additional \gls{dw} contributes one secondary peak. Given that the area between adjacent \gls{dw}s encloses $1/(N_{\text{DW}}+1)$ of the central region's area as these parallel \gls{dw}s are uniformly spaced, the interference period in these two figures is again reduced to approximately $\Phi_0$. By $N_{\text{DW}} = 19$, the pattern already resembles Fraunhofer diffraction associated with bulk conduction \cite{endnoteDWspacing}, as shown in Fig.~\ref{fig:dw_number_mono19}. This precise control over the \gls{dw} number quantitatively reveals the evolution from AB oscillations to Fraunhofer diffraction, consistent with the AB-to-Fraunhofer evolution observed in the experiments of Barrier et al. \cite{barrier2024One}. This systematic \gls{dw} number engineering quantitatively reveals the gradual evolution from sinusoidal \gls{ab} oscillations to Fraunhofer diffraction.

To gain deeper insight into the relation between the $I_{c}$--$B$ pattern and \gls{dw}s, the supercurrent distribution in the sample is investigated. Following the formalism of Refs.~\cite{dynes1971supercurrent, hart2014induced}, the $I_{c}$--$B$ pattern can be approximately described by $I_{c}(\beta) = \left|\int_{-W/2}^{+W/2} dy J_S(y) e^{i\beta y}\right|$, where $\beta = 2\pi BL/\Phi_{0}$, with $L$ and $W$ being the length (along $x$-axis) and width (along $y$-axis) of the central region, respectively. Here, $J_{S}(y)$ represents the spatial distribution of the supercurrent along the $y$-axis. By applying Fourier analysis to our numerical results for bilayer graphene with two \gls{dw}s [see Fig.~\ref{fig:dw_number_bilayer2}] and monolayer graphene with three \gls{dw}s [see Fig.~\ref{fig:dw_number_mono3}], we extract $J_{S}(y)$, as depicted in Fig.~\ref{fig:fourier_to_realspace}. The resulting distributions exhibit two and three distinct peaks for the bilayer and monolayer systems, respectively, with their number and spatial locations in excellent agreement with those of the \gls{dw}s in our numerical models. This directly demonstrates that the supercurrent is carried by topological kink states localized at these \gls{dw}s and that interference between such kink states gives rise to the observed $I_{c}$--$B$ patterns. By combining Fourier analysis with the quantitative evolution of the $I_{c}$--$B$ pattern with increasing \gls{dw} numbers, we establish a comprehensive theoretical description of the $I_{c}$--$B$ response in graphene-based \glspl{jj} hosting multiple \gls{dw}s, thereby providing a thorough understanding of the experimental observations \cite{barrier2024One}.

Our parallel \gls{dw} system also provides a platform for high-sensitivity magnetometers \cite{Crete2023Designing, Oppenlander2004Superconducting, Feynman1965Lectures}. We examine a series of \glspl{jj} with different \gls{dw} numbers, revealing the relation between $N_{\text{DW}}$ and the \gls{hwhm}, defined as the corresponding magnetic flux at which the supercurrent is reduced to half of its maximum value \cite{Cho2019Investigation}. The function of \gls{hwhm} versus $N_{\text{DW}}$ exhibits a power-law dependence, as plotted in Fig.~\ref{fig:squid_sensitivity}. The log--log plot gives a linear fit with slope $-1.13$ [see the inset in Fig.~\ref{fig:squid_sensitivity}], which closely matches the theoretical exponent $-1.25$ predicted for $N_{\mathrm{DW}}$ in Ref.~\cite{miller1991Enhanced}. This scaling relationship, together with recent advances in graphene-based \gls{jj} fabrication, suggests a viable route toward realizing high-sensitivity magnetometers by increasing $N_{\text{DW}}$ in practical applications. 

\subsection{\label{sec:dw_symmetry}DW Symmetry Engineering}

Beyond controlling the number of \gls{dw}s, the spatial arrangement of asymmetric \gls{dw}s can break inversion symmetry. Combined with the magnetic field, this \gls{dw} symmetry engineering strategy provides the ideal conditions for the Josephson diode effect (JDE), a phenomenon characterized by unidirectional and dissipationless superconducting transport that arises from the joint breaking of inversion and time-reversal symmetries in \glspl{jj} \cite{Yuan2022Supercurrent,Bauriedl2022Supercurrent}. The \gls{jde} has already been demonstrated in materials such as NiTe$_2$ and InSb nanosheets \cite{Pal2022Josephson,Turini2022Josephson, Lu2023Tunable}. In particular, graphene offers significant advantages for scalable \gls{jde} implementation owing to its versatile tunability, mature fabrication processes, and broad applicability \cite{Lin2022Zero, Hu2023Josephson, DiezMerida2023symmetry,Shen2025Josephson}.

Our proposed device setup consists of a monolayer graphene-based \gls{jj} that incorporates two linear \gls{dw}s. To achieve the \gls{jde}, one \gls{dw} is tilted, thereby introducing the required spatial asymmetry, as illustrated in Fig.~\ref{fig:diode_setup_and_results_a}. In the absence of \gls{dw}s, the system possesses inversion symmetry, which guarantees equal forward and backward critical currents \cite{Yuan2022Supercurrent}. Since a perpendicular magnetic field already breaks time-reversal symmetry, applying a non-zero $V_{\text{stag}}$ breaks inversion symmetry, fulfilling the necessary conditions for the Josephson diode effect \cite{Yuan2022Supercurrent, Bauriedl2022Supercurrent}. This symmetry breaking produces an asymmetric supercurrent distribution [see Fig.~\ref{fig:diode_setup_and_results_b}]. The asymmetric supercurrent is characterized by the unequal magnitudes of forward and backward critical currents $I_{c+} \neq I_{c-}$. Both $I_{c+}$ and $I_{c-}$ oscillate with magnetic field, similar to \glspl{jj} with two parallel \gls{dw}s, but here they exhibit a relative offset periodically along the $\Phi_B/\Phi_0$ axis, as shown in Fig.~\ref{fig:diode_setup_and_results_c}. This offset gives rise to a finite critical current difference $\Delta I_c = I_{c+} - I_{c-}$. At a fixed magnetic field, if the injected supercurrent amplitude falls within the range between $I_{c+}$ and $I_{c-}$, as shown by the current window [see Fig.~\ref{fig:diode_setup_and_results_d}], the supercurrent flows in a definite direction. This one-way conduction yields current rectification, enabling nonreciprocal transport in superconducting circuits. In this setup, $\Delta I_c$ can attain a maximum value of $\sim 0.25 \, \mu\text{A}$ at an optimal magnetic field, where $I_{c+} = 0.405 \, \mu\text{A}$ and $I_{c-} = 0.656 \, \mu\text{A}$, corresponding to a diode efficiency of $\eta = (I_{c+} - I_{c-})/(I_{c+} + I_{c-}) \approx -24\%$. The high diode efficiency $\eta$ together with the large $\Delta{I_c}$ underscores the potential of DW-induced graphene platforms for the JDE.

\begin{figure*}[!h]
  \centering
  \includegraphics[width=0.98\textwidth, trim=150 550 50 120, clip]{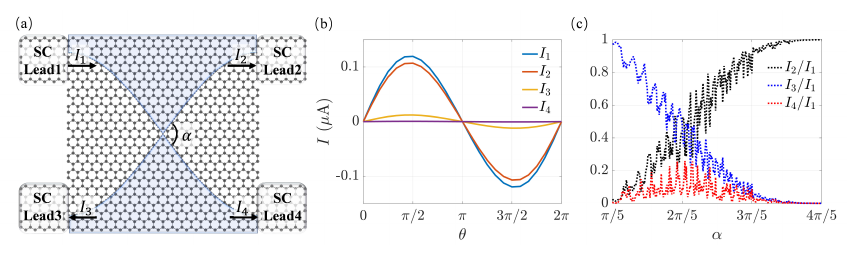}
  \vbox to 0pt {
      \raggedright
      \textcolor{white}{
          \subfloatlabel[1][fig:sine_domain_single_crossing_a]
          \subfloatlabel[2][fig:sine_domain_single_crossing_b]
          \subfloatlabel[3][fig:sine_domain_single_crossing_c]
      }
  }
  \caption{Monolayer graphene-based \gls{jj} with one crossing point. (a) Visualization of a \gls{jj} with two cosine-shaped \gls{dw}s intersecting at a single point with angle $\alpha$. The staggered potential $V_{\text{stag}}$ has opposite signs in the regions between and outside the two \gls{dw}s. The \gls{dw}s extend to the boundaries, defining four leads labeled 1 (upper-left), 2 (upper-right), 3 (lower-left), and 4 (lower-right). The arrows indicate the defined positive current direction for each lead, with the positive direction at lead 3 in the $-x$ direction. (b) Current-phase relations at zero magnetic field. The current splitting ratios $I_k/I_1$ ($k = 2,3,4$) are found to be nearly independent of the superconducting phase difference $\theta$, with values of $I_2/I_1 = 0.895$, $I_3/I_1 = 0.100$, and $I_4/I_1 = 0.005$. Parameters are $N_x = 101$, $N_y = 160$, $U_S = -1.2$ eV, $V_{\text{stag}} = \pm 0.6$ eV, $\alpha = 0.2\pi$. (c) Current splitting ratios $I_2/I_1$, $I_3/I_1$, $I_4/I_1$ versus angle $\alpha$. The observed fluctuations in these curves are attributed to finite-size effects. Parameters are $N_x = 141$, $N_y = 306$, $U_S = -1.2$ eV, $V_{\text{stag}} = \pm 0.6$ eV.}

  \label{fig:sine_domain_single_crossing}
  \end{figure*}

In summary, DW symmetry engineering is presented as an effective method for realizing high-performance superconducting diodes in graphene-based \glspl{jj}. The high tunability of these structures makes them promising candidates for ultra-low-power logic circuits and quantum processing units. Moreover, their compatibility with mature fabrication processes indicates a clear pathway toward practical implementation.

\subsection{\label{sec:dw_shape}DW Geometry Engineering}
Current splitters are crucial components in electrical circuits that are designed to divide a primary current into multiple branches at adjustable ratios, thereby enabling the distribution of energy and information \cite{Li2018Valley}. In conventional circuits, current division is typically achieved using resistors, which inevitably causes energy dissipation through Joule heating. To circumvent such losses, considerable research has focused on developing low-dissipation splitters based on topological materials \cite{Wieder2015Critical, 2021sinedomainwall, Romeo2023Experimental, Fang2021Thermal, Yan2024Rules, Ovchinnikov2022Topological}. These devices exploit topologically protected electronic states to realize dissipationless current splitting, thereby providing a route toward large-scale topological integrated circuits. Alternatively, the intrinsic zero resistance of superconductors offers a promising approach to realizing truly dissipationless splitters, which would facilitate the development of large-scale superconducting integrated circuits. Although this research area is still in its early stages, engineering kink states hosted on \gls{dw}s \cite{Hou2020Valley, Benchtaber2021Scattering, Jiang2020Soliton} provides a practical platform for realizing such supercurrent splitters.

\begin{figure*}[!htb]
\centering
\includegraphics[width=1.0\textwidth,trim=155 435 0 120,clip]{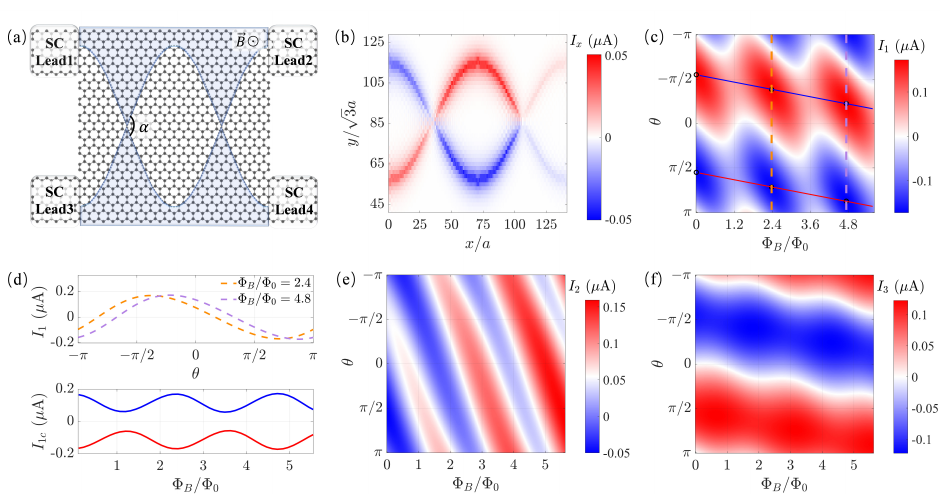}
\vbox to 0pt {
    \raggedright
    \textcolor{white}{
        \subfloatlabel[1][fig:sine_domain_double_crossing_a]
        \subfloatlabel[2][fig:sine_domain_double_crossing_b]
        \subfloatlabel[3][fig:sine_domain_double_crossing_c]
        \subfloatlabel[4][fig:sine_domain_double_crossing_d]
        \subfloatlabel[5][fig:sine_domain_double_crossing_e]
        \subfloatlabel[6][fig:sine_domain_double_crossing_f]
    }
}

\caption{Monolayer graphene-based \gls{jj} with double-crossing configuration. (a) Schematic of a \gls{jj} incorporating cosine-shaped \gls{dw}s, which generate two crossing points. Between and outside these two \gls{dw}s, $V_{\text{stag}}$ is applied with opposite signs. (b) The $x$-component of the supercurrent density, confined by the two \gls{dw}s, forms a closed loop. Here $\Phi_B/\Phi_0 = 1.7$ and $\theta = -3\pi/4$. Coordinates $x$ and $y$ denote lattice indices in units of $a$ and $\sqrt{3}a$, respectively. (c) $I_1$ in the $(\Phi_B/\Phi_0, \theta)$ plane. Blue (red) line labels the forward (backward) critical currents. The orange and purple dashed lines are $I_1$ at fixed magnetic flux $\Phi_B/\Phi_0 = 2.4$ and $4.8$, respectively. (d) Line cuts of $I_1$ from (c). Upper panel: $I_1$ versus $\theta$. Lower panel: critical current $I_{1c}$ versus $\Phi_B/\Phi_0$. (e), (f) $I_2$ and $I_3$ in the $(\Phi_B/\Phi_0, \theta)$ plane, respectively. Other parameters are $\alpha = 0.37\pi$, $N_x = 141$, $N_y = 168$, $U_S = -1.2$ eV and $V_{\text{stag}} = \pm 0.6$ eV.}

\label{fig:sine_domain_double_crossing}
\end{figure*}
In detail, our system consists of two cosine-shaped \gls{dw}s induced by a staggered onsite potential in monolayer graphene-based \glspl{jj}. These \gls{dw}s form either a single or double crossing within the central region, as illustrated in Figs.~\ref{fig:sine_domain_single_crossing_a} and \ref{fig:sine_domain_double_crossing_a}. The \gls{dw}s terminate at the sample boundaries, where four superconducting leads are attached as leads 1--4. A varying superconducting phase $\theta_1$ is applied to lead 1, while the phases of the other leads are maintained at zero. For the single-crossing geometry, the supercurrent splitting ratios between the leads, which govern transmission and reflection processes, are shown to be highly dependent on the \gls{dw} intersection angle. In the double-crossing geometry, a more versatile control of supercurrent splitting is demonstrated by utilizing both the superconducting phase difference $\theta = \theta_1$ and magnetic field.

\subsubsection{Single-Crossing Configuration}
In the single-crossing geometry, two \gls{dw}s intersect at a single point in the center of the \gls{jj}. Following the approach of Ref.~\cite{2021sinedomainwall}, in the absence of magnetic field, the dependence of the supercurrent splitting ratio on the intersection angle $\alpha$ is investigated, defined by the tangents to the two \gls{dw}s at their crossing point, as plotted in Fig.~\ref{fig:sine_domain_single_crossing_a}.

Numerical simulations show that the supercurrents at the four leads oscillate sinusoidally with $\theta$ and share the same period, consistent with the most common current-phase relation $I(\theta) = I_{c} \sin \theta$ \cite{Golubov2004Current}, as presented in Fig.~\ref{fig:sine_domain_single_crossing_b}. The supercurrent injected through lead 1 splits among leads 2, 3, and 4, with the supercurrent splitting ratios $I_2/I_1$, $I_3/I_1$ and $I_4/I_1$ remaining independent of $\theta$ as \gls{dw} angle $\alpha$ is fixed, demonstrating that $\alpha$ plays a crucial role in supercurrent splitting in the absence of magnetic field. These supercurrent splitting ratios satisfy the conservation law $I_2/I_1 + I_3/I_1 + I_4/I_1 = 1$. As demonstrated in Fig.~\ref{fig:sine_domain_single_crossing_c}, reflection is the dominant process for small $\alpha$, with $I_3/I_1 \approx 1$, while the other two ratios are suppressed. Conversely, as $\alpha$ increases, transmission becomes pronounced, with $I_2/I_1$ approaching unity while the other ratios diminishing. The supercurrent splitting ratio $I_4/I_1$ remains relatively small across the entire range of $\alpha$, indicating only weak dependence on $\alpha$. This behavior resembles the current splitting reported in Ref. \cite{2021sinedomainwall}, suggesting a shared underlying mechanism. The physical origin of the behavior of $I_2/I_1$ and $I_3/I_1$ can be understood as follows \cite{qiao2014Current}. For small $\alpha$, the two DWs leading to leads 1 and 3 are spatially close, which strengthens inter-DW coupling and enhances hopping, thereby increasing reflection into lead 3. With larger $\alpha$, the reduced coupling suppresses this hopping, transmission dominates, and $I_2/I_1$ increases.

\subsubsection{Double-Crossing Configuration}

We next study a DW-loop geometry in which two DWs cross to form a closed path [Fig.~\ref{fig:sine_domain_double_crossing_a}]. From a device-engineering perspective, the extension from a single-crossing DW configuration to the two-crossing loop entails no fundamental difficulty. The feasibility of such loops in graphene has been established experimentally \cite{jiang2018DWloopexperimental, Zhang2022Domino}. Building on these advances, the splitting of supercurrent among the leads is examined as a function of magnetic field and the superconducting phase difference, with the intersection angle held fixed throughout.

Figure~\ref{fig:sine_domain_double_crossing_b} shows the supercurrent density of the device. Supercurrent flows along the DW-loop, where the two possible paths around the loop acquire different phases determined by the enclosed magnetic flux and the superconducting phase difference, enabling tunable interference that controls the division of supercurrent. As a result, the lead currents $I_1$, $I_2$, and $I_3$ can be modulated by the magnetic flux $\Phi_B/\Phi_0$ and the superconducting phase difference $\theta$, as shown in Fig.~\ref{fig:sine_domain_double_crossing_c}, \ref{fig:sine_domain_double_crossing_e} and \ref{fig:sine_domain_double_crossing_f}. Specifically, $I_1$ exhibits periodically arranged maxima and minima [see Fig.~\ref{fig:sine_domain_double_crossing_c}]. $I_2$ is reflection-dominated and shows narrow, sign-alternating stripes [see Fig.~\ref{fig:sine_domain_double_crossing_e}], whereas $I_3$ is transmission-dominated and presents broader stripes with a different orientation [see Fig.~\ref{fig:sine_domain_double_crossing_f}].

Physically, $I_1$ corresponds to the total injected supercurrent. In Fig.~\ref{fig:sine_domain_double_crossing_d}, the upper panel plots $I_1$ versus $\theta$ by fixing $\Phi_B/\Phi_0$, and a sinusoidal current-phase relation is obtained. In the lower panel, the corresponding critical current of lead 1 along the red and blue lines in Fig.~\ref{fig:sine_domain_double_crossing_c} shows a regular periodic oscillation with $\Phi_B/\Phi_0$. These features demonstrate that the injected supercurrent can be effectively tuned by the DW-loop interference in this double-crossing configuration. Furthermore, $I_2$ and $I_3$ represent the splitting currents, and their distributions in the $(\Phi_B/\Phi_0, \theta)$ plane differ from that of $I_1$. Consequently, the supercurrent splitting ratios $I_2/I_1$ and $I_3/I_1$ are highly modulated by both magnetic field and superconducting phase difference, driven by the DW-loop interference.

These results demonstrate that the supercurrent splitting between different leads can be controlled through engineering the \gls{dw} geometry, enabling tunable splitting ratios. The intersection angle serves as the geometric parameter that governs the splitting behavior. Furthermore, magnetic field and the superconducting phase difference serve as control parameters that efficiently tune the performance of the supercurrent splitter.

\section{Conclusion}\label{sec:conclusions}
We have systematically investigated transport in graphene-based \glspl{jj} mediated by topological kink states and proposed three DW engineering strategies. First, DW number engineering reveals a continuous evolution from AB interference to Fraunhofer diffraction with increasing DW number, consistent with experimental observations and suggesting a route toward high-field-resolution magnetometry. Second, DW symmetry engineering generates a unidirectional critical current, providing an experimentally feasible pathway to realize the JDE. Third, DW geometry engineering enables controllable supercurrent splitting, highly tunable by intersection angle, magnetic field, and superconducting phase difference. Collectively, these strategies establish DW engineering as a versatile platform for next-generation superconducting circuits and low-dissipation quantum devices.

\section{Acknowledgments}
We thank Shu-Guang Cheng and Hailong Li for illuminating discussions. This work was financially supported by the National Key R\&D Program of China (Grant Nos.~2024YFA1409003 and 2022YFA1403700), the National Natural Science Foundation of China (Grant Nos.~12204053 and 12350401), and the Innovation Program for Quantum Science and Technology (Grant No.~2021ZD0302400). X.C.X. acknowledges additional support from the Innovation Program for Quantum Science and Technology.

\bibliography{graphene_superconductor_junction}

\providecommand{\noopsort}[1]{}\providecommand{\singleletter}[1]{#1}%
\begin{thebibliography}{97}%
\makeatletter
\providecommand \@ifxundefined [1]{%
 \@ifx{#1\undefined}
}%
\providecommand \@ifnum [1]{%
 \ifnum #1\expandafter \@firstoftwo
 \else \expandafter \@secondoftwo
 \fi
}%
\providecommand \@ifx [1]{%
 \ifx #1\expandafter \@firstoftwo
 \else \expandafter \@secondoftwo
 \fi
}%
\providecommand \natexlab [1]{#1}%
\providecommand \enquote  [1]{``#1''}%
\providecommand \bibnamefont  [1]{#1}%
\providecommand \bibfnamefont [1]{#1}%
\providecommand \citenamefont [1]{#1}%
\providecommand \href@noop [0]{\@secondoftwo}%
\providecommand \href [0]{\begingroup \@sanitize@url \@href}%
\providecommand \@href[1]{\@@startlink{#1}\@@href}%
\providecommand \@@href[1]{\endgroup#1\@@endlink}%
\providecommand \@sanitize@url [0]{\catcode `\\12\catcode `\$12\catcode `\&12\catcode `\#12\catcode `\^12\catcode `\_12\catcode `\%12\relax}%
\providecommand \@@startlink[1]{}%
\providecommand \@@endlink[0]{}%
\providecommand \url  [0]{\begingroup\@sanitize@url \@url }%
\providecommand \@url [1]{\endgroup\@href {#1}{\urlprefix }}%
\providecommand \urlprefix  [0]{URL }%
\providecommand \Eprint [0]{\href }%
\providecommand \doibase [0]{https://doi.org/}%
\providecommand \selectlanguage [0]{\@gobble}%
\providecommand \bibinfo  [0]{\@secondoftwo}%
\providecommand \bibfield  [0]{\@secondoftwo}%
\providecommand \translation [1]{[#1]}%
\providecommand \BibitemOpen [0]{}%
\providecommand \bibitemStop [0]{}%
\providecommand \bibitemNoStop [0]{.\EOS\space}%
\providecommand \EOS [0]{\spacefactor3000\relax}%
\providecommand \BibitemShut  [1]{\csname bibitem#1\endcsname}%
\let\auto@bib@innerbib\@empty
\bibitem [{\citenamefont {Rowell}(1963)}]{JJ_B_dependence_1963}%
  \BibitemOpen
  \bibfield  {author} {\bibinfo {author} {\bibfnamefont {J.~M.}\ \bibnamefont {Rowell}},\ }\bibfield  {title} {\bibinfo {title} {Magnetic field dependence of the josephson tunnel current},\ }\href {https://doi.org/10.1103/PhysRevLett.11.200} {\bibfield  {journal} {\bibinfo  {journal} {Phys. Rev. Lett.}\ }\textbf {\bibinfo {volume} {11}},\ \bibinfo {pages} {200} (\bibinfo {year} {1963})}\BibitemShut {NoStop}%
\bibitem [{\citenamefont {Tinkham}(1996)}]{michael1996introduction}%
  \BibitemOpen
  \bibfield  {author} {\bibinfo {author} {\bibfnamefont {M.}~\bibnamefont {Tinkham}},\ }\href {https://archive.org/details/introductiontosu0000mich} {\emph {\bibinfo {title} {Introduction to superconductivity}}},\ \bibinfo {edition} {2nd}\ ed.,\ International series in pure and applied physics\ (\bibinfo  {publisher} {McGraw-Hill},\ \bibinfo {address} {New York},\ \bibinfo {year} {1996})\BibitemShut {NoStop}%
\bibitem [{\citenamefont {Barone}\ and\ \citenamefont {Paterno}(1982)}]{barone1982magnitude}%
  \BibitemOpen
  \bibfield  {author} {\bibinfo {author} {\bibfnamefont {A.}~\bibnamefont {Barone}}\ and\ \bibinfo {author} {\bibfnamefont {G.}~\bibnamefont {Paterno}},\ }\bibinfo {title} {Magnitude and temperature dependence of the critical current},\ in\ \href {https://doi.org/10.1002/352760278X.ch3} {\emph {\bibinfo {booktitle} {Physics and Applications of the Josephson Effect}}}\ (\bibinfo  {publisher} {John Wiley \& Sons, Ltd},\ \bibinfo {year} {1982})\ Chap.~\bibinfo {chapter} {3}, pp.\ \bibinfo {pages} {50--68}\BibitemShut {NoStop}%
\bibitem [{\citenamefont {Pankratov}\ \emph {et~al.}(2025)\citenamefont {Pankratov}, \citenamefont {Gordeeva}, \citenamefont {Chiginev},\ and\ \citenamefont {et~al.}}]{Pankratov2025Detection}%
  \BibitemOpen
  \bibfield  {author} {\bibinfo {author} {\bibfnamefont {A.~L.}\ \bibnamefont {Pankratov}}, \bibinfo {author} {\bibfnamefont {A.~V.}\ \bibnamefont {Gordeeva}}, \bibinfo {author} {\bibfnamefont {A.~V.}\ \bibnamefont {Chiginev}},\ and\ \bibinfo {author} {\bibnamefont {et~al.}},\ }\bibfield  {title} {\bibinfo {title} {Detection of single-mode thermal microwave photons using an underdamped josephson junction},\ }\href {https://doi.org/10.1038/s41467-025-56040-4} {\bibfield  {journal} {\bibinfo  {journal} {Nat. Commun.}\ }\textbf {\bibinfo {volume} {16}},\ \bibinfo {pages} {3457} (\bibinfo {year} {2025})}\BibitemShut {NoStop}%
\bibitem [{\citenamefont {Calado}\ \emph {et~al.}(2015)\citenamefont {Calado}, \citenamefont {Goswami}, \citenamefont {Nanda}, \citenamefont {Diez}, \citenamefont {Akhmerov}, \citenamefont {Watanabe}, \citenamefont {Taniguchi}, \citenamefont {Klapwijk},\ and\ \citenamefont {Vandersypen}}]{calado2015ballistic_graphene_SC_coupling}%
  \BibitemOpen
  \bibfield  {author} {\bibinfo {author} {\bibfnamefont {V.~E.}\ \bibnamefont {Calado}}, \bibinfo {author} {\bibfnamefont {S.}~\bibnamefont {Goswami}}, \bibinfo {author} {\bibfnamefont {G.}~\bibnamefont {Nanda}}, \bibinfo {author} {\bibfnamefont {M.}~\bibnamefont {Diez}}, \bibinfo {author} {\bibfnamefont {A.~R.}\ \bibnamefont {Akhmerov}}, \bibinfo {author} {\bibfnamefont {K.}~\bibnamefont {Watanabe}}, \bibinfo {author} {\bibfnamefont {T.}~\bibnamefont {Taniguchi}}, \bibinfo {author} {\bibfnamefont {T.~M.}\ \bibnamefont {Klapwijk}},\ and\ \bibinfo {author} {\bibfnamefont {L.~M.~K.}\ \bibnamefont {Vandersypen}},\ }\bibfield  {title} {\bibinfo {title} {Ballistic {Josephson} junctions in edge-contacted graphene},\ }\href {https://doi.org/10.1038/nnano.2015.156} {\bibfield  {journal} {\bibinfo  {journal} {Nat. Nanotechnol.}\ }\textbf {\bibinfo {volume} {10}},\ \bibinfo {pages} {761} (\bibinfo {year} {2015})}\BibitemShut {NoStop}%
\bibitem [{\citenamefont {Heersche}\ \emph {et~al.}(2007)\citenamefont {Heersche}, \citenamefont {Jarillo-Herrero}, \citenamefont {Oostinga}, \citenamefont {Vandersypen},\ and\ \citenamefont {Morpurgo}}]{heersche2007bipolar2Dgatecontrol}%
  \BibitemOpen
  \bibfield  {author} {\bibinfo {author} {\bibfnamefont {H.~B.}\ \bibnamefont {Heersche}}, \bibinfo {author} {\bibfnamefont {P.}~\bibnamefont {Jarillo-Herrero}}, \bibinfo {author} {\bibfnamefont {J.~B.}\ \bibnamefont {Oostinga}}, \bibinfo {author} {\bibfnamefont {L.~M.}\ \bibnamefont {Vandersypen}},\ and\ \bibinfo {author} {\bibfnamefont {A.~F.}\ \bibnamefont {Morpurgo}},\ }\bibfield  {title} {\bibinfo {title} {Bipolar supercurrent in graphene},\ }\href {https://doi.org/10.1038/nature05555} {\bibfield  {journal} {\bibinfo  {journal} {Nature}\ }\textbf {\bibinfo {volume} {446}},\ \bibinfo {pages} {56} (\bibinfo {year} {2007})}\BibitemShut {NoStop}%
\bibitem [{\citenamefont {Ke}\ \emph {et~al.}(2019)\citenamefont {Ke}, \citenamefont {Moehle}, \citenamefont {de~Vries},\ and\ \citenamefont {et~al.}}]{Ke2019Ballistic}%
  \BibitemOpen
  \bibfield  {author} {\bibinfo {author} {\bibfnamefont {C.~T.}\ \bibnamefont {Ke}}, \bibinfo {author} {\bibfnamefont {C.~M.}\ \bibnamefont {Moehle}}, \bibinfo {author} {\bibfnamefont {F.~K.}\ \bibnamefont {de~Vries}},\ and\ \bibinfo {author} {\bibnamefont {et~al.}},\ }\bibfield  {title} {\bibinfo {title} {Ballistic superconductivity and tunable pi-junctions in insb quantum wells},\ }\href {https://doi.org/10.1038/s41467-019-11742-4} {\bibfield  {journal} {\bibinfo  {journal} {Nat. Commun.}\ }\textbf {\bibinfo {volume} {10}},\ \bibinfo {pages} {3764} (\bibinfo {year} {2019})}\BibitemShut {NoStop}%
\bibitem [{\citenamefont {Yuan}\ \emph {et~al.}(2021)\citenamefont {Yuan}, \citenamefont {Wickramasinghe}, \citenamefont {Strickland}, \citenamefont {Dartiailh}, \citenamefont {Sardashti}, \citenamefont {Hatefipour},\ and\ \citenamefont {Shabani}}]{Yuan2021Epitaxial2D}%
  \BibitemOpen
  \bibfield  {author} {\bibinfo {author} {\bibfnamefont {J.~O.}\ \bibnamefont {Yuan}}, \bibinfo {author} {\bibfnamefont {K.~S.}\ \bibnamefont {Wickramasinghe}}, \bibinfo {author} {\bibfnamefont {W.~M.}\ \bibnamefont {Strickland}}, \bibinfo {author} {\bibfnamefont {M.~C.}\ \bibnamefont {Dartiailh}}, \bibinfo {author} {\bibfnamefont {K.}~\bibnamefont {Sardashti}}, \bibinfo {author} {\bibfnamefont {M.}~\bibnamefont {Hatefipour}},\ and\ \bibinfo {author} {\bibfnamefont {J.}~\bibnamefont {Shabani}},\ }\bibfield  {title} {\bibinfo {title} {Epitaxial superconductor-semiconductor two-dimensional systems for superconducting quantum circuits},\ }\href {https://doi.org/10.1116/6.0000918} {\bibfield  {journal} {\bibinfo  {journal} {J. Vac. Sci. Technol. A}\ }\textbf {\bibinfo {volume} {39}},\ \bibinfo {pages} {033407} (\bibinfo {year} {2021})}\BibitemShut {NoStop}%
\bibitem [{\citenamefont {Wu}\ \emph {et~al.}(2021)\citenamefont {Wu}, \citenamefont {Wang}, \citenamefont {Zhang},\ and\ \citenamefont {Jiang}}]{Wu2021Programmable}%
  \BibitemOpen
  \bibfield  {author} {\bibinfo {author} {\bibfnamefont {B.-L.}\ \bibnamefont {Wu}}, \bibinfo {author} {\bibfnamefont {Z.-B.}\ \bibnamefont {Wang}}, \bibinfo {author} {\bibfnamefont {Z.-Q.}\ \bibnamefont {Zhang}},\ and\ \bibinfo {author} {\bibfnamefont {H.}~\bibnamefont {Jiang}},\ }\bibfield  {title} {\bibinfo {title} {Building programmable integrated circuits through disordered chern insulators},\ }\href {https://doi.org/10.1103/PhysRevB.104.195416} {\bibfield  {journal} {\bibinfo  {journal} {Phys. Rev. B}\ }\textbf {\bibinfo {volume} {104}},\ \bibinfo {pages} {195416} (\bibinfo {year} {2021})}\BibitemShut {NoStop}%
\bibitem [{\citenamefont {Fatemi}\ \emph {et~al.}(2018)\citenamefont {Fatemi}, \citenamefont {Wu}, \citenamefont {Cao}, \citenamefont {Bretheau}, \citenamefont {Gibson}, \citenamefont {Watanabe}, \citenamefont {Taniguchi}, \citenamefont {Cava},\ and\ \citenamefont {Jarillo-Herrero}}]{Fatemi2018Electrically}%
  \BibitemOpen
  \bibfield  {author} {\bibinfo {author} {\bibfnamefont {V.}~\bibnamefont {Fatemi}}, \bibinfo {author} {\bibfnamefont {S.}~\bibnamefont {Wu}}, \bibinfo {author} {\bibfnamefont {Y.}~\bibnamefont {Cao}}, \bibinfo {author} {\bibfnamefont {L.}~\bibnamefont {Bretheau}}, \bibinfo {author} {\bibfnamefont {Q.~D.}\ \bibnamefont {Gibson}}, \bibinfo {author} {\bibfnamefont {K.}~\bibnamefont {Watanabe}}, \bibinfo {author} {\bibfnamefont {T.}~\bibnamefont {Taniguchi}}, \bibinfo {author} {\bibfnamefont {R.~J.}\ \bibnamefont {Cava}},\ and\ \bibinfo {author} {\bibfnamefont {P.}~\bibnamefont {Jarillo-Herrero}},\ }\bibfield  {title} {\bibinfo {title} {Electrically tunable low-density superconductivity in a monolayer topological insulator},\ }\href {https://doi.org/10.1126/science.aar4642} {\bibfield  {journal} {\bibinfo  {journal} {Science}\ }\textbf {\bibinfo {volume} {362}},\ \bibinfo {pages} {926} (\bibinfo {year} {2018})}\BibitemShut {NoStop}%
\bibitem [{\citenamefont {Pribiag}\ \emph {et~al.}(2015{\natexlab{a}})\citenamefont {Pribiag}, \citenamefont {Beukman}, \citenamefont {Qu}, \citenamefont {Cassidy}, \citenamefont {Charpentier}, \citenamefont {Wegscheider},\ and\ \citenamefont {Kouwenhoven}}]{Pribiag2015Edge}%
  \BibitemOpen
  \bibfield  {author} {\bibinfo {author} {\bibfnamefont {V.~S.}\ \bibnamefont {Pribiag}}, \bibinfo {author} {\bibfnamefont {A.~J.~A.}\ \bibnamefont {Beukman}}, \bibinfo {author} {\bibfnamefont {F.}~\bibnamefont {Qu}}, \bibinfo {author} {\bibfnamefont {M.~C.}\ \bibnamefont {Cassidy}}, \bibinfo {author} {\bibfnamefont {C.}~\bibnamefont {Charpentier}}, \bibinfo {author} {\bibfnamefont {W.}~\bibnamefont {Wegscheider}},\ and\ \bibinfo {author} {\bibfnamefont {L.~P.}\ \bibnamefont {Kouwenhoven}},\ }\bibfield  {title} {\bibinfo {title} {Edge-mode superconductivity in a two-dimensional topological insulator},\ }\href {https://doi.org/10.1038/nnano.2015.86} {\bibfield  {journal} {\bibinfo  {journal} {Nat. Nanotechnol.}\ }\textbf {\bibinfo {volume} {10}},\ \bibinfo {pages} {593} (\bibinfo {year} {2015}{\natexlab{a}})}\BibitemShut {NoStop}%
\bibitem [{\citenamefont {guang Cheng}\ \emph {et~al.}(2020)\citenamefont {guang Cheng}, \citenamefont {Liu}, \citenamefont {Liu}, \citenamefont {Jiang}, \citenamefont {Sun},\ and\ \citenamefont {Xie}}]{Cheng2020Majorana}%
  \BibitemOpen
  \bibfield  {author} {\bibinfo {author} {\bibfnamefont {S.}~\bibnamefont {guang Cheng}}, \bibinfo {author} {\bibfnamefont {J.}~\bibnamefont {Liu}}, \bibinfo {author} {\bibfnamefont {H.}~\bibnamefont {Liu}}, \bibinfo {author} {\bibfnamefont {H.}~\bibnamefont {Jiang}}, \bibinfo {author} {\bibfnamefont {Q.-F.}\ \bibnamefont {Sun}},\ and\ \bibinfo {author} {\bibfnamefont {X.~C.}\ \bibnamefont {Xie}},\ }\bibfield  {title} {\bibinfo {title} {Majorana zero modes from topological kink states in the two-dimensional electron gas},\ }\href {https://doi.org/10.1103/PhysRevB.101.165420} {\bibfield  {journal} {\bibinfo  {journal} {Phys. Rev. B}\ }\textbf {\bibinfo {volume} {101}},\ \bibinfo {pages} {165420} (\bibinfo {year} {2020})}\BibitemShut {NoStop}%
\bibitem [{\citenamefont {Liu}\ \emph {et~al.}(2017)\citenamefont {Liu}, \citenamefont {Liu}, \citenamefont {Song}, \citenamefont {Sun},\ and\ \citenamefont {Xie}}]{Liu2017Superconductor}%
  \BibitemOpen
  \bibfield  {author} {\bibinfo {author} {\bibfnamefont {J.}~\bibnamefont {Liu}}, \bibinfo {author} {\bibfnamefont {H.}~\bibnamefont {Liu}}, \bibinfo {author} {\bibfnamefont {J.}~\bibnamefont {Song}}, \bibinfo {author} {\bibfnamefont {Q.-F.}\ \bibnamefont {Sun}},\ and\ \bibinfo {author} {\bibfnamefont {X.~C.}\ \bibnamefont {Xie}},\ }\bibfield  {title} {\bibinfo {title} {Superconductor-graphene-superconductor josephson junction in the quantum {Hall} regime},\ }\href {https://doi.org/10.1103/PhysRevB.96.045401} {\bibfield  {journal} {\bibinfo  {journal} {Phys. Rev. B}\ }\textbf {\bibinfo {volume} {96}},\ \bibinfo {pages} {045401} (\bibinfo {year} {2017})}\BibitemShut {NoStop}%
\bibitem [{\citenamefont {Vignaud}\ \emph {et~al.}(2023)\citenamefont {Vignaud}, \citenamefont {Perconte}, \citenamefont {Yang}, \citenamefont {Kousar}, \citenamefont {Wagner}, \citenamefont {Gay}, \citenamefont {Watanabe}, \citenamefont {Taniguchi}, \citenamefont {Courtois}, \citenamefont {Han}, \citenamefont {Sellier},\ and\ \citenamefont {Sacépé}}]{vignaud2023evidenceQH}%
  \BibitemOpen
  \bibfield  {author} {\bibinfo {author} {\bibfnamefont {H.}~\bibnamefont {Vignaud}}, \bibinfo {author} {\bibfnamefont {D.}~\bibnamefont {Perconte}}, \bibinfo {author} {\bibfnamefont {W.}~\bibnamefont {Yang}}, \bibinfo {author} {\bibfnamefont {B.}~\bibnamefont {Kousar}}, \bibinfo {author} {\bibfnamefont {E.}~\bibnamefont {Wagner}}, \bibinfo {author} {\bibfnamefont {F.}~\bibnamefont {Gay}}, \bibinfo {author} {\bibfnamefont {K.}~\bibnamefont {Watanabe}}, \bibinfo {author} {\bibfnamefont {T.}~\bibnamefont {Taniguchi}}, \bibinfo {author} {\bibfnamefont {H.}~\bibnamefont {Courtois}}, \bibinfo {author} {\bibfnamefont {Z.}~\bibnamefont {Han}}, \bibinfo {author} {\bibfnamefont {H.}~\bibnamefont {Sellier}},\ and\ \bibinfo {author} {\bibfnamefont {B.}~\bibnamefont {Sacépé}},\ }\bibfield  {title} {\bibinfo {title} {Evidence for chiral supercurrent in quantum {Hall} josephson junctions},\ }\href {https://doi.org/10.1038/s41586-023-06764-4} {\bibfield  {journal} {\bibinfo  {journal} {Nature}\ }\textbf {\bibinfo {volume} {624}},\ \bibinfo {pages} {545} (\bibinfo {year} {2023})}\BibitemShut {NoStop}%
\bibitem [{\citenamefont {Amet}\ \emph {et~al.}(2016)\citenamefont {Amet}, \citenamefont {Ke}, \citenamefont {Borzenets}, \citenamefont {Wang}, \citenamefont {Watanabe}, \citenamefont {Taniguchi}, \citenamefont {Deacon}, \citenamefont {Yamamoto}, \citenamefont {Bomze}, \citenamefont {Tarucha},\ and\ \citenamefont {Finkelstein}}]{Amet2016Supercurrent}%
  \BibitemOpen
  \bibfield  {author} {\bibinfo {author} {\bibfnamefont {F.}~\bibnamefont {Amet}}, \bibinfo {author} {\bibfnamefont {C.~T.}\ \bibnamefont {Ke}}, \bibinfo {author} {\bibfnamefont {I.~V.}\ \bibnamefont {Borzenets}}, \bibinfo {author} {\bibfnamefont {J.}~\bibnamefont {Wang}}, \bibinfo {author} {\bibfnamefont {K.}~\bibnamefont {Watanabe}}, \bibinfo {author} {\bibfnamefont {T.}~\bibnamefont {Taniguchi}}, \bibinfo {author} {\bibfnamefont {R.~S.}\ \bibnamefont {Deacon}}, \bibinfo {author} {\bibfnamefont {M.}~\bibnamefont {Yamamoto}}, \bibinfo {author} {\bibfnamefont {Y.}~\bibnamefont {Bomze}}, \bibinfo {author} {\bibfnamefont {S.}~\bibnamefont {Tarucha}},\ and\ \bibinfo {author} {\bibfnamefont {G.}~\bibnamefont {Finkelstein}},\ }\bibfield  {title} {\bibinfo {title} {Supercurrent in the quantum {Hall} regime},\ }\href {https://doi.org/10.1126/science.aad6203} {\bibfield  {journal} {\bibinfo  {journal} {Science}\ }\textbf {\bibinfo {volume} {352}},\ \bibinfo {pages} {966} (\bibinfo {year} {2016})}\BibitemShut {NoStop}%
\bibitem [{\citenamefont {Uday}\ \emph {et~al.}(2024)\citenamefont {Uday}, \citenamefont {Lippertz}, \citenamefont {Moors}, \citenamefont {Legg}, \citenamefont {Joris}, \citenamefont {Bliesener}, \citenamefont {Pereira}, \citenamefont {Taskin},\ and\ \citenamefont {Ando}}]{uday2024inducedQAH}%
  \BibitemOpen
  \bibfield  {author} {\bibinfo {author} {\bibfnamefont {A.}~\bibnamefont {Uday}}, \bibinfo {author} {\bibfnamefont {G.}~\bibnamefont {Lippertz}}, \bibinfo {author} {\bibfnamefont {K.}~\bibnamefont {Moors}}, \bibinfo {author} {\bibfnamefont {H.~F.}\ \bibnamefont {Legg}}, \bibinfo {author} {\bibfnamefont {R.}~\bibnamefont {Joris}}, \bibinfo {author} {\bibfnamefont {A.}~\bibnamefont {Bliesener}}, \bibinfo {author} {\bibfnamefont {L.~M.~C.}\ \bibnamefont {Pereira}}, \bibinfo {author} {\bibfnamefont {A.~A.}\ \bibnamefont {Taskin}},\ and\ \bibinfo {author} {\bibfnamefont {Y.}~\bibnamefont {Ando}},\ }\bibfield  {title} {\bibinfo {title} {Induced superconducting correlations in a quantum anomalous {Hall} insulator},\ }\href {https://doi.org/10.1038/s41567-024-02574-1} {\bibfield  {journal} {\bibinfo  {journal} {Nat. Phys.}\ }\textbf {\bibinfo {volume} {20}},\ \bibinfo {pages} {1589} (\bibinfo {year} {2024})}\BibitemShut {NoStop}%
\bibitem [{\citenamefont {Qi}\ \emph {et~al.}(2024)\citenamefont {Qi}, \citenamefont {Liu}, \citenamefont {Liu}, \citenamefont {Jiang}, \citenamefont {Liu}, \citenamefont {Chen}, \citenamefont {He},\ and\ \citenamefont {Xie}}]{Anomalous2024Qi}%
  \BibitemOpen
  \bibfield  {author} {\bibinfo {author} {\bibfnamefont {J.}~\bibnamefont {Qi}}, \bibinfo {author} {\bibfnamefont {H.}~\bibnamefont {Liu}}, \bibinfo {author} {\bibfnamefont {J.}~\bibnamefont {Liu}}, \bibinfo {author} {\bibfnamefont {H.}~\bibnamefont {Jiang}}, \bibinfo {author} {\bibfnamefont {D.~E.}\ \bibnamefont {Liu}}, \bibinfo {author} {\bibfnamefont {C.-Z.}\ \bibnamefont {Chen}}, \bibinfo {author} {\bibfnamefont {K.}~\bibnamefont {He}},\ and\ \bibinfo {author} {\bibfnamefont {X.~C.}\ \bibnamefont {Xie}},\ }\bibfield  {title} {\bibinfo {title} {Anomalous fraunhofer-like patterns in quantum anomalous {Hall} josephson junctions},\ }\href {https://doi.org/10.1103/PhysRevResearch.6.023293} {\bibfield  {journal} {\bibinfo  {journal} {Phys. Rev. Res.}\ }\textbf {\bibinfo {volume} {6}},\ \bibinfo {pages} {023293} (\bibinfo {year} {2024})}\BibitemShut {NoStop}%
\bibitem [{\citenamefont {Yan}\ \emph {et~al.}(2020)\citenamefont {Yan}, \citenamefont {Zhou},\ and\ \citenamefont {Sun}}]{Yan2020Anomalous}%
  \BibitemOpen
  \bibfield  {author} {\bibinfo {author} {\bibfnamefont {Q.}~\bibnamefont {Yan}}, \bibinfo {author} {\bibfnamefont {Y.-F.}\ \bibnamefont {Zhou}},\ and\ \bibinfo {author} {\bibfnamefont {Q.-F.}\ \bibnamefont {Sun}},\ }\bibfield  {title} {\bibinfo {title} {Anomalous josephson current in quantum anomalous {Hall} insulator-based superconducting junctions with a domain wall structure},\ }\href {https://doi.org/10.1088/1674-1056/aba272} {\bibfield  {journal} {\bibinfo  {journal} {Chin. Phys. B}\ }\textbf {\bibinfo {volume} {29}},\ \bibinfo {pages} {097401} (\bibinfo {year} {2020})}\BibitemShut {NoStop}%
\bibitem [{\citenamefont {Zhao}\ \emph {et~al.}(2023)\citenamefont {Zhao}, \citenamefont {Zhang}, \citenamefont {Cai}, \citenamefont {Zhuo}, \citenamefont {Zhou}, \citenamefont {Yan}, \citenamefont {Chan}, \citenamefont {Xu},\ and\ \citenamefont {Chang}}]{Zhao2023Chiral}%
  \BibitemOpen
  \bibfield  {author} {\bibinfo {author} {\bibfnamefont {Y.-F.}\ \bibnamefont {Zhao}}, \bibinfo {author} {\bibfnamefont {R.}~\bibnamefont {Zhang}}, \bibinfo {author} {\bibfnamefont {J.}~\bibnamefont {Cai}}, \bibinfo {author} {\bibfnamefont {D.}~\bibnamefont {Zhuo}}, \bibinfo {author} {\bibfnamefont {L.-J.}\ \bibnamefont {Zhou}}, \bibinfo {author} {\bibfnamefont {Z.-J.}\ \bibnamefont {Yan}}, \bibinfo {author} {\bibfnamefont {M.~H.~W.}\ \bibnamefont {Chan}}, \bibinfo {author} {\bibfnamefont {X.}~\bibnamefont {Xu}},\ and\ \bibinfo {author} {\bibfnamefont {C.-Z.}\ \bibnamefont {Chang}},\ }\bibfield  {title} {\bibinfo {title} {Creation of chiral interface channels for quantized transport in magnetic topological insulator multilayer heterostructures},\ }\href {https://doi.org/10.1038/s41467-023-36488-y} {\bibfield  {journal} {\bibinfo  {journal} {Nat. Commun.}\ }\textbf {\bibinfo {volume} {14}},\ \bibinfo {pages} {770} (\bibinfo {year} {2023})}\BibitemShut {NoStop}%
\bibitem [{\citenamefont {Chen}\ \emph {et~al.}(2018)\citenamefont {Chen}, \citenamefont {He}, \citenamefont {Xu},\ and\ \citenamefont {Law}}]{Chen2018Emergent}%
  \BibitemOpen
  \bibfield  {author} {\bibinfo {author} {\bibfnamefont {C.-Z.}\ \bibnamefont {Chen}}, \bibinfo {author} {\bibfnamefont {J.~J.}\ \bibnamefont {He}}, \bibinfo {author} {\bibfnamefont {D.-H.}\ \bibnamefont {Xu}},\ and\ \bibinfo {author} {\bibfnamefont {K.~T.}\ \bibnamefont {Law}},\ }\bibfield  {title} {\bibinfo {title} {Emergent josephson current of $n=1$ chiral topological superconductor in quantum anomalous {Hall} insulator/superconductor heterostructures},\ }\href {https://doi.org/10.1103/PhysRevB.98.165439} {\bibfield  {journal} {\bibinfo  {journal} {Phys. Rev. B}\ }\textbf {\bibinfo {volume} {98}},\ \bibinfo {pages} {165439} (\bibinfo {year} {2018})}\BibitemShut {NoStop}%
\bibitem [{\citenamefont {Bocquillon}\ \emph {et~al.}(2017)\citenamefont {Bocquillon}, \citenamefont {Deacon}, \citenamefont {Wiedenmann}, \citenamefont {Leubner}, \citenamefont {Klapwijk}, \citenamefont {Brüne}, \citenamefont {Ishibashi}, \citenamefont {Buhmann},\ and\ \citenamefont {Molenkamp}}]{bocquillon2017gaplessQSH}%
  \BibitemOpen
  \bibfield  {author} {\bibinfo {author} {\bibfnamefont {E.}~\bibnamefont {Bocquillon}}, \bibinfo {author} {\bibfnamefont {R.~S.}\ \bibnamefont {Deacon}}, \bibinfo {author} {\bibfnamefont {J.}~\bibnamefont {Wiedenmann}}, \bibinfo {author} {\bibfnamefont {P.}~\bibnamefont {Leubner}}, \bibinfo {author} {\bibfnamefont {T.~M.}\ \bibnamefont {Klapwijk}}, \bibinfo {author} {\bibfnamefont {C.}~\bibnamefont {Brüne}}, \bibinfo {author} {\bibfnamefont {K.}~\bibnamefont {Ishibashi}}, \bibinfo {author} {\bibfnamefont {H.}~\bibnamefont {Buhmann}},\ and\ \bibinfo {author} {\bibfnamefont {L.~W.}\ \bibnamefont {Molenkamp}},\ }\bibfield  {title} {\bibinfo {title} {Gapless andreev bound states in the quantum spin {Hall} insulator hgte},\ }\href {https://doi.org/10.1038/nnano.2016.159} {\bibfield  {journal} {\bibinfo  {journal} {Nat. Nanotechnol.}\ }\textbf {\bibinfo {volume} {12}},\ \bibinfo {pages} {137} (\bibinfo {year} {2017})}\BibitemShut {NoStop}%
\bibitem [{\citenamefont {Pribiag}\ \emph {et~al.}(2015{\natexlab{b}})\citenamefont {Pribiag}, \citenamefont {Beukman}, \citenamefont {Qu}, \citenamefont {Cassidy}, \citenamefont {Charpentier}, \citenamefont {Wegscheider},\ and\ \citenamefont {Kouwenhoven}}]{Pribiag2015}%
  \BibitemOpen
  \bibfield  {author} {\bibinfo {author} {\bibfnamefont {V.~S.}\ \bibnamefont {Pribiag}}, \bibinfo {author} {\bibfnamefont {A.~J.~A.}\ \bibnamefont {Beukman}}, \bibinfo {author} {\bibfnamefont {F.}~\bibnamefont {Qu}}, \bibinfo {author} {\bibfnamefont {M.~C.}\ \bibnamefont {Cassidy}}, \bibinfo {author} {\bibfnamefont {C.}~\bibnamefont {Charpentier}}, \bibinfo {author} {\bibfnamefont {W.}~\bibnamefont {Wegscheider}},\ and\ \bibinfo {author} {\bibfnamefont {L.~P.}\ \bibnamefont {Kouwenhoven}},\ }\bibfield  {title} {\bibinfo {title} {Edge-mode superconductivity in a two-dimensional topological insulator},\ }\href {https://doi.org/10.1038/nnano.2015.86} {\bibfield  {journal} {\bibinfo  {journal} {Nature Nanotechnology}\ }\textbf {\bibinfo {volume} {10}},\ \bibinfo {pages} {593} (\bibinfo {year} {2015}{\natexlab{b}})}\BibitemShut {NoStop}%
\bibitem [{\citenamefont {Hart}\ \emph {et~al.}(2014)\citenamefont {Hart}, \citenamefont {Ren}, \citenamefont {Wagner}, \citenamefont {Leubner}, \citenamefont {Mühlbauer}, \citenamefont {Brüne}, \citenamefont {Buhmann}, \citenamefont {Molenkamp},\ and\ \citenamefont {Yacoby}}]{hart2014induced}%
  \BibitemOpen
  \bibfield  {author} {\bibinfo {author} {\bibfnamefont {S.}~\bibnamefont {Hart}}, \bibinfo {author} {\bibfnamefont {H.}~\bibnamefont {Ren}}, \bibinfo {author} {\bibfnamefont {T.}~\bibnamefont {Wagner}}, \bibinfo {author} {\bibfnamefont {P.}~\bibnamefont {Leubner}}, \bibinfo {author} {\bibfnamefont {M.}~\bibnamefont {Mühlbauer}}, \bibinfo {author} {\bibfnamefont {C.}~\bibnamefont {Brüne}}, \bibinfo {author} {\bibfnamefont {H.}~\bibnamefont {Buhmann}}, \bibinfo {author} {\bibfnamefont {L.~W.}\ \bibnamefont {Molenkamp}},\ and\ \bibinfo {author} {\bibfnamefont {A.}~\bibnamefont {Yacoby}},\ }\bibfield  {title} {\bibinfo {title} {Induced superconductivity in the quantum spin {Hall} edge},\ }\href {https://doi.org/10.1038/nphys3036} {\bibfield  {journal} {\bibinfo  {journal} {Nat. Phys.}\ }\textbf {\bibinfo {volume} {10}},\ \bibinfo {pages} {638} (\bibinfo {year} {2014})}\BibitemShut {NoStop}%
\bibitem [{\citenamefont {Martin}\ \emph {et~al.}(2008)\citenamefont {Martin}, \citenamefont {Blanter},\ and\ \citenamefont {Morpurgo}}]{PhysRevLett.100.036804}%
  \BibitemOpen
  \bibfield  {author} {\bibinfo {author} {\bibfnamefont {I.}~\bibnamefont {Martin}}, \bibinfo {author} {\bibfnamefont {Y.~M.}\ \bibnamefont {Blanter}},\ and\ \bibinfo {author} {\bibfnamefont {A.~F.}\ \bibnamefont {Morpurgo}},\ }\bibfield  {title} {\bibinfo {title} {Topological confinement in bilayer graphene},\ }\href {https://doi.org/10.1103/PhysRevLett.100.036804} {\bibfield  {journal} {\bibinfo  {journal} {Phys. Rev. Lett.}\ }\textbf {\bibinfo {volume} {100}},\ \bibinfo {pages} {036804} (\bibinfo {year} {2008})}\BibitemShut {NoStop}%
\bibitem [{\citenamefont {Ju}\ \emph {et~al.}(2015)\citenamefont {Ju}, \citenamefont {Shi}, \citenamefont {Nair}, \citenamefont {Lv}, \citenamefont {Jin}, \citenamefont {Velasco~Jr}, \citenamefont {Ojeda-Aristizabal}, \citenamefont {Bechtel}, \citenamefont {Martin}, \citenamefont {Zettl}, \citenamefont {Analytis},\ and\ \citenamefont {Wang}}]{Ju2015Topological}%
  \BibitemOpen
  \bibfield  {author} {\bibinfo {author} {\bibfnamefont {L.}~\bibnamefont {Ju}}, \bibinfo {author} {\bibfnamefont {Z.}~\bibnamefont {Shi}}, \bibinfo {author} {\bibfnamefont {N.}~\bibnamefont {Nair}}, \bibinfo {author} {\bibfnamefont {Y.}~\bibnamefont {Lv}}, \bibinfo {author} {\bibfnamefont {C.}~\bibnamefont {Jin}}, \bibinfo {author} {\bibfnamefont {J.}~\bibnamefont {Velasco~Jr}}, \bibinfo {author} {\bibfnamefont {C.}~\bibnamefont {Ojeda-Aristizabal}}, \bibinfo {author} {\bibfnamefont {H.~A.}\ \bibnamefont {Bechtel}}, \bibinfo {author} {\bibfnamefont {M.~C.}\ \bibnamefont {Martin}}, \bibinfo {author} {\bibfnamefont {A.}~\bibnamefont {Zettl}}, \bibinfo {author} {\bibfnamefont {J.}~\bibnamefont {Analytis}},\ and\ \bibinfo {author} {\bibfnamefont {F.}~\bibnamefont {Wang}},\ }\bibfield  {title} {\bibinfo {title} {Topological valley transport at bilayer graphene domain walls},\ }\href {https://doi.org/10.1038/nature14364} {\bibfield  {journal} {\bibinfo  {journal} {Nature}\ }\textbf {\bibinfo {volume} {520}},\ \bibinfo {pages} {650} (\bibinfo {year} {2015})}\BibitemShut {NoStop}%
\bibitem [{\citenamefont {Wang}\ \emph {et~al.}(2021)\citenamefont {Wang}, \citenamefont {Cheng}, \citenamefont {Liu},\ and\ \citenamefont {Jiang}}]{Wang2021TopologicalKink}%
  \BibitemOpen
  \bibfield  {author} {\bibinfo {author} {\bibfnamefont {Z.}~\bibnamefont {Wang}}, \bibinfo {author} {\bibfnamefont {S.}~\bibnamefont {Cheng}}, \bibinfo {author} {\bibfnamefont {X.}~\bibnamefont {Liu}},\ and\ \bibinfo {author} {\bibfnamefont {H.}~\bibnamefont {Jiang}},\ }\bibfield  {title} {\bibinfo {title} {Topological kink states in graphene},\ }\href {https://doi.org/10.1088/1361-6528/ac0dd8} {\bibfield  {journal} {\bibinfo  {journal} {Nanotechnology}\ }\textbf {\bibinfo {volume} {32}},\ \bibinfo {pages} {402001} (\bibinfo {year} {2021})}\BibitemShut {NoStop}%
\bibitem [{\citenamefont {Huang}\ \emph {et~al.}(2024)\citenamefont {Huang}, \citenamefont {Fu}, \citenamefont {Watanabe}, \citenamefont {Taniguchi},\ and\ \citenamefont {Zhu}}]{Huang2024QVH}%
  \BibitemOpen
  \bibfield  {author} {\bibinfo {author} {\bibfnamefont {K.}~\bibnamefont {Huang}}, \bibinfo {author} {\bibfnamefont {H.}~\bibnamefont {Fu}}, \bibinfo {author} {\bibfnamefont {K.}~\bibnamefont {Watanabe}}, \bibinfo {author} {\bibfnamefont {T.}~\bibnamefont {Taniguchi}},\ and\ \bibinfo {author} {\bibfnamefont {J.}~\bibnamefont {Zhu}},\ }\bibfield  {title} {\bibinfo {title} {High-temperature quantum valley {Hall} effect with quantized resistance and a topological switch},\ }\href {https://doi.org/10.1126/science.adj3742} {\bibfield  {journal} {\bibinfo  {journal} {Science}\ }\textbf {\bibinfo {volume} {385}},\ \bibinfo {pages} {657} (\bibinfo {year} {2024})}\BibitemShut {NoStop}%
\bibitem [{\citenamefont {Rickhaus}\ \emph {et~al.}(2018)\citenamefont {Rickhaus}, \citenamefont {Wallbank}, \citenamefont {Slizovskiy}, \citenamefont {Pisoni}, \citenamefont {Overweg}, \citenamefont {Lee}, \citenamefont {Eich}, \citenamefont {Liu}, \citenamefont {Watanabe}, \citenamefont {Taniguchi}, \citenamefont {Ihn},\ and\ \citenamefont {Ensslin}}]{Rickhaus2018Transport}%
  \BibitemOpen
  \bibfield  {author} {\bibinfo {author} {\bibfnamefont {P.}~\bibnamefont {Rickhaus}}, \bibinfo {author} {\bibfnamefont {J.}~\bibnamefont {Wallbank}}, \bibinfo {author} {\bibfnamefont {S.}~\bibnamefont {Slizovskiy}}, \bibinfo {author} {\bibfnamefont {R.}~\bibnamefont {Pisoni}}, \bibinfo {author} {\bibfnamefont {H.}~\bibnamefont {Overweg}}, \bibinfo {author} {\bibfnamefont {Y.}~\bibnamefont {Lee}}, \bibinfo {author} {\bibfnamefont {M.}~\bibnamefont {Eich}}, \bibinfo {author} {\bibfnamefont {M.-H.}\ \bibnamefont {Liu}}, \bibinfo {author} {\bibfnamefont {K.}~\bibnamefont {Watanabe}}, \bibinfo {author} {\bibfnamefont {T.}~\bibnamefont {Taniguchi}}, \bibinfo {author} {\bibfnamefont {T.}~\bibnamefont {Ihn}},\ and\ \bibinfo {author} {\bibfnamefont {K.}~\bibnamefont {Ensslin}},\ }\bibfield  {title} {\bibinfo {title} {Transport through a network of topological channels in twisted bilayer graphene},\ }\href {https://doi.org/10.1021/acs.nanolett.8b02387} {\bibfield  {journal} {\bibinfo  {journal} {Nano Lett.}\ }\textbf {\bibinfo {volume} {18}},\ \bibinfo {pages} {6725} (\bibinfo {year} {2018})}\BibitemShut {NoStop}%
\bibitem [{\citenamefont {Yin}\ \emph {et~al.}(2016)\citenamefont {Yin}, \citenamefont {Jiang}, \citenamefont {Qiao},\ and\ \citenamefont {He}}]{Yin2016Direct}%
  \BibitemOpen
  \bibfield  {author} {\bibinfo {author} {\bibfnamefont {L.-J.}\ \bibnamefont {Yin}}, \bibinfo {author} {\bibfnamefont {H.}~\bibnamefont {Jiang}}, \bibinfo {author} {\bibfnamefont {J.-B.}\ \bibnamefont {Qiao}},\ and\ \bibinfo {author} {\bibfnamefont {L.}~\bibnamefont {He}},\ }\bibfield  {title} {\bibinfo {title} {Direct imaging of topological edge states at a bilayer graphene domain wall},\ }\href {https://doi.org/10.1038/ncomms11760} {\bibfield  {journal} {\bibinfo  {journal} {Nat. Commun.}\ }\textbf {\bibinfo {volume} {7}},\ \bibinfo {pages} {11760} (\bibinfo {year} {2016})}\BibitemShut {NoStop}%
\bibitem [{\citenamefont {Barrier}\ \emph {et~al.}(2024)\citenamefont {Barrier}, \citenamefont {Kim}, \citenamefont {Kumar}, \citenamefont {Xin}, \citenamefont {Kumaravadivel}, \citenamefont {Hague}, \citenamefont {Nguyen}, \citenamefont {Berdyugin}, \citenamefont {Moulsdale}, \citenamefont {Enaldiev}, \citenamefont {Prance}, \citenamefont {Koppens}, \citenamefont {Gorbachev}, \citenamefont {Watanabe}, \citenamefont {Taniguchi}, \citenamefont {Glazman}, \citenamefont {Grigorieva}, \citenamefont {Fal’ko},\ and\ \citenamefont {Geim}}]{barrier2024One}%
  \BibitemOpen
  \bibfield  {author} {\bibinfo {author} {\bibfnamefont {J.}~\bibnamefont {Barrier}}, \bibinfo {author} {\bibfnamefont {M.}~\bibnamefont {Kim}}, \bibinfo {author} {\bibfnamefont {R.~K.}\ \bibnamefont {Kumar}}, \bibinfo {author} {\bibfnamefont {N.}~\bibnamefont {Xin}}, \bibinfo {author} {\bibfnamefont {P.}~\bibnamefont {Kumaravadivel}}, \bibinfo {author} {\bibfnamefont {L.}~\bibnamefont {Hague}}, \bibinfo {author} {\bibfnamefont {E.}~\bibnamefont {Nguyen}}, \bibinfo {author} {\bibfnamefont {A.~I.}\ \bibnamefont {Berdyugin}}, \bibinfo {author} {\bibfnamefont {C.}~\bibnamefont {Moulsdale}}, \bibinfo {author} {\bibfnamefont {V.~V.}\ \bibnamefont {Enaldiev}}, \bibinfo {author} {\bibfnamefont {J.~R.}\ \bibnamefont {Prance}}, \bibinfo {author} {\bibfnamefont {F.~H.~L.}\ \bibnamefont {Koppens}}, \bibinfo {author} {\bibfnamefont {R.~V.}\ \bibnamefont {Gorbachev}}, \bibinfo {author} {\bibfnamefont {K.}~\bibnamefont {Watanabe}}, \bibinfo {author} {\bibfnamefont {T.}~\bibnamefont {Taniguchi}}, \bibinfo {author} {\bibfnamefont {L.~I.}\ \bibnamefont {Glazman}}, \bibinfo {author} {\bibfnamefont {I.~V.}\ \bibnamefont {Grigorieva}}, \bibinfo {author} {\bibfnamefont {V.~I.}\ \bibnamefont {Fal’ko}},\ and\ \bibinfo {author} {\bibfnamefont {A.~K.}\ \bibnamefont {Geim}},\ }\bibfield  {title} {\bibinfo {title} {{One-dimensional proximity superconductivity in the quantum {Hall} regime}},\ }\href {https://doi.org/10.1038/s41586-024-07271-w} {\bibfield  {journal} {\bibinfo  {journal} {Nature}\ }\textbf {\bibinfo {volume} {628}},\ \bibinfo {pages} {741} (\bibinfo {year} {2024})}\BibitemShut {NoStop}%
\bibitem [{\citenamefont {Yoo}\ \emph {et~al.}(2019)\citenamefont {Yoo}, \citenamefont {Engelke}, \citenamefont {Carr}, \citenamefont {Fang}, \citenamefont {Zhang}, \citenamefont {Cazeaux}, \citenamefont {Sung}, \citenamefont {Hovden}, \citenamefont {Tsen}, \citenamefont {Taniguchi}, \citenamefont {Watanabe}, \citenamefont {Yi}, \citenamefont {Kim}, \citenamefont {Luskin}, \citenamefont {Tadmor}, \citenamefont {Kaxiras},\ and\ \citenamefont {Kim}}]{yoo2019TBGmanufacture}%
  \BibitemOpen
  \bibfield  {author} {\bibinfo {author} {\bibfnamefont {H.}~\bibnamefont {Yoo}}, \bibinfo {author} {\bibfnamefont {R.}~\bibnamefont {Engelke}}, \bibinfo {author} {\bibfnamefont {S.}~\bibnamefont {Carr}}, \bibinfo {author} {\bibfnamefont {S.}~\bibnamefont {Fang}}, \bibinfo {author} {\bibfnamefont {K.}~\bibnamefont {Zhang}}, \bibinfo {author} {\bibfnamefont {P.}~\bibnamefont {Cazeaux}}, \bibinfo {author} {\bibfnamefont {S.~H.}\ \bibnamefont {Sung}}, \bibinfo {author} {\bibfnamefont {R.}~\bibnamefont {Hovden}}, \bibinfo {author} {\bibfnamefont {A.~W.}\ \bibnamefont {Tsen}}, \bibinfo {author} {\bibfnamefont {T.}~\bibnamefont {Taniguchi}}, \bibinfo {author} {\bibfnamefont {K.}~\bibnamefont {Watanabe}}, \bibinfo {author} {\bibfnamefont {G.~C.}\ \bibnamefont {Yi}}, \bibinfo {author} {\bibfnamefont {M.}~\bibnamefont {Kim}}, \bibinfo {author} {\bibfnamefont {M.}~\bibnamefont {Luskin}}, \bibinfo {author} {\bibfnamefont {E.~B.}\ \bibnamefont {Tadmor}}, \bibinfo {author} {\bibfnamefont {E.}~\bibnamefont {Kaxiras}},\ and\ \bibinfo {author} {\bibfnamefont {P.}~\bibnamefont {Kim}},\ }\bibfield  {title} {\bibinfo {title} {Atomic and electronic reconstruction at the van der waals interface in twisted bilayer graphene},\ }\href {https://doi.org/10.1038/s41563-019-0346-z} {\bibfield  {journal} {\bibinfo  {journal} {Nat. Mater.}\ }\textbf {\bibinfo {volume} {18}},\ \bibinfo {pages} {448} (\bibinfo {year} {2019})}\BibitemShut {NoStop}%
\bibitem [{\citenamefont {Lebedeva}\ and\ \citenamefont {Popov}(2020)}]{Lebedeva2020Energetics}%
  \BibitemOpen
  \bibfield  {author} {\bibinfo {author} {\bibfnamefont {I.~V.}\ \bibnamefont {Lebedeva}}\ and\ \bibinfo {author} {\bibfnamefont {A.~M.}\ \bibnamefont {Popov}},\ }\bibfield  {title} {\bibinfo {title} {Energetics and structure of domain wall networks in minimally twisted bilayer graphene under strain},\ }\href {https://doi.org/10.1021/acs.jpcc.9b08306} {\bibfield  {journal} {\bibinfo  {journal} {J. Phys. Chem. C}\ }\textbf {\bibinfo {volume} {124}},\ \bibinfo {pages} {2120} (\bibinfo {year} {2020})}\BibitemShut {NoStop}%
\bibitem [{\citenamefont {de~Vries}\ \emph {et~al.}(2021)\citenamefont {de~Vries}, \citenamefont {Portolés}, \citenamefont {Zheng}, \citenamefont {Taniguchi}, \citenamefont {Watanabe}, \citenamefont {Ihn}, \citenamefont {Ensslin},\ and\ \citenamefont {Rickhaus}}]{deVries2021GateDefined}%
  \BibitemOpen
  \bibfield  {author} {\bibinfo {author} {\bibfnamefont {F.~K.}\ \bibnamefont {de~Vries}}, \bibinfo {author} {\bibfnamefont {E.}~\bibnamefont {Portolés}}, \bibinfo {author} {\bibfnamefont {G.}~\bibnamefont {Zheng}}, \bibinfo {author} {\bibfnamefont {T.}~\bibnamefont {Taniguchi}}, \bibinfo {author} {\bibfnamefont {K.}~\bibnamefont {Watanabe}}, \bibinfo {author} {\bibfnamefont {T.}~\bibnamefont {Ihn}}, \bibinfo {author} {\bibfnamefont {K.}~\bibnamefont {Ensslin}},\ and\ \bibinfo {author} {\bibfnamefont {P.}~\bibnamefont {Rickhaus}},\ }\bibfield  {title} {\bibinfo {title} {Gate-defined josephson junctions in magic-angle twisted bilayer graphene},\ }\href {https://doi.org/10.1038/s41565-021-00896-2} {\bibfield  {journal} {\bibinfo  {journal} {Nat. Nanotechnol.}\ }\textbf {\bibinfo {volume} {16}},\ \bibinfo {pages} {760} (\bibinfo {year} {2021})}\BibitemShut {NoStop}%
\bibitem [{\citenamefont {Li}\ \emph {et~al.}(2016)\citenamefont {Li}, \citenamefont {Wang}, \citenamefont {McFaul}, \citenamefont {Zern}, \citenamefont {Ren}, \citenamefont {Watanabe}, \citenamefont {Taniguchi}, \citenamefont {Qiao},\ and\ \citenamefont {Zhu}}]{bilayerchannel2016gate}%
  \BibitemOpen
  \bibfield  {author} {\bibinfo {author} {\bibfnamefont {J.}~\bibnamefont {Li}}, \bibinfo {author} {\bibfnamefont {K.}~\bibnamefont {Wang}}, \bibinfo {author} {\bibfnamefont {K.~J.}\ \bibnamefont {McFaul}}, \bibinfo {author} {\bibfnamefont {Z.}~\bibnamefont {Zern}}, \bibinfo {author} {\bibfnamefont {Y.}~\bibnamefont {Ren}}, \bibinfo {author} {\bibfnamefont {K.}~\bibnamefont {Watanabe}}, \bibinfo {author} {\bibfnamefont {T.}~\bibnamefont {Taniguchi}}, \bibinfo {author} {\bibfnamefont {Z.}~\bibnamefont {Qiao}},\ and\ \bibinfo {author} {\bibfnamefont {J.}~\bibnamefont {Zhu}},\ }\bibfield  {title} {\bibinfo {title} {Gate-controlled topological conducting channels in bilayer graphene},\ }\href {https://doi.org/10.1038/nnano.2016.158} {\bibfield  {journal} {\bibinfo  {journal} {Nat. Nanotechnol.}\ }\textbf {\bibinfo {volume} {11}},\ \bibinfo {pages} {1060} (\bibinfo {year} {2016})}\BibitemShut {NoStop}%
\bibitem [{\citenamefont {Lee}\ \emph {et~al.}(2017)\citenamefont {Lee}, \citenamefont {Watanabe}, \citenamefont {Taniguchi},\ and\ \citenamefont {Lee}}]{Lee2017Realisation}%
  \BibitemOpen
  \bibfield  {author} {\bibinfo {author} {\bibfnamefont {J.}~\bibnamefont {Lee}}, \bibinfo {author} {\bibfnamefont {K.}~\bibnamefont {Watanabe}}, \bibinfo {author} {\bibfnamefont {T.}~\bibnamefont {Taniguchi}},\ and\ \bibinfo {author} {\bibfnamefont {H.-J.}\ \bibnamefont {Lee}},\ }\bibfield  {title} {\bibinfo {title} {Realisation of topological zero-energy mode in bilayer graphene in zero magnetic field},\ }\href {https://doi.org/10.1038/s41598-017-06902-9} {\bibfield  {journal} {\bibinfo  {journal} {Sci. Rep.}\ }\textbf {\bibinfo {volume} {7}},\ \bibinfo {pages} {6466} (\bibinfo {year} {2017})}\BibitemShut {NoStop}%
\bibitem [{\citenamefont {Chen}\ \emph {et~al.}(2020)\citenamefont {Chen}, \citenamefont {Zhou}, \citenamefont {Liu}, \citenamefont {Liu}, \citenamefont {Zhang}, \citenamefont {Wan}, \citenamefont {Guo}, \citenamefont {Guo}, \citenamefont {Lu}, \citenamefont {Gao}, \citenamefont {Guo},\ and\ \citenamefont {Liu}}]{Chen2020Gate}%
  \BibitemOpen
  \bibfield  {author} {\bibinfo {author} {\bibfnamefont {H.}~\bibnamefont {Chen}}, \bibinfo {author} {\bibfnamefont {P.}~\bibnamefont {Zhou}}, \bibinfo {author} {\bibfnamefont {J.}~\bibnamefont {Liu}}, \bibinfo {author} {\bibfnamefont {J.}~\bibnamefont {Liu}}, \bibinfo {author} {\bibfnamefont {J.}~\bibnamefont {Zhang}}, \bibinfo {author} {\bibfnamefont {J.}~\bibnamefont {Wan}}, \bibinfo {author} {\bibfnamefont {L.}~\bibnamefont {Guo}}, \bibinfo {author} {\bibfnamefont {X.}~\bibnamefont {Guo}}, \bibinfo {author} {\bibfnamefont {J.}~\bibnamefont {Lu}}, \bibinfo {author} {\bibfnamefont {J.}~\bibnamefont {Gao}}, \bibinfo {author} {\bibfnamefont {G.}~\bibnamefont {Guo}},\ and\ \bibinfo {author} {\bibfnamefont {X.}~\bibnamefont {Liu}},\ }\bibfield  {title} {\bibinfo {title} {Gate controlled valley polarizer in bilayer graphene},\ }\href {https://doi.org/10.1038/s41467-020-15117-y} {\bibfield  {journal} {\bibinfo  {journal} {Nat. Commun.}\ }\textbf {\bibinfo {volume} {11}},\ \bibinfo {pages} {1202} (\bibinfo {year} {2020})}\BibitemShut {NoStop}%
\bibitem [{\citenamefont {Kim}\ \emph {et~al.}(2016)\citenamefont {Kim}, \citenamefont {Choi}, \citenamefont {Lee}, \citenamefont {Watanabe}, \citenamefont {Taniguchi}, \citenamefont {Jhi},\ and\ \citenamefont {Lee}}]{Kim2016ValleySymmetry}%
  \BibitemOpen
  \bibfield  {author} {\bibinfo {author} {\bibfnamefont {M.}~\bibnamefont {Kim}}, \bibinfo {author} {\bibfnamefont {J.-H.}\ \bibnamefont {Choi}}, \bibinfo {author} {\bibfnamefont {S.-H.}\ \bibnamefont {Lee}}, \bibinfo {author} {\bibfnamefont {K.}~\bibnamefont {Watanabe}}, \bibinfo {author} {\bibfnamefont {T.}~\bibnamefont {Taniguchi}}, \bibinfo {author} {\bibfnamefont {S.-H.}\ \bibnamefont {Jhi}},\ and\ \bibinfo {author} {\bibfnamefont {H.-J.}\ \bibnamefont {Lee}},\ }\bibfield  {title} {\bibinfo {title} {Valley-symmetry-preserved transport in ballistic graphene with gate-defined carrier guiding},\ }\href {https://doi.org/10.1038/nphys3804} {\bibfield  {journal} {\bibinfo  {journal} {Nat. Phys.}\ }\textbf {\bibinfo {volume} {12}},\ \bibinfo {pages} {1022} (\bibinfo {year} {2016})}\BibitemShut {NoStop}%
\bibitem [{\citenamefont {Hou}\ \emph {et~al.}(2020{\natexlab{a}})\citenamefont {Hou}, \citenamefont {Ren}, \citenamefont {Quan}, \citenamefont {Jung}, \citenamefont {Ren},\ and\ \citenamefont {Qiao}}]{Hou2020Metallic}%
  \BibitemOpen
  \bibfield  {author} {\bibinfo {author} {\bibfnamefont {T.}~\bibnamefont {Hou}}, \bibinfo {author} {\bibfnamefont {Y.}~\bibnamefont {Ren}}, \bibinfo {author} {\bibfnamefont {Y.}~\bibnamefont {Quan}}, \bibinfo {author} {\bibfnamefont {J.}~\bibnamefont {Jung}}, \bibinfo {author} {\bibfnamefont {W.}~\bibnamefont {Ren}},\ and\ \bibinfo {author} {\bibfnamefont {Z.}~\bibnamefont {Qiao}},\ }\bibfield  {title} {\bibinfo {title} {Metallic network of topological domain walls},\ }\href {https://doi.org/10.1103/PhysRevB.101.201403} {\bibfield  {journal} {\bibinfo  {journal} {Phys. Rev. B}\ }\textbf {\bibinfo {volume} {101}},\ \bibinfo {pages} {201403} (\bibinfo {year} {2020}{\natexlab{a}})}\BibitemShut {NoStop}%
\bibitem [{\citenamefont {Pan}\ \emph {et~al.}(2025)\citenamefont {Pan}, \citenamefont {Wang}, \citenamefont {Zou}, \citenamefont {Wang}, \citenamefont {Zhang}, \citenamefont {Dong}, \citenamefont {Xie}, \citenamefont {Ding}, \citenamefont {Zhang}, \citenamefont {Taniguchi}, \citenamefont {Watanabe}, \citenamefont {Wang},\ and\ \citenamefont {Wang}}]{Pan2025Topological}%
  \BibitemOpen
  \bibfield  {author} {\bibinfo {author} {\bibfnamefont {J.}~\bibnamefont {Pan}}, \bibinfo {author} {\bibfnamefont {H.}~\bibnamefont {Wang}}, \bibinfo {author} {\bibfnamefont {L.}~\bibnamefont {Zou}}, \bibinfo {author} {\bibfnamefont {X.}~\bibnamefont {Wang}}, \bibinfo {author} {\bibfnamefont {L.}~\bibnamefont {Zhang}}, \bibinfo {author} {\bibfnamefont {X.}~\bibnamefont {Dong}}, \bibinfo {author} {\bibfnamefont {H.}~\bibnamefont {Xie}}, \bibinfo {author} {\bibfnamefont {Y.}~\bibnamefont {Ding}}, \bibinfo {author} {\bibfnamefont {Y.}~\bibnamefont {Zhang}}, \bibinfo {author} {\bibfnamefont {T.}~\bibnamefont {Taniguchi}}, \bibinfo {author} {\bibfnamefont {K.}~\bibnamefont {Watanabe}}, \bibinfo {author} {\bibfnamefont {S.}~\bibnamefont {Wang}},\ and\ \bibinfo {author} {\bibfnamefont {Z.}~\bibnamefont {Wang}},\ }\bibfield  {title} {\bibinfo {title} {Topological valley transport in bilayer graphene induced by interlayer sliding},\ }\bibfield  {journal} {\bibinfo  {journal} {Phys. Rev. Lett.}\ }\href {https://doi.org/10.1103/26q7-dsm1} {10.1103/26q7-dsm1} (\bibinfo {year} {2025})\BibitemShut {NoStop}%
\bibitem [{\citenamefont {Li}\ \emph {et~al.}(2018)\citenamefont {Li}, \citenamefont {Wen}, \citenamefont {Watanabe}, \citenamefont {Taniguchi},\ and\ \citenamefont {Zhu}}]{Li2018Valley}%
  \BibitemOpen
  \bibfield  {author} {\bibinfo {author} {\bibfnamefont {J.}~\bibnamefont {Li}}, \bibinfo {author} {\bibfnamefont {H.}~\bibnamefont {Wen}}, \bibinfo {author} {\bibfnamefont {K.}~\bibnamefont {Watanabe}}, \bibinfo {author} {\bibfnamefont {T.}~\bibnamefont {Taniguchi}},\ and\ \bibinfo {author} {\bibfnamefont {J.}~\bibnamefont {Zhu}},\ }\bibfield  {title} {\bibinfo {title} {A valley valve and electron beam splitter},\ }\href {https://doi.org/10.1126/science.aao5989} {\bibfield  {journal} {\bibinfo  {journal} {Science}\ }\textbf {\bibinfo {volume} {362}},\ \bibinfo {pages} {1149} (\bibinfo {year} {2018})}\BibitemShut {NoStop}%
\bibitem [{\citenamefont {Qiao}\ \emph {et~al.}(2011)\citenamefont {Qiao}, \citenamefont {Jung}, \citenamefont {Niu},\ and\ \citenamefont {MacDonald}}]{Qiao2011Electronic}%
  \BibitemOpen
  \bibfield  {author} {\bibinfo {author} {\bibfnamefont {Z.}~\bibnamefont {Qiao}}, \bibinfo {author} {\bibfnamefont {J.}~\bibnamefont {Jung}}, \bibinfo {author} {\bibfnamefont {Q.}~\bibnamefont {Niu}},\ and\ \bibinfo {author} {\bibfnamefont {A.~H.}\ \bibnamefont {MacDonald}},\ }\bibfield  {title} {\bibinfo {title} {Electronic highways in bilayer graphene},\ }\href {https://doi.org/10.1021/nl201941f} {\bibfield  {journal} {\bibinfo  {journal} {Nano Lett.}\ }\textbf {\bibinfo {volume} {11}},\ \bibinfo {pages} {3453} (\bibinfo {year} {2011})}\BibitemShut {NoStop}%
\bibitem [{\citenamefont {Yao}\ \emph {et~al.}(2009)\citenamefont {Yao}, \citenamefont {Yang},\ and\ \citenamefont {Niu}}]{Yao2009Edge}%
  \BibitemOpen
  \bibfield  {author} {\bibinfo {author} {\bibfnamefont {W.}~\bibnamefont {Yao}}, \bibinfo {author} {\bibfnamefont {S.~A.}\ \bibnamefont {Yang}},\ and\ \bibinfo {author} {\bibfnamefont {Q.}~\bibnamefont {Niu}},\ }\bibfield  {title} {\bibinfo {title} {Edge states in graphene: From gapped flat-band to gapless chiral modes},\ }\href {https://doi.org/10.1103/PhysRevLett.102.096801} {\bibfield  {journal} {\bibinfo  {journal} {Phys. Rev. Lett.}\ }\textbf {\bibinfo {volume} {102}},\ \bibinfo {pages} {096801} (\bibinfo {year} {2009})}\BibitemShut {NoStop}%
\bibitem [{\citenamefont {Hunt}\ \emph {et~al.}(2013)\citenamefont {Hunt}, \citenamefont {Sanchez-Yamagishi}, \citenamefont {Young}, \citenamefont {Yankowitz}, \citenamefont {LeRoy}, \citenamefont {Watanabe}, \citenamefont {Taniguchi}, \citenamefont {Moon}, \citenamefont {Koshino}, \citenamefont {Jarillo-Herrero},\ and\ \citenamefont {Ashoori}}]{Hunt2013Massive}%
  \BibitemOpen
  \bibfield  {author} {\bibinfo {author} {\bibfnamefont {B.}~\bibnamefont {Hunt}}, \bibinfo {author} {\bibfnamefont {J.~D.}\ \bibnamefont {Sanchez-Yamagishi}}, \bibinfo {author} {\bibfnamefont {A.~F.}\ \bibnamefont {Young}}, \bibinfo {author} {\bibfnamefont {M.}~\bibnamefont {Yankowitz}}, \bibinfo {author} {\bibfnamefont {B.~J.}\ \bibnamefont {LeRoy}}, \bibinfo {author} {\bibfnamefont {K.}~\bibnamefont {Watanabe}}, \bibinfo {author} {\bibfnamefont {T.}~\bibnamefont {Taniguchi}}, \bibinfo {author} {\bibfnamefont {P.}~\bibnamefont {Moon}}, \bibinfo {author} {\bibfnamefont {M.}~\bibnamefont {Koshino}}, \bibinfo {author} {\bibfnamefont {P.}~\bibnamefont {Jarillo-Herrero}},\ and\ \bibinfo {author} {\bibfnamefont {R.~C.}\ \bibnamefont {Ashoori}},\ }\bibfield  {title} {\bibinfo {title} {Massive {Dirac} fermions and {Hofstadter} butterfly in a van der {Waals} heterostructure},\ }\href {https://doi.org/10.1126/science.1237240} {\bibfield  {journal} {\bibinfo  {journal} {Science}\ }\textbf {\bibinfo {volume} {340}},\ \bibinfo {pages} {1427} (\bibinfo {year} {2013})}\BibitemShut {NoStop}%
\bibitem [{\citenamefont {Woods}\ \emph {et~al.}(2014)\citenamefont {Woods}, \citenamefont {Britnell}, \citenamefont {Eckmann}, \citenamefont {Ma}, \citenamefont {Lu}, \citenamefont {Guo}, \citenamefont {Lin}, \citenamefont {Yu}, \citenamefont {Cao}, \citenamefont {Gorbachev}, \citenamefont {Kretinin}, \citenamefont {Park}, \citenamefont {Ponomarenko}, \citenamefont {Katsnelson}, \citenamefont {Gornostyrev}, \citenamefont {Watanabe}, \citenamefont {Taniguchi}, \citenamefont {Casiraghi}, \citenamefont {Gao}, \citenamefont {Geim},\ and\ \citenamefont {Novoselov}}]{Woods2014Commensurate}%
  \BibitemOpen
  \bibfield  {author} {\bibinfo {author} {\bibfnamefont {C.~R.}\ \bibnamefont {Woods}}, \bibinfo {author} {\bibfnamefont {L.}~\bibnamefont {Britnell}}, \bibinfo {author} {\bibfnamefont {A.}~\bibnamefont {Eckmann}}, \bibinfo {author} {\bibfnamefont {R.~S.}\ \bibnamefont {Ma}}, \bibinfo {author} {\bibfnamefont {J.~C.}\ \bibnamefont {Lu}}, \bibinfo {author} {\bibfnamefont {H.~M.}\ \bibnamefont {Guo}}, \bibinfo {author} {\bibfnamefont {X.}~\bibnamefont {Lin}}, \bibinfo {author} {\bibfnamefont {G.~L.}\ \bibnamefont {Yu}}, \bibinfo {author} {\bibfnamefont {Y.}~\bibnamefont {Cao}}, \bibinfo {author} {\bibfnamefont {R.~V.}\ \bibnamefont {Gorbachev}}, \bibinfo {author} {\bibfnamefont {A.~V.}\ \bibnamefont {Kretinin}}, \bibinfo {author} {\bibfnamefont {J.}~\bibnamefont {Park}}, \bibinfo {author} {\bibfnamefont {L.~A.}\ \bibnamefont {Ponomarenko}}, \bibinfo {author} {\bibfnamefont {M.~I.}\ \bibnamefont {Katsnelson}}, \bibinfo {author} {\bibfnamefont {Y.~N.}\ \bibnamefont {Gornostyrev}}, \bibinfo {author} {\bibfnamefont {K.}~\bibnamefont {Watanabe}}, \bibinfo {author} {\bibfnamefont {T.}~\bibnamefont {Taniguchi}}, \bibinfo {author} {\bibfnamefont {C.}~\bibnamefont {Casiraghi}}, \bibinfo {author} {\bibfnamefont {H.-M.}\ \bibnamefont {Gao}}, \bibinfo {author} {\bibfnamefont {A.~K.}\ \bibnamefont {Geim}},\ and\ \bibinfo {author} {\bibfnamefont {K.~S.}\ \bibnamefont {Novoselov}},\ }\bibfield  {title} {\bibinfo {title} {Commensurate--incommensurate transition in graphene on hexagonal boron nitride},\ }\href {https://doi.org/10.1038/nphys2954} {\bibfield  {journal} {\bibinfo  {journal} {Nat. Phys.}\ }\textbf {\bibinfo {volume} {10}},\ \bibinfo {pages} {451} (\bibinfo {year} {2014})}\BibitemShut {NoStop}%
\bibitem [{\citenamefont {Xiao}\ \emph {et~al.}(2007)\citenamefont {Xiao}, \citenamefont {Yao},\ and\ \citenamefont {Niu}}]{Xiao2007Valley}%
  \BibitemOpen
  \bibfield  {author} {\bibinfo {author} {\bibfnamefont {D.}~\bibnamefont {Xiao}}, \bibinfo {author} {\bibfnamefont {W.}~\bibnamefont {Yao}},\ and\ \bibinfo {author} {\bibfnamefont {Q.}~\bibnamefont {Niu}},\ }\bibfield  {title} {\bibinfo {title} {Valley-contrasting physics in graphene: magnetic moment and topological transport},\ }\href {https://doi.org/10.1103/PhysRevLett.99.236809} {\bibfield  {journal} {\bibinfo  {journal} {Phys. Rev. Lett.}\ }\textbf {\bibinfo {volume} {99}},\ \bibinfo {pages} {236809} (\bibinfo {year} {2007})}\BibitemShut {NoStop}%
\bibitem [{\citenamefont {Zhou}\ \emph {et~al.}(2007)\citenamefont {Zhou}, \citenamefont {Gweon}, \citenamefont {Fedorov}, \citenamefont {First}, \citenamefont {de~Heer}, \citenamefont {Lee}, \citenamefont {Guinea}, \citenamefont {Castro~Neto},\ and\ \citenamefont {Lanzara}}]{Zhou2007Substrate}%
  \BibitemOpen
  \bibfield  {author} {\bibinfo {author} {\bibfnamefont {S.~Y.}\ \bibnamefont {Zhou}}, \bibinfo {author} {\bibfnamefont {G.-H.}\ \bibnamefont {Gweon}}, \bibinfo {author} {\bibfnamefont {A.~V.}\ \bibnamefont {Fedorov}}, \bibinfo {author} {\bibfnamefont {P.~N.}\ \bibnamefont {First}}, \bibinfo {author} {\bibfnamefont {W.~A.}\ \bibnamefont {de~Heer}}, \bibinfo {author} {\bibfnamefont {D.-H.}\ \bibnamefont {Lee}}, \bibinfo {author} {\bibfnamefont {F.}~\bibnamefont {Guinea}}, \bibinfo {author} {\bibfnamefont {A.~H.}\ \bibnamefont {Castro~Neto}},\ and\ \bibinfo {author} {\bibfnamefont {A.}~\bibnamefont {Lanzara}},\ }\bibfield  {title} {\bibinfo {title} {Substrate-induced bandgap opening in epitaxial graphene},\ }\href {https://doi.org/10.1038/nmat2003} {\bibfield  {journal} {\bibinfo  {journal} {Nat. Mater.}\ }\textbf {\bibinfo {volume} {6}},\ \bibinfo {pages} {770} (\bibinfo {year} {2007})}\BibitemShut {NoStop}%
\bibitem [{\citenamefont {Wang}\ \emph {et~al.}(2016)\citenamefont {Wang}, \citenamefont {Lu}, \citenamefont {Ding}, \citenamefont {Yao}, \citenamefont {Yan}, \citenamefont {Wan}, \citenamefont {Deng}, \citenamefont {Wang}, \citenamefont {Chen}, \citenamefont {Ma}, \citenamefont {Jung}, \citenamefont {Fedorov}, \citenamefont {Zhang}, \citenamefont {Zhang},\ and\ \citenamefont {Zhou}}]{Wang2016Gaps}%
  \BibitemOpen
  \bibfield  {author} {\bibinfo {author} {\bibfnamefont {E.}~\bibnamefont {Wang}}, \bibinfo {author} {\bibfnamefont {X.}~\bibnamefont {Lu}}, \bibinfo {author} {\bibfnamefont {S.}~\bibnamefont {Ding}}, \bibinfo {author} {\bibfnamefont {W.}~\bibnamefont {Yao}}, \bibinfo {author} {\bibfnamefont {M.}~\bibnamefont {Yan}}, \bibinfo {author} {\bibfnamefont {G.}~\bibnamefont {Wan}}, \bibinfo {author} {\bibfnamefont {K.}~\bibnamefont {Deng}}, \bibinfo {author} {\bibfnamefont {S.}~\bibnamefont {Wang}}, \bibinfo {author} {\bibfnamefont {G.}~\bibnamefont {Chen}}, \bibinfo {author} {\bibfnamefont {L.}~\bibnamefont {Ma}}, \bibinfo {author} {\bibfnamefont {J.}~\bibnamefont {Jung}}, \bibinfo {author} {\bibfnamefont {A.~V.}\ \bibnamefont {Fedorov}}, \bibinfo {author} {\bibfnamefont {Y.}~\bibnamefont {Zhang}}, \bibinfo {author} {\bibfnamefont {G.}~\bibnamefont {Zhang}},\ and\ \bibinfo {author} {\bibfnamefont {S.}~\bibnamefont {Zhou}},\ }\bibfield  {title} {\bibinfo {title} {Gaps induced by inversion symmetry breaking and second-generation dirac cones in graphene hexagonal boron nitride},\ }\href {https://doi.org/10.1038/nphys3856} {\bibfield  {journal} {\bibinfo  {journal} {Nat. Phys.}\ }\textbf {\bibinfo {volume} {12}},\ \bibinfo {pages} {1111} (\bibinfo {year} {2016})}\BibitemShut {NoStop}%
\bibitem [{\citenamefont {Bi}\ \emph {et~al.}(2015)\citenamefont {Bi}, \citenamefont {Jung},\ and\ \citenamefont {Qiao}}]{stagsublattice2015role}%
  \BibitemOpen
  \bibfield  {author} {\bibinfo {author} {\bibfnamefont {X.}~\bibnamefont {Bi}}, \bibinfo {author} {\bibfnamefont {J.}~\bibnamefont {Jung}},\ and\ \bibinfo {author} {\bibfnamefont {Z.}~\bibnamefont {Qiao}},\ }\bibfield  {title} {\bibinfo {title} {Role of geometry and topological defects in the one-dimensional zero-line modes of graphene},\ }\href {https://doi.org/10.1103/PhysRevB.92.235421} {\bibfield  {journal} {\bibinfo  {journal} {Phys. Rev. B}\ }\textbf {\bibinfo {volume} {92}},\ \bibinfo {pages} {235421} (\bibinfo {year} {2015})}\BibitemShut {NoStop}%
\bibitem [{\citenamefont {Gorbachev}\ \emph {et~al.}(2014)\citenamefont {Gorbachev}, \citenamefont {Song}, \citenamefont {Yu}, \citenamefont {Kretinin}, \citenamefont {Withers}, \citenamefont {Cao}, \citenamefont {Mishchenko}, \citenamefont {Grigorieva}, \citenamefont {Novoselov}, \citenamefont {Levitov},\ and\ \citenamefont {Geim}}]{Gorbachev2014Detecting}%
  \BibitemOpen
  \bibfield  {author} {\bibinfo {author} {\bibfnamefont {R.~V.}\ \bibnamefont {Gorbachev}}, \bibinfo {author} {\bibfnamefont {J.~C.~W.}\ \bibnamefont {Song}}, \bibinfo {author} {\bibfnamefont {G.~L.}\ \bibnamefont {Yu}}, \bibinfo {author} {\bibfnamefont {A.~V.}\ \bibnamefont {Kretinin}}, \bibinfo {author} {\bibfnamefont {F.}~\bibnamefont {Withers}}, \bibinfo {author} {\bibfnamefont {Y.}~\bibnamefont {Cao}}, \bibinfo {author} {\bibfnamefont {A.}~\bibnamefont {Mishchenko}}, \bibinfo {author} {\bibfnamefont {I.~V.}\ \bibnamefont {Grigorieva}}, \bibinfo {author} {\bibfnamefont {K.~S.}\ \bibnamefont {Novoselov}}, \bibinfo {author} {\bibfnamefont {L.~S.}\ \bibnamefont {Levitov}},\ and\ \bibinfo {author} {\bibfnamefont {A.~K.}\ \bibnamefont {Geim}},\ }\bibfield  {title} {\bibinfo {title} {Detecting topological currents in graphene superlattices},\ }\href {https://doi.org/10.1126/science.1254966} {\bibfield  {journal} {\bibinfo  {journal} {Science}\ }\textbf {\bibinfo {volume} {346}},\ \bibinfo {pages} {448} (\bibinfo {year} {2014})}\BibitemShut {NoStop}%
\bibitem [{\citenamefont {He}\ \emph {et~al.}(2022)\citenamefont {He}, \citenamefont {Koon}, \citenamefont {Isobe}, \citenamefont {Tan}, \citenamefont {Hu}, \citenamefont {Castro~Neto}, \citenamefont {Fu},\ and\ \citenamefont {Yang}}]{He2022Graphene}%
  \BibitemOpen
  \bibfield  {author} {\bibinfo {author} {\bibfnamefont {P.}~\bibnamefont {He}}, \bibinfo {author} {\bibfnamefont {G.~K.~W.}\ \bibnamefont {Koon}}, \bibinfo {author} {\bibfnamefont {H.}~\bibnamefont {Isobe}}, \bibinfo {author} {\bibfnamefont {J.~Y.}\ \bibnamefont {Tan}}, \bibinfo {author} {\bibfnamefont {J.}~\bibnamefont {Hu}}, \bibinfo {author} {\bibfnamefont {A.~H.}\ \bibnamefont {Castro~Neto}}, \bibinfo {author} {\bibfnamefont {L.}~\bibnamefont {Fu}},\ and\ \bibinfo {author} {\bibfnamefont {H.}~\bibnamefont {Yang}},\ }\bibfield  {title} {\bibinfo {title} {Graphene moiré superlattices with giant quantum nonlinearity of chiral bloch electrons},\ }\href {https://doi.org/10.1038/s41565-021-01062-4} {\bibfield  {journal} {\bibinfo  {journal} {Nat. Nanotechnol.}\ }\textbf {\bibinfo {volume} {17}},\ \bibinfo {pages} {378} (\bibinfo {year} {2022})}\BibitemShut {NoStop}%
\bibitem [{\citenamefont {Song}\ \emph {et~al.}(2015)\citenamefont {Song}, \citenamefont {Samutpraphoot},\ and\ \citenamefont {Levitov}}]{Song2015Topological}%
  \BibitemOpen
  \bibfield  {author} {\bibinfo {author} {\bibfnamefont {J.~C.~W.}\ \bibnamefont {Song}}, \bibinfo {author} {\bibfnamefont {P.}~\bibnamefont {Samutpraphoot}},\ and\ \bibinfo {author} {\bibfnamefont {L.~S.}\ \bibnamefont {Levitov}},\ }\bibfield  {title} {\bibinfo {title} {Topological bloch bands in graphene superlattices},\ }\href {https://doi.org/10.1073/pnas.1424760112} {\bibfield  {journal} {\bibinfo  {journal} {Proc. Natl. Acad. Sci. U.S.A.}\ }\textbf {\bibinfo {volume} {112}},\ \bibinfo {pages} {10879} (\bibinfo {year} {2015})}\BibitemShut {NoStop}%
\bibitem [{\citenamefont {Han}\ \emph {et~al.}(2021)\citenamefont {Han}, \citenamefont {Yang}, \citenamefont {Zhang}, \citenamefont {Wang}, \citenamefont {Watanabe}, \citenamefont {Taniguchi}, \citenamefont {McEuen},\ and\ \citenamefont {Ju}}]{Han2021Accurate}%
  \BibitemOpen
  \bibfield  {author} {\bibinfo {author} {\bibfnamefont {T.}~\bibnamefont {Han}}, \bibinfo {author} {\bibfnamefont {J.}~\bibnamefont {Yang}}, \bibinfo {author} {\bibfnamefont {Q.}~\bibnamefont {Zhang}}, \bibinfo {author} {\bibfnamefont {L.}~\bibnamefont {Wang}}, \bibinfo {author} {\bibfnamefont {K.}~\bibnamefont {Watanabe}}, \bibinfo {author} {\bibfnamefont {T.}~\bibnamefont {Taniguchi}}, \bibinfo {author} {\bibfnamefont {P.~L.}\ \bibnamefont {McEuen}},\ and\ \bibinfo {author} {\bibfnamefont {L.}~\bibnamefont {Ju}},\ }\bibfield  {title} {\bibinfo {title} {Accurate measurement of the gap of graphene/h-bn moiré superlattice through photocurrent spectroscopy},\ }\href {https://doi.org/10.1103/PhysRevLett.126.146402} {\bibfield  {journal} {\bibinfo  {journal} {Phys. Rev. Lett.}\ }\textbf {\bibinfo {volume} {126}},\ \bibinfo {pages} {146402} (\bibinfo {year} {2021})}\BibitemShut {NoStop}%
\bibitem [{\citenamefont {Jiang}\ \emph {et~al.}(2018)\citenamefont {Jiang}, \citenamefont {Wang}, \citenamefont {Shi}, \citenamefont {Jin}, \citenamefont {Utama}, \citenamefont {Zhao}, \citenamefont {Shen}, \citenamefont {Gao}, \citenamefont {Zhang},\ and\ \citenamefont {Wang}}]{jiang2018DWloopexperimental}%
  \BibitemOpen
  \bibfield  {author} {\bibinfo {author} {\bibfnamefont {L.}~\bibnamefont {Jiang}}, \bibinfo {author} {\bibfnamefont {S.}~\bibnamefont {Wang}}, \bibinfo {author} {\bibfnamefont {Z.}~\bibnamefont {Shi}}, \bibinfo {author} {\bibfnamefont {C.}~\bibnamefont {Jin}}, \bibinfo {author} {\bibfnamefont {M.~I.~B.}\ \bibnamefont {Utama}}, \bibinfo {author} {\bibfnamefont {S.}~\bibnamefont {Zhao}}, \bibinfo {author} {\bibfnamefont {Y.-R.}\ \bibnamefont {Shen}}, \bibinfo {author} {\bibfnamefont {H.-J.}\ \bibnamefont {Gao}}, \bibinfo {author} {\bibfnamefont {G.}~\bibnamefont {Zhang}},\ and\ \bibinfo {author} {\bibfnamefont {F.}~\bibnamefont {Wang}},\ }\bibfield  {title} {\bibinfo {title} {Manipulation of domain-wall solitons in bi- and trilayer graphene},\ }\href {https://doi.org/10.1038/s41565-017-0042-6} {\bibfield  {journal} {\bibinfo  {journal} {Nat. Nanotechnol.}\ }\textbf {\bibinfo {volume} {13}},\ \bibinfo {pages} {204} (\bibinfo {year} {2018})}\BibitemShut {NoStop}%
\bibitem [{\citenamefont {Mania}\ \emph {et~al.}(2019)\citenamefont {Mania}, \citenamefont {Cadore}, \citenamefont {Taniguchi}, \citenamefont {Watanabe},\ and\ \citenamefont {Campos}}]{Mania2019Topological}%
  \BibitemOpen
  \bibfield  {author} {\bibinfo {author} {\bibfnamefont {E.}~\bibnamefont {Mania}}, \bibinfo {author} {\bibfnamefont {A.~R.}\ \bibnamefont {Cadore}}, \bibinfo {author} {\bibfnamefont {T.}~\bibnamefont {Taniguchi}}, \bibinfo {author} {\bibfnamefont {K.}~\bibnamefont {Watanabe}},\ and\ \bibinfo {author} {\bibfnamefont {L.~C.}\ \bibnamefont {Campos}},\ }\bibfield  {title} {\bibinfo {title} {Topological valley transport at the curved boundary of a folded bilayer graphene},\ }\href {https://doi.org/10.1038/s42005-018-0106-4} {\bibfield  {journal} {\bibinfo  {journal} {Commun. Phys.}\ }\textbf {\bibinfo {volume} {2}},\ \bibinfo {pages} {6} (\bibinfo {year} {2019})}\BibitemShut {NoStop}%
\bibitem [{\citenamefont {Zhang}\ \emph {et~al.}(2022)\citenamefont {Zhang}, \citenamefont {Xu}, \citenamefont {Hou}, \citenamefont {Song}, \citenamefont {Ma}, \citenamefont {Gao}, \citenamefont {Zhu}, \citenamefont {Ma}, \citenamefont {Liu}, \citenamefont {Feng},\ and\ \citenamefont {Li}}]{Zhang2022Domino}%
  \BibitemOpen
  \bibfield  {author} {\bibinfo {author} {\bibfnamefont {S.}~\bibnamefont {Zhang}}, \bibinfo {author} {\bibfnamefont {Q.}~\bibnamefont {Xu}}, \bibinfo {author} {\bibfnamefont {Y.}~\bibnamefont {Hou}}, \bibinfo {author} {\bibfnamefont {A.}~\bibnamefont {Song}}, \bibinfo {author} {\bibfnamefont {Y.}~\bibnamefont {Ma}}, \bibinfo {author} {\bibfnamefont {L.}~\bibnamefont {Gao}}, \bibinfo {author} {\bibfnamefont {M.}~\bibnamefont {Zhu}}, \bibinfo {author} {\bibfnamefont {T.}~\bibnamefont {Ma}}, \bibinfo {author} {\bibfnamefont {L.}~\bibnamefont {Liu}}, \bibinfo {author} {\bibfnamefont {X.-Q.}\ \bibnamefont {Feng}},\ and\ \bibinfo {author} {\bibfnamefont {Q.}~\bibnamefont {Li}},\ }\bibfield  {title} {\bibinfo {title} {Domino-like stacking order switching in twisted monolayer--multilayer graphene},\ }\href {https://doi.org/10.1038/s41563-022-01232-2} {\bibfield  {journal} {\bibinfo  {journal} {Nat. Mater.}\ }\textbf {\bibinfo {volume} {21}},\ \bibinfo {pages} {621} (\bibinfo {year} {2022})}\BibitemShut {NoStop}%
\bibitem [{\citenamefont {Ben~Shalom}\ \emph {et~al.}(2016)\citenamefont {Ben~Shalom}, \citenamefont {Zhu}, \citenamefont {Fal'ko}, \citenamefont {Mishchenko}, \citenamefont {Kretinin}, \citenamefont {Novoselov}, \citenamefont {Woods}, \citenamefont {Watanabe}, \citenamefont {Taniguchi}, \citenamefont {Geim},\ and\ \citenamefont {Prance}}]{BenShalom2016Quantum}%
  \BibitemOpen
  \bibfield  {author} {\bibinfo {author} {\bibfnamefont {M.}~\bibnamefont {Ben~Shalom}}, \bibinfo {author} {\bibfnamefont {M.-J.}\ \bibnamefont {Zhu}}, \bibinfo {author} {\bibfnamefont {V.~I.}\ \bibnamefont {Fal'ko}}, \bibinfo {author} {\bibfnamefont {A.}~\bibnamefont {Mishchenko}}, \bibinfo {author} {\bibfnamefont {A.~V.}\ \bibnamefont {Kretinin}}, \bibinfo {author} {\bibfnamefont {K.~S.}\ \bibnamefont {Novoselov}}, \bibinfo {author} {\bibfnamefont {C.~R.}\ \bibnamefont {Woods}}, \bibinfo {author} {\bibfnamefont {K.}~\bibnamefont {Watanabe}}, \bibinfo {author} {\bibfnamefont {T.}~\bibnamefont {Taniguchi}}, \bibinfo {author} {\bibfnamefont {A.~K.}\ \bibnamefont {Geim}},\ and\ \bibinfo {author} {\bibfnamefont {J.~R.}\ \bibnamefont {Prance}},\ }\bibfield  {title} {\bibinfo {title} {Quantum oscillations of the critical current and high-field superconducting proximity in ballistic graphene},\ }\href {https://doi.org/10.1038/nphys3592} {\bibfield  {journal} {\bibinfo  {journal} {Nat. Phys.}\ }\textbf {\bibinfo {volume} {12}},\ \bibinfo {pages} {318} (\bibinfo {year} {2016})}\BibitemShut {NoStop}%
\bibitem [{\citenamefont {Turini}\ \emph {et~al.}(2022)\citenamefont {Turini}, \citenamefont {Salimian}, \citenamefont {Carrega}, \citenamefont {Iorio}, \citenamefont {Strambini}, \citenamefont {Giazotto}, \citenamefont {Zannier}, \citenamefont {Sorba},\ and\ \citenamefont {Heun}}]{Turini2022Josephson}%
  \BibitemOpen
  \bibfield  {author} {\bibinfo {author} {\bibfnamefont {B.}~\bibnamefont {Turini}}, \bibinfo {author} {\bibfnamefont {S.}~\bibnamefont {Salimian}}, \bibinfo {author} {\bibfnamefont {M.}~\bibnamefont {Carrega}}, \bibinfo {author} {\bibfnamefont {A.}~\bibnamefont {Iorio}}, \bibinfo {author} {\bibfnamefont {E.}~\bibnamefont {Strambini}}, \bibinfo {author} {\bibfnamefont {F.}~\bibnamefont {Giazotto}}, \bibinfo {author} {\bibfnamefont {V.}~\bibnamefont {Zannier}}, \bibinfo {author} {\bibfnamefont {L.}~\bibnamefont {Sorba}},\ and\ \bibinfo {author} {\bibfnamefont {S.}~\bibnamefont {Heun}},\ }\bibfield  {title} {\bibinfo {title} {{Josephson} diode effect in high-mobility {InSb} nanoflags},\ }\href {https://doi.org/10.1021/acs.nanolett.2c02899} {\bibfield  {journal} {\bibinfo  {journal} {Nano Lett.}\ }\textbf {\bibinfo {volume} {22}},\ \bibinfo {pages} {8502} (\bibinfo {year} {2022})}\BibitemShut {NoStop}%
\bibitem [{\citenamefont {Yuan}\ and\ \citenamefont {Fu}(2022)}]{Yuan2022Supercurrent}%
  \BibitemOpen
  \bibfield  {author} {\bibinfo {author} {\bibfnamefont {N.~F.~Q.}\ \bibnamefont {Yuan}}\ and\ \bibinfo {author} {\bibfnamefont {L.}~\bibnamefont {Fu}},\ }\bibfield  {title} {\bibinfo {title} {Supercurrent diode effect and finite-momentum superconductors},\ }\href {https://doi.org/10.1073/pnas.2119548119} {\bibfield  {journal} {\bibinfo  {journal} {Proc. Natl. Acad. Sci. U.S.A.}\ }\textbf {\bibinfo {volume} {119}},\ \bibinfo {pages} {e2119548119} (\bibinfo {year} {2022})}\BibitemShut {NoStop}%
\bibitem [{\citenamefont {Davydova}\ \emph {et~al.}(2022)\citenamefont {Davydova}, \citenamefont {Prembabu},\ and\ \citenamefont {Fu}}]{Davydova2022Universal}%
  \BibitemOpen
  \bibfield  {author} {\bibinfo {author} {\bibfnamefont {M.}~\bibnamefont {Davydova}}, \bibinfo {author} {\bibfnamefont {S.}~\bibnamefont {Prembabu}},\ and\ \bibinfo {author} {\bibfnamefont {L.}~\bibnamefont {Fu}},\ }\bibfield  {title} {\bibinfo {title} {Universal {Josephson} diode effect},\ }\href {https://doi.org/10.1126/sciadv.abo0309} {\bibfield  {journal} {\bibinfo  {journal} {Sci. Adv.}\ }\textbf {\bibinfo {volume} {8}},\ \bibinfo {pages} {eabo0309} (\bibinfo {year} {2022})}\BibitemShut {NoStop}%
\bibitem [{\citenamefont {Semenoff}\ \emph {et~al.}(2008)\citenamefont {Semenoff}, \citenamefont {Semenoff},\ and\ \citenamefont {Zhou}}]{semenoff2008Domain}%
  \BibitemOpen
  \bibfield  {author} {\bibinfo {author} {\bibfnamefont {G.~W.}\ \bibnamefont {Semenoff}}, \bibinfo {author} {\bibfnamefont {V.}~\bibnamefont {Semenoff}},\ and\ \bibinfo {author} {\bibfnamefont {F.}~\bibnamefont {Zhou}},\ }\bibfield  {title} {\bibinfo {title} {{Domain Walls in Gapped Graphene}},\ }\href {https://doi.org/10.1103/physrevlett.101.087204} {\bibfield  {journal} {\bibinfo  {journal} {Phys. Rev. Lett.}\ }\textbf {\bibinfo {volume} {101}},\ \bibinfo {pages} {087204} (\bibinfo {year} {2008})}\BibitemShut {NoStop}%
\bibitem [{\citenamefont {Perfetto}\ \emph {et~al.}(2009)\citenamefont {Perfetto}, \citenamefont {Stefanucci},\ and\ \citenamefont {Cini}}]{Perfetto2009Equilibrium}%
  \BibitemOpen
  \bibfield  {author} {\bibinfo {author} {\bibfnamefont {E.}~\bibnamefont {Perfetto}}, \bibinfo {author} {\bibfnamefont {G.}~\bibnamefont {Stefanucci}},\ and\ \bibinfo {author} {\bibfnamefont {M.}~\bibnamefont {Cini}},\ }\bibfield  {title} {\bibinfo {title} {Equilibrium and time-dependent josephson current in one-dimensional superconducting junctions},\ }\href {https://doi.org/10.1103/PhysRevB.80.205408} {\bibfield  {journal} {\bibinfo  {journal} {Phys. Rev. B}\ }\textbf {\bibinfo {volume} {80}},\ \bibinfo {pages} {205408} (\bibinfo {year} {2009})}\BibitemShut {NoStop}%
\bibitem [{\citenamefont {Kamenev}(2011)}]{Kamenev2011Field}%
  \BibitemOpen
  \bibfield  {author} {\bibinfo {author} {\bibfnamefont {A.}~\bibnamefont {Kamenev}},\ }\href {https://www.cambridge.org/core/books/field-theory-of-nonequilibrium-systems/CEA995D5C5C7E043E9BAAE6DCA282354} {\emph {\bibinfo {title} {Field Theory of Non-Equilibrium Systems}}}\ (\bibinfo  {publisher} {Cambridge University Press},\ \bibinfo {year} {2011})\BibitemShut {NoStop}%
\bibitem [{\citenamefont {Furusaki}\ and\ \citenamefont {Tsukada}(1991)}]{Furusaki1991Current}%
  \BibitemOpen
  \bibfield  {author} {\bibinfo {author} {\bibfnamefont {A.}~\bibnamefont {Furusaki}}\ and\ \bibinfo {author} {\bibfnamefont {M.}~\bibnamefont {Tsukada}},\ }\bibfield  {title} {\bibinfo {title} {Current-carrying states in josephson junctions},\ }\href {https://doi.org/10.1103/PhysRevB.43.10164} {\bibfield  {journal} {\bibinfo  {journal} {Phys. Rev. B}\ }\textbf {\bibinfo {volume} {43}},\ \bibinfo {pages} {10164} (\bibinfo {year} {1991})}\BibitemShut {NoStop}%
\bibitem [{\citenamefont {Haug}\ and\ \citenamefont {Jauho}(2008)}]{Haug2008Quantum}%
  \BibitemOpen
  \bibfield  {author} {\bibinfo {author} {\bibfnamefont {H.}~\bibnamefont {Haug}}\ and\ \bibinfo {author} {\bibfnamefont {A.-P.}\ \bibnamefont {Jauho}},\ }\href {https://link.springer.com/book/10.1007/978-3-540-73564-9} {\emph {\bibinfo {title} {Quantum Kinetics in Transport and Optics of Semiconductors}}}\ (\bibinfo  {publisher} {Springer Berlin Heidelberg},\ \bibinfo {address} {Berlin, Heidelberg},\ \bibinfo {year} {2008})\BibitemShut {NoStop}%
\bibitem [{\citenamefont {Ryndyk}(2017)}]{Ryndyk2017Theory}%
  \BibitemOpen
  \bibfield  {author} {\bibinfo {author} {\bibfnamefont {D.~A.}\ \bibnamefont {Ryndyk}},\ }\bibfield  {title} {\bibinfo {title} {Theory of quantum transport at nanoscale: An introduction}\ }(\bibinfo {year} {2017})\ pp.\ \bibinfo {pages} {74--75}\BibitemShut {NoStop}%
\bibitem [{\citenamefont {Asano}(2001)}]{Asano2001Numerical}%
  \BibitemOpen
  \bibfield  {author} {\bibinfo {author} {\bibfnamefont {Y.}~\bibnamefont {Asano}},\ }\bibfield  {title} {\bibinfo {title} {Numerical method for dc josephson current between d-wave superconductors},\ }\href {https://doi.org/10.1103/PhysRevB.63.052512} {\bibfield  {journal} {\bibinfo  {journal} {Phys. Rev. B}\ }\textbf {\bibinfo {volume} {63}},\ \bibinfo {pages} {052512} (\bibinfo {year} {2001})}\BibitemShut {NoStop}%
\bibitem [{\citenamefont {Song}\ \emph {et~al.}(2016)\citenamefont {Song}, \citenamefont {Liu}, \citenamefont {Liu}, \citenamefont {Li}, \citenamefont {Joynt}, \citenamefont {Sun},\ and\ \citenamefont {Xie}}]{song2016Quantum}%
  \BibitemOpen
  \bibfield  {author} {\bibinfo {author} {\bibfnamefont {J.}~\bibnamefont {Song}}, \bibinfo {author} {\bibfnamefont {H.}~\bibnamefont {Liu}}, \bibinfo {author} {\bibfnamefont {J.}~\bibnamefont {Liu}}, \bibinfo {author} {\bibfnamefont {Y.-X.}\ \bibnamefont {Li}}, \bibinfo {author} {\bibfnamefont {R.}~\bibnamefont {Joynt}}, \bibinfo {author} {\bibfnamefont {Q.-f.}\ \bibnamefont {Sun}},\ and\ \bibinfo {author} {\bibfnamefont {X.~C.}\ \bibnamefont {Xie}},\ }\bibfield  {title} {\bibinfo {title} {{Quantum interference in topological insulator Josephson junctions}},\ }\href {https://doi.org/10.1103/physrevb.93.195302} {\bibfield  {journal} {\bibinfo  {journal} {Phys. Rev. B}\ }\textbf {\bibinfo {volume} {93}},\ \bibinfo {pages} {195302} (\bibinfo {year} {2016})}\BibitemShut {NoStop}%
\bibitem [{\citenamefont {Golubov}\ \emph {et~al.}(2004)\citenamefont {Golubov}, \citenamefont {Kupriyanov},\ and\ \citenamefont {Il'ichev}}]{Golubov2004Current}%
  \BibitemOpen
  \bibfield  {author} {\bibinfo {author} {\bibfnamefont {A.~A.}\ \bibnamefont {Golubov}}, \bibinfo {author} {\bibfnamefont {M.~Y.}\ \bibnamefont {Kupriyanov}},\ and\ \bibinfo {author} {\bibfnamefont {E.}~\bibnamefont {Il'ichev}},\ }\bibfield  {title} {\bibinfo {title} {The current-phase relation in josephson junctions},\ }\href {https://doi.org/10.1103/RevModPhys.76.411} {\bibfield  {journal} {\bibinfo  {journal} {Rev. Mod. Phys.}\ }\textbf {\bibinfo {volume} {76}},\ \bibinfo {pages} {411} (\bibinfo {year} {2004})}\BibitemShut {NoStop}%
\bibitem [{\citenamefont {Li}\ and\ \citenamefont {Cheng}(2022)}]{Li2022Quantum}%
  \BibitemOpen
  \bibfield  {author} {\bibinfo {author} {\bibfnamefont {Y.-H.}\ \bibnamefont {Li}}\ and\ \bibinfo {author} {\bibfnamefont {R.}~\bibnamefont {Cheng}},\ }\bibfield  {title} {\bibinfo {title} {Quantum interference in a superconductor-mnbi2te4-superconductor josephson junction},\ }\href {https://doi.org/10.1103/PhysRevResearch.4.033227} {\bibfield  {journal} {\bibinfo  {journal} {Phys. Rev. Res.}\ }\textbf {\bibinfo {volume} {4}},\ \bibinfo {pages} {033227} (\bibinfo {year} {2022})}\BibitemShut {NoStop}%
\bibitem [{\citenamefont {Beenakker}(2024)}]{Beenakker2024Josephson}%
  \BibitemOpen
  \bibfield  {author} {\bibinfo {author} {\bibfnamefont {C.~W.~J.}\ \bibnamefont {Beenakker}},\ }\bibfield  {title} {\bibinfo {title} {Josephson effect in a junction coupled to an electron reservoir},\ }\href {https://doi.org/10.1063/5.0215522} {\bibfield  {journal} {\bibinfo  {journal} {Appl. Phys. Lett.}\ }\textbf {\bibinfo {volume} {125}},\ \bibinfo {pages} {122601} (\bibinfo {year} {2024})}\BibitemShut {NoStop}%
\bibitem [{\citenamefont {Oostinga}\ \emph {et~al.}(2008)\citenamefont {Oostinga}, \citenamefont {Heersche}, \citenamefont {Liu}, \citenamefont {Morpurgo},\ and\ \citenamefont {Vandersypen}}]{Oostinga2008}%
  \BibitemOpen
  \bibfield  {author} {\bibinfo {author} {\bibfnamefont {J.~B.}\ \bibnamefont {Oostinga}}, \bibinfo {author} {\bibfnamefont {H.~B.}\ \bibnamefont {Heersche}}, \bibinfo {author} {\bibfnamefont {X.}~\bibnamefont {Liu}}, \bibinfo {author} {\bibfnamefont {A.~F.}\ \bibnamefont {Morpurgo}},\ and\ \bibinfo {author} {\bibfnamefont {L.~M.~K.}\ \bibnamefont {Vandersypen}},\ }\bibfield  {title} {\bibinfo {title} {Gate-induced insulating state in bilayer graphene devices},\ }\href {https://doi.org/10.1038/nmat2082} {\bibfield  {journal} {\bibinfo  {journal} {Nat. Mater.}\ }\textbf {\bibinfo {volume} {7}},\ \bibinfo {pages} {151} (\bibinfo {year} {2008})}\BibitemShut {NoStop}%
\bibitem [{\citenamefont {Alden}\ \emph {et~al.}(2013)\citenamefont {Alden}, \citenamefont {Tsen}, \citenamefont {Huang}, \citenamefont {Hovden}, \citenamefont {Brown}, \citenamefont {Park}, \citenamefont {Muller},\ and\ \citenamefont {McEuen}}]{Alden2013}%
  \BibitemOpen
  \bibfield  {author} {\bibinfo {author} {\bibfnamefont {J.~S.}\ \bibnamefont {Alden}}, \bibinfo {author} {\bibfnamefont {A.~W.}\ \bibnamefont {Tsen}}, \bibinfo {author} {\bibfnamefont {P.~Y.}\ \bibnamefont {Huang}}, \bibinfo {author} {\bibfnamefont {R.}~\bibnamefont {Hovden}}, \bibinfo {author} {\bibfnamefont {L.}~\bibnamefont {Brown}}, \bibinfo {author} {\bibfnamefont {J.}~\bibnamefont {Park}}, \bibinfo {author} {\bibfnamefont {D.~A.}\ \bibnamefont {Muller}},\ and\ \bibinfo {author} {\bibfnamefont {P.~L.}\ \bibnamefont {McEuen}},\ }\bibfield  {title} {\bibinfo {title} {Strain solitons and topological defects in bilayer graphene},\ }\href {https://doi.org/10.1073/pnas.1309394110} {\bibfield  {journal} {\bibinfo  {journal} {Proc. Natl. Acad. Sci. U.S.A.}\ }\textbf {\bibinfo {volume} {110}},\ \bibinfo {pages} {11256} (\bibinfo {year} {2013})}\BibitemShut {NoStop}%
\bibitem [{\citenamefont {Qi}\ \emph {et~al.}(2025)\citenamefont {Qi}, \citenamefont {Chen}, \citenamefont {Song}, \citenamefont {Liu}, \citenamefont {He}, \citenamefont {Sun},\ and\ \citenamefont {Xie}}]{Qi2025EdgeSupercurrent}%
  \BibitemOpen
  \bibfield  {author} {\bibinfo {author} {\bibfnamefont {J.}~\bibnamefont {Qi}}, \bibinfo {author} {\bibfnamefont {C.-Z.}\ \bibnamefont {Chen}}, \bibinfo {author} {\bibfnamefont {J.}~\bibnamefont {Song}}, \bibinfo {author} {\bibfnamefont {J.}~\bibnamefont {Liu}}, \bibinfo {author} {\bibfnamefont {K.}~\bibnamefont {He}}, \bibinfo {author} {\bibfnamefont {Q.-F.}\ \bibnamefont {Sun}},\ and\ \bibinfo {author} {\bibfnamefont {X.~C.}\ \bibnamefont {Xie}},\ }\bibfield  {title} {\bibinfo {title} {Edge supercurrent in josephson junctions based on topological materials},\ }\href {https://doi.org/10.1007/s11433-024-2520-9} {\bibfield  {journal} {\bibinfo  {journal} {Sci. China Phys. Mech. Astron.}\ }\textbf {\bibinfo {volume} {68}},\ \bibinfo {pages} {227401} (\bibinfo {year} {2025})}\BibitemShut {NoStop}%
\bibitem [{end()}]{endnoteDWspacing}%
  \BibitemOpen
  \href@noop {} {}\bibinfo {note} {In Fig.~2(i), the maximum critical current is relatively small because, as the number of domain walls (DWs) increases while maintaining a constant $N_y$, the spacing between DWs decreases. This reduction in spacing increases self-inductance and scattering among DWs, resulting in a decrease in total current, which is actually lower than the current observed when DW = 2, 3, or 5. If the distance between adjacent DWs is kept constant, as in Fig.~4, the critical current would increase with the number of DWs.}\BibitemShut {Stop}%
\bibitem [{\citenamefont {Dynes}\ and\ \citenamefont {Fulton}(1971)}]{dynes1971supercurrent}%
  \BibitemOpen
  \bibfield  {author} {\bibinfo {author} {\bibfnamefont {R.~C.}\ \bibnamefont {Dynes}}\ and\ \bibinfo {author} {\bibfnamefont {T.~A.}\ \bibnamefont {Fulton}},\ }\bibfield  {title} {\bibinfo {title} {Supercurrent density distribution in josephson junctions},\ }\href {https://doi.org/10.1103/PhysRevB.3.3015} {\bibfield  {journal} {\bibinfo  {journal} {Phys. Rev. B}\ }\textbf {\bibinfo {volume} {3}},\ \bibinfo {pages} {3015} (\bibinfo {year} {1971})}\BibitemShut {NoStop}%
\bibitem [{\citenamefont {Cr{\'e}t{\'e}}\ \emph {et~al.}(2023)\citenamefont {Cr{\'e}t{\'e}}, \citenamefont {Menouni}, \citenamefont {Trastoy}, \citenamefont {Mesoraca}, \citenamefont {Kermorvant}, \citenamefont {Lema{\^i}tre}, \citenamefont {Marcilhac},\ and\ \citenamefont {Ulysse}}]{Crete2023Designing}%
  \BibitemOpen
  \bibfield  {author} {\bibinfo {author} {\bibfnamefont {D.-G.}\ \bibnamefont {Cr{\'e}t{\'e}}}, \bibinfo {author} {\bibfnamefont {S.}~\bibnamefont {Menouni}}, \bibinfo {author} {\bibfnamefont {J.}~\bibnamefont {Trastoy}}, \bibinfo {author} {\bibfnamefont {S.}~\bibnamefont {Mesoraca}}, \bibinfo {author} {\bibfnamefont {J.}~\bibnamefont {Kermorvant}}, \bibinfo {author} {\bibfnamefont {Y.}~\bibnamefont {Lema{\^i}tre}}, \bibinfo {author} {\bibfnamefont {B.}~\bibnamefont {Marcilhac}},\ and\ \bibinfo {author} {\bibfnamefont {C.}~\bibnamefont {Ulysse}},\ }\bibfield  {title} {\bibinfo {title} {Designing large two-dimensional arrays of {Josephson} junctions for {RF} magnetic field detection},\ }\href {https://doi.org/10.3390/electronics12153239} {\bibfield  {journal} {\bibinfo  {journal} {Electronics}\ }\textbf {\bibinfo {volume} {12}},\ \bibinfo {pages} {3239} (\bibinfo {year} {2023})}\BibitemShut {NoStop}%
\bibitem [{\citenamefont {Oppenl{\"a}nder}(2004)}]{Oppenlander2004Superconducting}%
  \BibitemOpen
  \bibfield  {author} {\bibinfo {author} {\bibfnamefont {J.}~\bibnamefont {Oppenl{\"a}nder}},\ }\bibfield  {title} {\bibinfo {title} {Superconducting quantum interference filters},\ }in\ \href {https://doi.org/10.1007/978-3-540-44838-9_52} {\emph {\bibinfo {booktitle} {Advances in Solid State Physics}}},\ Vol.~\bibinfo {volume} {43},\ \bibinfo {editor} {edited by\ \bibinfo {editor} {\bibfnamefont {B.}~\bibnamefont {Kramer}}}\ (\bibinfo  {publisher} {Springer},\ \bibinfo {address} {Berlin, Heidelberg},\ \bibinfo {year} {2004})\ Chap.~\bibinfo {chapter} {4}\BibitemShut {NoStop}%
\bibitem [{\citenamefont {Feynman}\ \emph {et~al.}(1965)\citenamefont {Feynman}, \citenamefont {Leighton},\ and\ \citenamefont {Sands}}]{Feynman1965Lectures}%
  \BibitemOpen
  \bibfield  {author} {\bibinfo {author} {\bibfnamefont {R.~P.}\ \bibnamefont {Feynman}}, \bibinfo {author} {\bibfnamefont {R.~B.}\ \bibnamefont {Leighton}},\ and\ \bibinfo {author} {\bibfnamefont {M.}~\bibnamefont {Sands}},\ }\bibfield  {title} {\bibinfo {title} {The feynman lectures on physics},\ }in\ \href@noop {} {\emph {\bibinfo {booktitle} {The Feynman Lectures on Physics}}},\ Vol.\ \bibinfo {volume} {III}\ (\bibinfo  {publisher} {Addison-Wesley},\ \bibinfo {address} {New York},\ \bibinfo {year} {1965})\ Chap.~\bibinfo {chapter} {21}, pp.\ \bibinfo {pages} {21--18}\BibitemShut {NoStop}%
\bibitem [{\citenamefont {Cho}\ \emph {et~al.}(2019)\citenamefont {Cho}, \citenamefont {Zhou}, \citenamefont {Khapaev},\ and\ \citenamefont {Cybart}}]{Cho2019Investigation}%
  \BibitemOpen
  \bibfield  {author} {\bibinfo {author} {\bibfnamefont {E.~Y.}\ \bibnamefont {Cho}}, \bibinfo {author} {\bibfnamefont {Y.~W.}\ \bibnamefont {Zhou}}, \bibinfo {author} {\bibfnamefont {M.~M.}\ \bibnamefont {Khapaev}},\ and\ \bibinfo {author} {\bibfnamefont {S.~A.}\ \bibnamefont {Cybart}},\ }\bibfield  {title} {\bibinfo {title} {Investigation of arrays of two-dimensional high-$t_c$ squids for optimization of electrical properties},\ }\href {https://doi.org/10.1109/TASC.2019.2904481} {\bibfield  {journal} {\bibinfo  {journal} {IEEE Trans. Appl. Supercond.}\ }\textbf {\bibinfo {volume} {29}},\ \bibinfo {pages} {1501605} (\bibinfo {year} {2019})}\BibitemShut {NoStop}%
\bibitem [{\citenamefont {Miller}\ \emph {et~al.}(1991)\citenamefont {Miller}, \citenamefont {Gunaratne}, \citenamefont {Huang},\ and\ \citenamefont {Golding}}]{miller1991Enhanced}%
  \BibitemOpen
  \bibfield  {author} {\bibinfo {author} {\bibfnamefont {J.~H.}\ \bibnamefont {Miller}}, \bibinfo {author} {\bibfnamefont {G.~H.}\ \bibnamefont {Gunaratne}}, \bibinfo {author} {\bibfnamefont {J.}~\bibnamefont {Huang}},\ and\ \bibinfo {author} {\bibfnamefont {T.~D.}\ \bibnamefont {Golding}},\ }\bibfield  {title} {\bibinfo {title} {{Enhanced quantum interference effects in parallel Josephson junction arrays}},\ }\href {https://doi.org/10.1063/1.105721} {\bibfield  {journal} {\bibinfo  {journal} {Appl. Phys. Lett.}\ }\textbf {\bibinfo {volume} {59}},\ \bibinfo {pages} {3330} (\bibinfo {year} {1991})}\BibitemShut {NoStop}%
\bibitem [{\citenamefont {Bauriedl}\ \emph {et~al.}(2022)\citenamefont {Bauriedl}, \citenamefont {B{\"a}uml}, \citenamefont {Fuchs}, \citenamefont {Baumgartner}, \citenamefont {Paulik}, \citenamefont {Bauer}, \citenamefont {Lin}, \citenamefont {Lupton}, \citenamefont {Taniguchi}, \citenamefont {Watanabe}, \citenamefont {Strunk},\ and\ \citenamefont {Paradiso}}]{Bauriedl2022Supercurrent}%
  \BibitemOpen
  \bibfield  {author} {\bibinfo {author} {\bibfnamefont {L.}~\bibnamefont {Bauriedl}}, \bibinfo {author} {\bibfnamefont {C.}~\bibnamefont {B{\"a}uml}}, \bibinfo {author} {\bibfnamefont {L.}~\bibnamefont {Fuchs}}, \bibinfo {author} {\bibfnamefont {C.}~\bibnamefont {Baumgartner}}, \bibinfo {author} {\bibfnamefont {N.}~\bibnamefont {Paulik}}, \bibinfo {author} {\bibfnamefont {J.~M.}\ \bibnamefont {Bauer}}, \bibinfo {author} {\bibfnamefont {K.-Q.}\ \bibnamefont {Lin}}, \bibinfo {author} {\bibfnamefont {J.~M.}\ \bibnamefont {Lupton}}, \bibinfo {author} {\bibfnamefont {T.}~\bibnamefont {Taniguchi}}, \bibinfo {author} {\bibfnamefont {K.}~\bibnamefont {Watanabe}}, \bibinfo {author} {\bibfnamefont {C.}~\bibnamefont {Strunk}},\ and\ \bibinfo {author} {\bibfnamefont {N.}~\bibnamefont {Paradiso}},\ }\bibfield  {title} {\bibinfo {title} {Supercurrent diode effect and magnetochiral anisotropy in few-layer nbse2},\ }\href {https://doi.org/10.1038/s41467-022-31954-5} {\bibfield  {journal} {\bibinfo  {journal} {Nat. Commun.}\ }\textbf {\bibinfo {volume} {13}},\ \bibinfo {pages} {4266} (\bibinfo {year} {2022})}\BibitemShut {NoStop}%
\bibitem [{\citenamefont {Pal}\ \emph {et~al.}(2022)\citenamefont {Pal}, \citenamefont {Chakraborty}, \citenamefont {Sivakumar}, \citenamefont {Anil}, \citenamefont {Sharma}, \citenamefont {Sharma}, \citenamefont {Gayen}, \citenamefont {Guptasarma},\ and\ \citenamefont {Sheet}}]{Pal2022Josephson}%
  \BibitemOpen
  \bibfield  {author} {\bibinfo {author} {\bibfnamefont {B.}~\bibnamefont {Pal}}, \bibinfo {author} {\bibfnamefont {A.}~\bibnamefont {Chakraborty}}, \bibinfo {author} {\bibfnamefont {P.~K.}\ \bibnamefont {Sivakumar}}, \bibinfo {author} {\bibfnamefont {M.~A.}\ \bibnamefont {Anil}}, \bibinfo {author} {\bibfnamefont {G.}~\bibnamefont {Sharma}}, \bibinfo {author} {\bibfnamefont {Y.}~\bibnamefont {Sharma}}, \bibinfo {author} {\bibfnamefont {S.}~\bibnamefont {Gayen}}, \bibinfo {author} {\bibfnamefont {P.}~\bibnamefont {Guptasarma}},\ and\ \bibinfo {author} {\bibfnamefont {G.}~\bibnamefont {Sheet}},\ }\bibfield  {title} {\bibinfo {title} {{Josephson} diode effect from {Cooper} pair momentum in a topological semimetal},\ }\href {https://doi.org/10.1038/s41567-022-01699-5} {\bibfield  {journal} {\bibinfo  {journal} {Nat. Phys.}\ }\textbf {\bibinfo {volume} {18}},\ \bibinfo {pages} {1228} (\bibinfo {year} {2022})}\BibitemShut {NoStop}%
\bibitem [{\citenamefont {Lu}\ \emph {et~al.}(2023)\citenamefont {Lu}, \citenamefont {Ikegaya}, \citenamefont {Burset}, \citenamefont {Tanaka},\ and\ \citenamefont {Nagaosa}}]{Lu2023Tunable}%
  \BibitemOpen
  \bibfield  {author} {\bibinfo {author} {\bibfnamefont {B.}~\bibnamefont {Lu}}, \bibinfo {author} {\bibfnamefont {S.}~\bibnamefont {Ikegaya}}, \bibinfo {author} {\bibfnamefont {P.}~\bibnamefont {Burset}}, \bibinfo {author} {\bibfnamefont {Y.}~\bibnamefont {Tanaka}},\ and\ \bibinfo {author} {\bibfnamefont {N.}~\bibnamefont {Nagaosa}},\ }\bibfield  {title} {\bibinfo {title} {Tunable josephson diode effect on the surface of topological insulators},\ }\href {https://doi.org/10.1103/PhysRevLett.131.096001} {\bibfield  {journal} {\bibinfo  {journal} {Phys. Rev. Lett.}\ }\textbf {\bibinfo {volume} {131}},\ \bibinfo {pages} {096001} (\bibinfo {year} {2023})}\BibitemShut {NoStop}%
\bibitem [{\citenamefont {Lin}\ \emph {et~al.}(2022)\citenamefont {Lin}, \citenamefont {Siriviboon}, \citenamefont {Scammell}, \citenamefont {Liu}, \citenamefont {Rhodes}, \citenamefont {Watanabe}, \citenamefont {Taniguchi}, \citenamefont {Hone}, \citenamefont {Scheurer},\ and\ \citenamefont {Li}}]{Lin2022Zero}%
  \BibitemOpen
  \bibfield  {author} {\bibinfo {author} {\bibfnamefont {J.-X.}\ \bibnamefont {Lin}}, \bibinfo {author} {\bibfnamefont {P.}~\bibnamefont {Siriviboon}}, \bibinfo {author} {\bibfnamefont {H.~D.}\ \bibnamefont {Scammell}}, \bibinfo {author} {\bibfnamefont {S.}~\bibnamefont {Liu}}, \bibinfo {author} {\bibfnamefont {D.}~\bibnamefont {Rhodes}}, \bibinfo {author} {\bibfnamefont {K.}~\bibnamefont {Watanabe}}, \bibinfo {author} {\bibfnamefont {T.}~\bibnamefont {Taniguchi}}, \bibinfo {author} {\bibfnamefont {J.}~\bibnamefont {Hone}}, \bibinfo {author} {\bibfnamefont {M.~S.}\ \bibnamefont {Scheurer}},\ and\ \bibinfo {author} {\bibfnamefont {J.~I.~A.}\ \bibnamefont {Li}},\ }\bibfield  {title} {\bibinfo {title} {Zero-field superconducting diode effect in small-twist-angle trilayer graphene},\ }\href {https://doi.org/10.1038/s41567-022-01700-1} {\bibfield  {journal} {\bibinfo  {journal} {Nat. Phys.}\ }\textbf {\bibinfo {volume} {18}},\ \bibinfo {pages} {1221} (\bibinfo {year} {2022})}\BibitemShut {NoStop}%
\bibitem [{\citenamefont {Hu}\ \emph {et~al.}(2023)\citenamefont {Hu}, \citenamefont {Sun}, \citenamefont {Xie},\ and\ \citenamefont {Law}}]{Hu2023Josephson}%
  \BibitemOpen
  \bibfield  {author} {\bibinfo {author} {\bibfnamefont {J.-X.}\ \bibnamefont {Hu}}, \bibinfo {author} {\bibfnamefont {Z.-T.}\ \bibnamefont {Sun}}, \bibinfo {author} {\bibfnamefont {Y.-M.}\ \bibnamefont {Xie}},\ and\ \bibinfo {author} {\bibfnamefont {K.~T.}\ \bibnamefont {Law}},\ }\bibfield  {title} {\bibinfo {title} {Josephson diode effect induced by valley polarization in twisted bilayer graphene},\ }\href {https://doi.org/10.1103/PhysRevLett.130.266003} {\bibfield  {journal} {\bibinfo  {journal} {Phys. Rev. Lett.}\ }\textbf {\bibinfo {volume} {130}},\ \bibinfo {pages} {266003} (\bibinfo {year} {2023})}\BibitemShut {NoStop}%
\bibitem [{\citenamefont {D\'{\i}ez-M\'erida}\ \emph {et~al.}(2023)\citenamefont {D\'{\i}ez-M\'erida}, \citenamefont {D\'{\i}ez-Carl\'on}, \citenamefont {Yang}, \citenamefont {Xie}, \citenamefont {Gao}, \citenamefont {Senior}, \citenamefont {Watanabe}, \citenamefont {Taniguchi}, \citenamefont {Lu}, \citenamefont {Higginbotham}, \citenamefont {Law},\ and\ \citenamefont {Efetov}}]{DiezMerida2023symmetry}%
  \BibitemOpen
  \bibfield  {author} {\bibinfo {author} {\bibfnamefont {J.}~\bibnamefont {D\'{\i}ez-M\'erida}}, \bibinfo {author} {\bibfnamefont {A.}~\bibnamefont {D\'{\i}ez-Carl\'on}}, \bibinfo {author} {\bibfnamefont {S.~Y.}\ \bibnamefont {Yang}}, \bibinfo {author} {\bibfnamefont {Y.-M.}\ \bibnamefont {Xie}}, \bibinfo {author} {\bibfnamefont {X.-J.}\ \bibnamefont {Gao}}, \bibinfo {author} {\bibfnamefont {J.}~\bibnamefont {Senior}}, \bibinfo {author} {\bibfnamefont {K.}~\bibnamefont {Watanabe}}, \bibinfo {author} {\bibfnamefont {T.}~\bibnamefont {Taniguchi}}, \bibinfo {author} {\bibfnamefont {X.}~\bibnamefont {Lu}}, \bibinfo {author} {\bibfnamefont {A.~P.}\ \bibnamefont {Higginbotham}}, \bibinfo {author} {\bibfnamefont {K.~T.}\ \bibnamefont {Law}},\ and\ \bibinfo {author} {\bibfnamefont {D.~K.}\ \bibnamefont {Efetov}},\ }\bibfield  {title} {\bibinfo {title} {Symmetry-broken josephson junctions and superconducting diodes in magic-angle twisted bilayer graphene},\ }\href {https://doi.org/10.1038/s41467-023-38005-7} {\bibfield  {journal} {\bibinfo  {journal} {Nat. Commun.}\ }\textbf {\bibinfo {volume} {14}},\ \bibinfo {pages} {2396} (\bibinfo {year} {2023})}\BibitemShut {NoStop}%
\bibitem [{\citenamefont {Shen}\ and\ \citenamefont {Zhang}(2025)}]{Shen2025Josephson}%
  \BibitemOpen
  \bibfield  {author} {\bibinfo {author} {\bibfnamefont {Q.-K.}\ \bibnamefont {Shen}}\ and\ \bibinfo {author} {\bibfnamefont {Y.}~\bibnamefont {Zhang}},\ }\bibfield  {title} {\bibinfo {title} {Josephson diodes induced by loop current states},\ }\href {https://doi.org/10.1103/PhysRevB.111.174515} {\bibfield  {journal} {\bibinfo  {journal} {Phys. Rev. B}\ }\textbf {\bibinfo {volume} {111}},\ \bibinfo {pages} {174515} (\bibinfo {year} {2025})}\BibitemShut {NoStop}%
\bibitem [{\citenamefont {Wieder}\ \emph {et~al.}(2015)\citenamefont {Wieder}, \citenamefont {Zhang},\ and\ \citenamefont {Kane}}]{Wieder2015Critical}%
  \BibitemOpen
  \bibfield  {author} {\bibinfo {author} {\bibfnamefont {B.~J.}\ \bibnamefont {Wieder}}, \bibinfo {author} {\bibfnamefont {F.}~\bibnamefont {Zhang}},\ and\ \bibinfo {author} {\bibfnamefont {C.~L.}\ \bibnamefont {Kane}},\ }\bibfield  {title} {\bibinfo {title} {Critical behavior of four-terminal conductance of bilayer graphene domain walls},\ }\href {https://doi.org/10.1103/PhysRevB.92.085425} {\bibfield  {journal} {\bibinfo  {journal} {Phys. Rev. B}\ }\textbf {\bibinfo {volume} {92}},\ \bibinfo {pages} {085425} (\bibinfo {year} {2015})}\BibitemShut {NoStop}%
\bibitem [{\citenamefont {Cheng}\ \emph {et~al.}(2018)\citenamefont {Cheng}, \citenamefont {Liu}, \citenamefont {Jiang}, \citenamefont {Sun},\ and\ \citenamefont {Xie}}]{2021sinedomainwall}%
  \BibitemOpen
  \bibfield  {author} {\bibinfo {author} {\bibfnamefont {S.-g.}\ \bibnamefont {Cheng}}, \bibinfo {author} {\bibfnamefont {H.}~\bibnamefont {Liu}}, \bibinfo {author} {\bibfnamefont {H.}~\bibnamefont {Jiang}}, \bibinfo {author} {\bibfnamefont {Q.-F.}\ \bibnamefont {Sun}},\ and\ \bibinfo {author} {\bibfnamefont {X.~C.}\ \bibnamefont {Xie}},\ }\bibfield  {title} {\bibinfo {title} {Manipulation and characterization of the valley-polarized topological kink states in graphene-based interferometers},\ }\href {https://doi.org/10.1103/PhysRevLett.121.156801} {\bibfield  {journal} {\bibinfo  {journal} {Phys. Rev. Lett.}\ }\textbf {\bibinfo {volume} {121}},\ \bibinfo {pages} {156801} (\bibinfo {year} {2018})}\BibitemShut {NoStop}%
\bibitem [{\citenamefont {Romeo}\ and\ \citenamefont {Di~Bartolomeo}(2023)}]{Romeo2023Experimental}%
  \BibitemOpen
  \bibfield  {author} {\bibinfo {author} {\bibfnamefont {F.}~\bibnamefont {Romeo}}\ and\ \bibinfo {author} {\bibfnamefont {A.}~\bibnamefont {Di~Bartolomeo}},\ }\bibfield  {title} {\bibinfo {title} {The experimental demonstration of a topological current divider},\ }\href {https://doi.org/10.1038/s41467-023-39503-4} {\bibfield  {journal} {\bibinfo  {journal} {Nat. Commun.}\ }\textbf {\bibinfo {volume} {14}},\ \bibinfo {pages} {3709} (\bibinfo {year} {2023})}\BibitemShut {NoStop}%
\bibitem [{\citenamefont {Fang}\ \emph {et~al.}(2021)\citenamefont {Fang}, \citenamefont {Yang}, \citenamefont {Yan}, \citenamefont {Guo},\ and\ \citenamefont {Sun}}]{Fang2021Thermal}%
  \BibitemOpen
  \bibfield  {author} {\bibinfo {author} {\bibfnamefont {J.-Y.}\ \bibnamefont {Fang}}, \bibinfo {author} {\bibfnamefont {N.-X.}\ \bibnamefont {Yang}}, \bibinfo {author} {\bibfnamefont {Q.}~\bibnamefont {Yan}}, \bibinfo {author} {\bibfnamefont {A.-M.}\ \bibnamefont {Guo}},\ and\ \bibinfo {author} {\bibfnamefont {Q.-F.}\ \bibnamefont {Sun}},\ }\bibfield  {title} {\bibinfo {title} {Thermal dissipation in the quantum {Hall} regime in graphene},\ }\href {https://doi.org/10.1103/PhysRevB.104.115411} {\bibfield  {journal} {\bibinfo  {journal} {Phys. Rev. B}\ }\textbf {\bibinfo {volume} {104}},\ \bibinfo {pages} {115411} (\bibinfo {year} {2021})}\BibitemShut {NoStop}%
\bibitem [{\citenamefont {Yan}\ \emph {et~al.}(2024)\citenamefont {Yan}, \citenamefont {Li}, \citenamefont {Jiang}, \citenamefont {Sun},\ and\ \citenamefont {Xie}}]{Yan2024Rules}%
  \BibitemOpen
  \bibfield  {author} {\bibinfo {author} {\bibfnamefont {Q.}~\bibnamefont {Yan}}, \bibinfo {author} {\bibfnamefont {H.}~\bibnamefont {Li}}, \bibinfo {author} {\bibfnamefont {H.}~\bibnamefont {Jiang}}, \bibinfo {author} {\bibfnamefont {Q.-F.}\ \bibnamefont {Sun}},\ and\ \bibinfo {author} {\bibfnamefont {X.~C.}\ \bibnamefont {Xie}},\ }\bibfield  {title} {\bibinfo {title} {Rules for dissipationless topotronics},\ }\href {https://doi.org/10.1126/sciadv.ado4756} {\bibfield  {journal} {\bibinfo  {journal} {Sci. Adv.}\ }\textbf {\bibinfo {volume} {10}},\ \bibinfo {pages} {eado4756} (\bibinfo {year} {2024})}\BibitemShut {NoStop}%
\bibitem [{\citenamefont {Ovchinnikov}\ \emph {et~al.}(2022)\citenamefont {Ovchinnikov}, \citenamefont {Cai}, \citenamefont {Lin}, \citenamefont {Fei}, \citenamefont {Liu}, \citenamefont {Cui}, \citenamefont {Cobden}, \citenamefont {Chu}, \citenamefont {Chang}, \citenamefont {Xiao}, \citenamefont {Yan},\ and\ \citenamefont {Xu}}]{Ovchinnikov2022Topological}%
  \BibitemOpen
  \bibfield  {author} {\bibinfo {author} {\bibfnamefont {D.}~\bibnamefont {Ovchinnikov}}, \bibinfo {author} {\bibfnamefont {J.}~\bibnamefont {Cai}}, \bibinfo {author} {\bibfnamefont {Z.}~\bibnamefont {Lin}}, \bibinfo {author} {\bibfnamefont {Z.}~\bibnamefont {Fei}}, \bibinfo {author} {\bibfnamefont {Z.}~\bibnamefont {Liu}}, \bibinfo {author} {\bibfnamefont {Y.-T.}\ \bibnamefont {Cui}}, \bibinfo {author} {\bibfnamefont {D.~H.}\ \bibnamefont {Cobden}}, \bibinfo {author} {\bibfnamefont {J.-H.}\ \bibnamefont {Chu}}, \bibinfo {author} {\bibfnamefont {C.-Z.}\ \bibnamefont {Chang}}, \bibinfo {author} {\bibfnamefont {D.}~\bibnamefont {Xiao}}, \bibinfo {author} {\bibfnamefont {J.}~\bibnamefont {Yan}},\ and\ \bibinfo {author} {\bibfnamefont {X.}~\bibnamefont {Xu}},\ }\bibfield  {title} {\bibinfo {title} {Topological current divider in a chern insulator junction},\ }\href {https://doi.org/10.1038/s41467-022-33645-7} {\bibfield  {journal} {\bibinfo  {journal} {Nat. Commun.}\ }\textbf {\bibinfo {volume} {13}},\ \bibinfo {pages} {5967} (\bibinfo {year} {2022})}\BibitemShut {NoStop}%
\bibitem [{\citenamefont {Hou}\ \emph {et~al.}(2020{\natexlab{b}})\citenamefont {Hou}, \citenamefont {Ren}, \citenamefont {Quan}, \citenamefont {Jung}, \citenamefont {Ren},\ and\ \citenamefont {Qiao}}]{Hou2020Valley}%
  \BibitemOpen
  \bibfield  {author} {\bibinfo {author} {\bibfnamefont {T.}~\bibnamefont {Hou}}, \bibinfo {author} {\bibfnamefont {Y.}~\bibnamefont {Ren}}, \bibinfo {author} {\bibfnamefont {Y.}~\bibnamefont {Quan}}, \bibinfo {author} {\bibfnamefont {J.}~\bibnamefont {Jung}}, \bibinfo {author} {\bibfnamefont {W.}~\bibnamefont {Ren}},\ and\ \bibinfo {author} {\bibfnamefont {Z.}~\bibnamefont {Qiao}},\ }\bibfield  {title} {\bibinfo {title} {Valley current splitter in minimally twisted bilayer graphene},\ }\href {https://doi.org/10.1103/PhysRevB.102.085433} {\bibfield  {journal} {\bibinfo  {journal} {Phys. Rev. B}\ }\textbf {\bibinfo {volume} {102}},\ \bibinfo {pages} {085433} (\bibinfo {year} {2020}{\natexlab{b}})}\BibitemShut {NoStop}%
\bibitem [{\citenamefont {Benchtaber}\ \emph {et~al.}(2021)\citenamefont {Benchtaber}, \citenamefont {S{\'a}nchez},\ and\ \citenamefont {Serra}}]{Benchtaber2021Scattering}%
  \BibitemOpen
  \bibfield  {author} {\bibinfo {author} {\bibfnamefont {N.}~\bibnamefont {Benchtaber}}, \bibinfo {author} {\bibfnamefont {D.}~\bibnamefont {S{\'a}nchez}},\ and\ \bibinfo {author} {\bibfnamefont {L.}~\bibnamefont {Serra}},\ }\bibfield  {title} {\bibinfo {title} {Scattering of topological kink-antikink states in bilayer graphene structures},\ }\href {https://doi.org/10.1103/PhysRevB.104.155303} {\bibfield  {journal} {\bibinfo  {journal} {Phys. Rev. B}\ }\textbf {\bibinfo {volume} {104}},\ \bibinfo {pages} {155303} (\bibinfo {year} {2021})}\BibitemShut {NoStop}%
\bibitem [{\citenamefont {Jiang}\ \emph {et~al.}(2020)\citenamefont {Jiang}, \citenamefont {Wang}, \citenamefont {Zhao}, \citenamefont {Crommie},\ and\ \citenamefont {Wang}}]{Jiang2020Soliton}%
  \BibitemOpen
  \bibfield  {author} {\bibinfo {author} {\bibfnamefont {L.}~\bibnamefont {Jiang}}, \bibinfo {author} {\bibfnamefont {S.}~\bibnamefont {Wang}}, \bibinfo {author} {\bibfnamefont {S.}~\bibnamefont {Zhao}}, \bibinfo {author} {\bibfnamefont {M.}~\bibnamefont {Crommie}},\ and\ \bibinfo {author} {\bibfnamefont {F.}~\bibnamefont {Wang}},\ }\bibfield  {title} {\bibinfo {title} {Soliton-dependent electronic transport across bilayer graphene domain wall},\ }\href {https://doi.org/10.1021/acs.nanolett.0c01911} {\bibfield  {journal} {\bibinfo  {journal} {Nano Lett.}\ }\textbf {\bibinfo {volume} {20}},\ \bibinfo {pages} {5936} (\bibinfo {year} {2020})}\BibitemShut {NoStop}%
\bibitem [{\citenamefont {Qiao}\ \emph {et~al.}(2014)\citenamefont {Qiao}, \citenamefont {Jung}, \citenamefont {Lin}, \citenamefont {Ren}, \citenamefont {MacDonald},\ and\ \citenamefont {Niu}}]{qiao2014Current}%
  \BibitemOpen
  \bibfield  {author} {\bibinfo {author} {\bibfnamefont {Z.}~\bibnamefont {Qiao}}, \bibinfo {author} {\bibfnamefont {J.}~\bibnamefont {Jung}}, \bibinfo {author} {\bibfnamefont {C.}~\bibnamefont {Lin}}, \bibinfo {author} {\bibfnamefont {Y.}~\bibnamefont {Ren}}, \bibinfo {author} {\bibfnamefont {A.~H.}\ \bibnamefont {MacDonald}},\ and\ \bibinfo {author} {\bibfnamefont {Q.}~\bibnamefont {Niu}},\ }\bibfield  {title} {\bibinfo {title} {Current partition at topological channel intersections},\ }\href {https://doi.org/10.1103/PhysRevLett.112.206601} {\bibfield  {journal} {\bibinfo  {journal} {Phys. Rev. Lett.}\ }\textbf {\bibinfo {volume} {112}},\ \bibinfo {pages} {206601} (\bibinfo {year} {2014})}\BibitemShut {NoStop}%
\end{thebibliography}%

\end{document}